\documentclass[reprint,
superscriptaddress,
 amsmath,amssymb,
 aps,
prb,
]{revtex4-2}

\usepackage{graphicx}% Include figure files
\usepackage{dcolumn}% Align table columns on decimal point
\usepackage{bm}% bold math
\usepackage{siunitx}
\usepackage{physics}
\usepackage{amsmath}
\usepackage{bm}
\usepackage{dsfont}
\usepackage[hidelinks]{hyperref}% add hypertext capabilities
\usepackage[nameinlink,capitalise]{cleveref}
\usepackage{csquotes}
\usepackage{color,soul}
\usepackage{bbm}
\usepackage{comment}
\usepackage[T1]{fontenc}

\newcommand{\red}[1]{\textcolor{red}{#1}}
\newcommand{\blue}[1]{\textcolor{blue}{#1}}

\begin{document}

\preprint{APS/123-QED}

%\title{Electron Shuttling in Silicon}% Force line breaks with \\
%\title{Simulation of valley splitting and spin shuttling in Si/SiGe heterostructures}
%\title{High fidelity spin shuttling in Si/SiGe heterostructures with random valley splitting due to alloy disorder}
%\title{Simulations of spin shuttling in Si/SiGe heterostructures with random valley splitting due to alloy disorder}
\title{Strategies for enhancing spin-shuttling fidelities in Si/SiGe quantum wells with random-alloy disorder}

\author{Merritt P. \surname{Losert}}
\altaffiliation{These two authors contributed equally to this work.}
\affiliation{Department of Physics, University of Wisconsin-Madison, Madison, Wisconsin 53706, USA}
\author{Max Oberl\"ander}
\altaffiliation{These two authors contributed equally to this work.}
\author{Julian D. \surname{Teske}}
\affiliation{%
 JARA-FIT Institute for Quantum Information, Forschungszentrum J\"ulich GmbH and RWTH Aachen University, 52074 Aachen, Germany
}%
\author{Mats \surname{Volmer}}
\affiliation{%
 JARA-FIT Institute for Quantum Information, Forschungszentrum J\"ulich GmbH and RWTH Aachen University, 52074 Aachen, Germany
}%
\author{Lars R. \surname{Schreiber}}
\affiliation{%
 JARA-FIT Institute for Quantum Information, Forschungszentrum J\"ulich GmbH and RWTH Aachen University, 52074 Aachen, Germany
}%
\affiliation{ARQUE Systems GmbH, 52074 Aachen, Germany}
\author{Hendrik Bluhm}
\altaffiliation{bluhm@physik.rwth-aachen.de}%
\affiliation{%
 JARA-FIT Institute for Quantum Information, Forschungszentrum J\"ulich GmbH and RWTH Aachen University, 52074 Aachen, Germany
}%
\affiliation{ARQUE Systems GmbH, 52074 Aachen, Germany}
\author{S. N. \surname{Coppersmith}}
\affiliation{School of Physics, University of New South Wales, Sydney, New South Wales 2052, Australia}
\author{Mark Friesen}
\altaffiliation{friesen@physics.wisc.edu}
\affiliation{Department of Physics, University of Wisconsin-Madison, Madison, Wisconsin 53706, USA}
%\affiliation{
% need to include julian's new address?
%}%
\date{\today}% It is always \today, today,

\begin{abstract}
Coherent coupling between distant qubits is needed for many scalable quantum computing schemes. In quantum dot systems, one proposal for long-distance coupling is to coherently transfer electron spins across a chip in a moving dot potential. 
Here, we use simulations to study challenges for spin shuttling in Si/SiGe heterostructures caused by the valley degree of freedom. 
We show that for devices with valley splitting dominated by alloy disorder, one can expect to encounter pockets of low valley splitting, given a long-enough shuttling path. At such locations, inter-valley tunneling leads to dephasing of the spin wavefunction, substantially reducing the shuttling fidelity.
We show how to mitigate this problem by modifying the heterostructure composition, or by varying the vertical electric field, the shuttling velocity, the shape and size of the dot, or the shuttling path. We further show that combinations of these strategies can reduce the shuttling infidelity by several orders of magnitude, putting shuttling fidelities sufficient for error correction within reach.
\end{abstract}

\begin{comment}
\begin{abstract}
This file currently contains the brainstorming for a possible publication of Merritt Losert, Max Oberländer, Julian Teske, Hendrik Bluhm and possibly more coauthors. The basis for the publication can contain the valley splitting calculations of Merritt and the quantum dynamical simulations of Max for the shuttling of an electron spin qubit in silicon based materials. The focus lies on the influence of the valley degree of freedom on the shuttling fidelity
\end{abstract}
\end{comment}
\maketitle

%\tableofcontents

\section{Introduction}

%\blue{MO: I'll begin to add comments of LS in the comments tab. MV has also sent some remarks, which I'll also paraphrase in the comments tab (and add exemplary resolutions, which may be either accepted or rejected). All other changes prior to uploading on arxiv are now accepted.}

Quantum dots formed in Si/SiGe heterostructures are a promising technology for scalable quantum computing.
Their strengths include the fact that silicon and germanium both have abundant zero-spin isotopes and are compatible with existing semiconductor fabrication technologies.
Moreover, one- and two-qubit gate fidelities in Si/SiGe have now exceeded 99\% 
\cite{noiri2022fast, Xue:2022p343, mills2022two}. 
%\cite{philips2022universal,mills2022two,takeda2022quantum,noiri2022fast,yoneda2018quantum}. 
However, scalable quantum computing also requires the coupling of distant qubits, which is not possible via short-ranged exchange interactions.
Coupling qubits beyond the nearest neighbor is therefore a topic of great current interest \cite{Trifunovic:2012p011006, Braakman:2013p432, Serina:2017p245422, Tosi:2017p450, samkharadze2018strong, Warren:2019p161303, Qiao:2021p017701, Holman:2021p137, Wang:2023p2208557}.
%\red{(Here, I think we should include more references, making sure to cite all the well-known groups. And below, for shuttling, we should include \emph{all} references, or nearly all.)}
Among other approaches \cite{samkharadze2018strong,landig2018coherent,sigillito2019coherent}, one promising strategy consists of physically shuttling the qubits over distances of one or more microns~\cite{Fujita:2017p22, mills2019shuttling, yoneda2021coherent, Jadot:2021p570, seidler2022conveyor, noiri2022shuttling, boter2022spiderweb, Kuenne2023, zwerver2023shuttling, langrock2023blueprint, struck2024spin, xue2024si}.

\begin{figure*}
    \centering
    \includegraphics[width=1.0\textwidth]{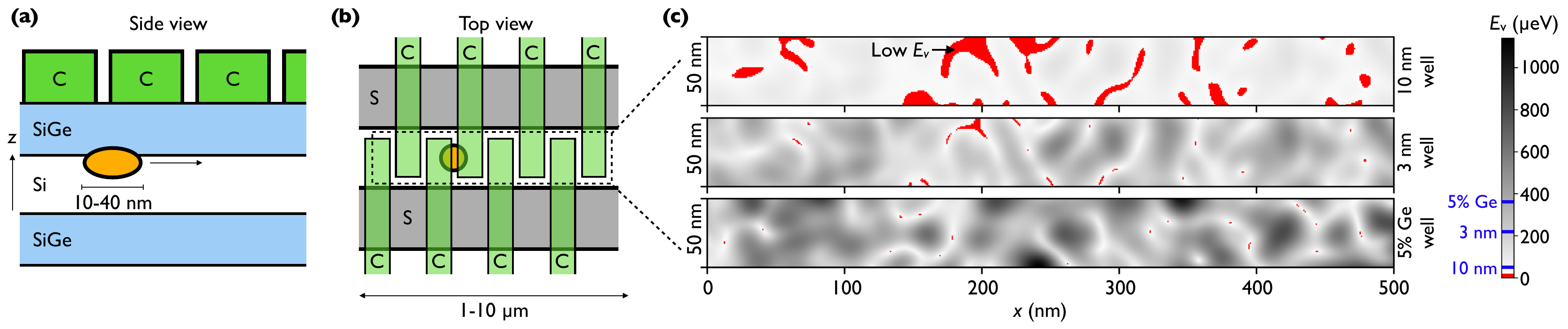}
    \caption{Schematic illustration of a conveyor-mode spin-shuttling device: (a) side view; (b) top view. A quantum dot in a Si quantum well is confined vertically by the SiGe barriers, and laterally by the electrostatic potential from screening gates (S) and stacked clavier (C) gates. Phase-shifted sinusoidal ac fields are applied to the C gates, yielding a moving potential minimum that carries the electron across the device~\cite{langrock2023blueprint}. 
    (c) Simulated valley-splitting landscapes (top view), as described in Sec.~\ref{sec:landscape}, for quantum wells of width \SI{10}{\nano\meter} (top) or \SI{3}{\nano\meter} (middle), and a \SI{10}{\nano\meter} quantum well with a uniform 5\% concentration of Ge (bottom). Results are shown for a \qtyproduct{500x50}{\nano\meter\squared} shuttling channel, where ``dangerous'' regions with $E_v\leq\SI{20}{\micro\electronvolt}$ are highlighted in red. We also indicate the average valley splittings for these three quantum wells on the colorbar.
    }
    \label{fig:intro_fig}
\end{figure*}

Two main shuttling schemes have been proposed for quantum dot qubits: the \textit{bucket-brigade} mode and the \textit{conveyor} mode. In the bucket-brigade mode, the electron is moved serially along a line of quantum dots, by modulating the detuning potential between nearest neighbors \cite{mills2019shuttling, langrock2023blueprint}.
In the conveyor mode, which is the topic of this work, phase-shifted sinusoidal potentials are applied to interleaved clavier gates along a channel defined by two screening gates, yielding a moving potential well that carries the electron across a device~\cite{langrock2023blueprint}.
A schematic illustration of a conveyor-mode device is shown in Figs.~\ref{fig:intro_fig}(a) and \ref{fig:intro_fig}(b).

Experimentally, high-fidelity charge shuttling of electrons has now been demonstrated in silicon over distances of $\sim \SI{20}{\micro\meter}$~\cite{xue2024si,seidler2022conveyor,mills2019shuttling}, while phase-coherent shuttling has been demonstrated over distances of $\sim \SI{400}{\nano\meter}$~\cite{struck2024spin} and over a cumulative distance of \SI{10}{\micro\meter} using four of the dots in a 6-dot device \cite{desmet2024high}.
Other experiments have demonstrated transfer across a double dot \cite{noiri2022shuttling}, and repeated transport of spins, without spin flips, through a short dot array \cite{zwerver2023shuttling}. 
However, an important question remains: what are the dominant limitations to coherent spin transfer over extended distances?

One of the main challenges for Si/SiGe qubits, which also affects spin shuttling, is the near-degeneracy of the two low-lying valley states \cite{zwanenburg2013review,burkard2023semiconductor}. 
The energy spacing between these states, known as the valley splitting, can be as large as $\SI{300}{\micro\electronvolt}$ in some cases, but can also be lower than $\SI{30}{\micro\electronvolt}$~\cite{borselli2011measurement, shi2011tunable, zajac2015reconfigurable, scarlino2017dressed, hollmann2020large, mi2017valleyspectroscopy, ferdous2018valley, mi2018landauzener, neyens2018substrate, borjans2019relaxation, oh2021cryogenfree}. 
Recent theoretical advancements have shown that, for current state-of-the-art heterostructures, random-alloy disorder in the quantum well barriers is the source of the wide valley-splitting variability \cite{paquelet2022atomic, losert2023practical, lima2023valley, pena2023utilizing}.
Crucially, large valley-splitting fluctuations are even observed in neighboring dots formed on the same chip \cite{mcjunkin2022sige, losert2023practical, lima2023valley2}.
These fluctuations inevitably lead to local regions with relatively low valley splittings \cite{volmer2023mapping}.
While a stationary spin qubit can potentially be shifted away from such a region~\cite{hollmann2020large, shi2011tunable,Dodson2022}, such fluctuations pose a greater challenge for spin shuttling experiments, where a quantum dot is rapidly shuttled across an extended and highly variable valley-splitting landscape.

Previous theoretical work has considered both the bucket-brigade and conveyor modes of shuttling~\cite{krzywda2024decoherence,bosco2024high,zhao2018coherent, ginzel2020spin,buonacorsi2020simulated,li2017intrinsic,langrock2023blueprint}.
Detailed models have been employed to study spin transfer; however at present, an extensive analysis of valley-splitting variations is still lacking for realistic heterostructures.
In the current work, we incorporate a realistic statistical description of valley-splitting variations caused by alloy disorder, and we apply this to a conveyor-mode shuttling process.
In particular, we apply the valley-splitting theory derived in Refs.~\cite{paquelet2022atomic, losert2023practical} to an effective dynamical model that captures the relevant effects in the parameter regime of interest. 
Our results show that large valley-splitting variations can cause leakage to the excited valley state, posing a significant challenge for conveyor-mode shuttling architectures where the valley splitting is dominated by alloy disorder.
To address this problem, we consider variations of the conventional quantum-well heterostructure that reduce the size of regions with low valley splitting.
We also propose and investigate several control strategies to suppress valley-state excitations. By combining these strategies, we show that shuttling fidelities can be improved by several orders of magnitude, enabling high-fidelity shuttling over distances of $\SI{10}{\micro m}$.
A summary description of these strategies is given in Table~\ref{tab:1}.

\begin{table}[b]
\caption{\label{tab:1}%
Summary of the heterostructure modifications and control strategies used here to boost the shuttling fidelity.
}
{\renewcommand{\arraystretch}{1.3}%
\begin{ruledtabular}
\begin{tabular}{p{25mm}|p{60mm}}
\textrm{Heterostructure modification}&
\textrm{Effect}\\
\colrule
Narrow quantum well & Increase wavefunction overlap  with Ge in the barrier regions.  \\
High-Ge quantum well & Increase wavefunction overlap with Ge in the quantum well. \\
%\end{tabular}
\colrule
%\begin{tabular}{p{25mm}|p{55mm}}
\textrm{Control Strategy}&
\textrm{Effect}\\
\colrule
Channel shift & Steer around low-$E_v$ regions using screening gates. (Most effective single strategy.) \\
$E_z$ modulation & Modulate vertical electric field to tune $E_v$. Requires high-Ge quantum wells, for which $E_v$ is more tunable. \\
Velocity modulation & Shuttle slowly in regions of low $E_v$ to reduce Landau-Zener excitations. (Most effective in combination with other strategies.) \\
Dot elongation & Elongate the dot in the shuttling direction to reduce the number of local minima and the magnitude of $E_v$ fluctuations. %Squeezing the dot transversally maintains same average $E_v$, and increases tunability in combination with channel shifting. 
(Most effective in combination with other strategies.)
\end{tabular}
\end{ruledtabular}}
\end{table}

The paper is organized as follows.
In Sec.~\ref{sec:vs_challenges}, we provide an intuitive explanation for the dangers of low valley splitting in shuttling experiments.
In Sec.~\ref{sec:vs_model}, we review the theory of valley splitting in the presence of alloy disorder. 
\Cref{sec:sim_dyn} describes the numerical model we use to simulate spin shuttling, and outlines possible tuning strategies that mitigate the effects of valley excitations. 
In Sec.~\ref{sec:results}, we describe the results of shuttling simulations across a \SI{10}{\micro\meter} device, while employing different mitigation strategies. 
In Sec.~\ref{sec:scalability}, we comment on the potential scalability of these schemes.
Finally in Sec.~\ref{sec:conclusion}, we summarize our findings and discuss future paths for spin shuttling.
Additional details are provided in the Appendices.

\section{Effects of valley leakage on spin shuttling} \label{sec:vs_challenges}

In this section, we outline the problems caused by small valley splittings in spin shuttling experiments, leaving mathematical and computational details for later sections.
The main problem is leakage outside the computational subspace caused by Landau-Zener transitions from the valley ground state to the valley excited state. 
Since the Land\'e~$g$~factor differs by a small amount $\delta g$ for these two states, valley excitations cause undesired spin rotations and dephasing.
For example, for an external magnetic field of 0.5-\SI{1}{\tesla} \cite{ferdous2018valley,ruskov2018electron}, the inter-valley Zeeman energy difference can be of order $\Delta E_B/h=\SI{10}{\mega\hertz}$, yielding spin rotation errors in about \SI{100}{\nano\second}. 
In principle, fast valley relaxation processes could mitigate this problem, as we discuss briefly in a later section.
%However, without applying special procedures to leverage this effect, it should not significantly improve the shuttling fidelity.
However, we do not attempt to leverage this effect here, and we simply consider valley excitations as errors.
%In this work, we therefore treat valley excitations as errors.

Recent theoretical work has identified two valley-splitting regimes: \textit{disordered} vs \textit{deterministic} \cite{losert2023practical}.
Since deterministically enhanced valley splittings are extremely difficult to achieve in the laboratory, we focus mainly on the disordered case here.

In the disordered regime, valley splitting variability is attributed mainly to alloy disorder, due to the electron overlap with the SiGe alloy.
As the Ge exposure increases (for example, by adding Ge to the quantum well), the variability and average value of the valley splitting $E_v$ also increase.
This trend is evident in Fig.~\ref{fig:intro_fig}(c), where we show results of valley-splitting simulations for three different quantum-well profiles, all in the disordered regime. 
Here in red, we highlight regions where $E_v <$ \SI{20}{\micro\electronvolt}, which pose a significant risk for shuttling at speeds of a few meters per second, due to enhanced Landau-Zener tunneling into the excited valley state. (See \cref{app:app_sweeps} for details.)
In the top panel, we consider a conventional quantum well of width \SI{10}{\nano\meter}, and top and bottom interface widths of \SI{1}{\nano\meter}.
Here, large portions of the device exhibit dangerously low valley splittings.
In the lower two panels, the Ge exposure is further enhanced: the middle panel shows a narrow \SI{3}{\nano\meter} quantum well with \SI{1}{\nano\meter} interfaces, while the lower panel shows a \SI{10}{\nano\meter} quantum well, with a uniform 5\% Ge concentration inside the quantum well. 
As consistent with our expectations, the size of the dangerous regions decreases in these examples.\
However, regions of low $E_v$ still persist.
Indeed, as shown in Sec.~\ref{sec:landscape}, such regions are statistically \textit{guaranteed} to exist in the disordered regime.
For a long-enough shuttling trajectory, a dot is very likely to encounter at least one such region, resulting in valley excitations and subsequent phase errors.
In the disordered regime, additional tuning strategies are therefore needed to achieve high shuttling fidelities, as described in Sec.~\ref{sec:results}.

In Sec.~\ref{sec:steps}, we also briefly consider the possibility of valley excitations in the deterministic regime.
In this case, the valley splitting can be made uniformly large, with no randomly small values.
Inter-valley leakage is then strongly reduced, even in the presence of interfacial disorder, so that high-fidelity spin shuttling is relatively easy to achieve.
(See Fig.~\ref{fig:fig_app_step}, below, and  \cref{app:perfect_step} for further details.)
However, as noted above, this regime is very difficult to reach experimentally, since it requires the presence of very abrupt features in the quantum-well profile (e.g., super-sharp interfaces, narrower than three atomic monolayers, or \SI{0.4}{\nano\meter}~\cite{losert2023practical}).
State-of-the-art growth processes have been shown to produce quantum well interfaces with characteristic widths of \SI{0.8}{\nano\meter}, which do not fall into the deterministic regime \cite{paquelet2022atomic}.

Recent work has therefore suggested alternative strategies for achieving consistently large valley splittings in Si/SiGe systems.
For example, shear-strain is known to affect valley splitting in Si systems \cite{Ungersboeck2007effect, Sverdlov2008subband, adelsberger2023valleyfree}, and recent theories have proposed to use shear strain to boost valley splittings in Si/SiGe quantum dots \cite{woods2023coupling}.
However, some of these techniques may likewise be challenging to implement in the laboratory. Consequently, we expect the great majority of Si/SiGe devices should fall into the disordered regime, which is more consistent with current fabrication techniques. We therefore focus mainly on the disordered regime in this work.

\section{Valley-splitting model} \label{sec:vs_model}

\subsection{Effective-mass theory}
In this work, we adopt an effective-mass envelope-function formalism to study the valley states, as outlined in Refs.~\cite{paquelet2022atomic, losert2023practical}.
In this model, the $\pm z$ valley wavefunctions are approximated by
\begin{equation}
    \psi_\pm (\mathbf{r}) = e^{\pm i k_0 z} \psi_\text{env}(\mathbf{r}) ,
\end{equation}
where $k_0 = 0.82( 2 \pi/a_0)$ is the position of the valley minimum in the first Brillouin zone and $a_0 = $~\SI{0.543}{\nano\meter} is the size of the conventional Si unit cell. 
For our purposes, the envelope function $\psi_\text{env}$ is approximately identical for both valleys.
The inter-valley coupling matrix element is defined as
\begin{equation} \label{eq:delta3d}
    \Delta = \langle \psi_- | H | \psi_+ \rangle = \int d^3r \; e^{-2 i k_0 z} U_\text{qw}(\mathbf{r}) |\psi_\text{env}(\mathbf{r})|^2 ,
\end{equation}
where the quantum-well confinement potential $U_\text{qw}$ is the only term in the Hamiltonian $H$ that significantly couples the two valley states.
Since the Ge concentration of the atomic layers along $\hat z$ plays an important role in determining the valley splitting, we may discretize the integral in \cref{eq:delta3d} as follows: 
\begin{equation} \label{eq:delta1d}
    \Delta = \frac{a_0}{4} \sum_l e^{-2 i k_0 z_l} U_\text{qw}(z_l) |\psi_\text{env}(z_l)|^2,
\end{equation}
where $l$ is the atomic layer index.
The resulting valley splitting is given by $E_v = 2|\Delta|$.

The principle observation of Refs.~\cite{losert2023practical, paquelet2022atomic} is that alloy disorder partially randomizes $\Delta$.
We therefore write $\Delta = \Delta_0 + \delta \Delta$, where $\Delta_0$ is the deterministic contribution to $\Delta$, and $\delta \Delta$ arises from random variations of the Ge concentration.
To compute these quantities, we define the Si concentration in layer $l$ as $X_l$, where $X_l$ is averaged over the area of a quantum dot, while the mean concentration $\bar X_l$ is averaged over the whole atomic layer.
In Appendix~\ref{app:vs_model_supp}, we provide more precise definitions of these quantities, and we describe the relation between $X_l$ and $U_\text{qw}(z_l)$, where the latter also depends on the dot size and location.

\begin{figure*}
    \centering
    \includegraphics[width=1.0\textwidth]{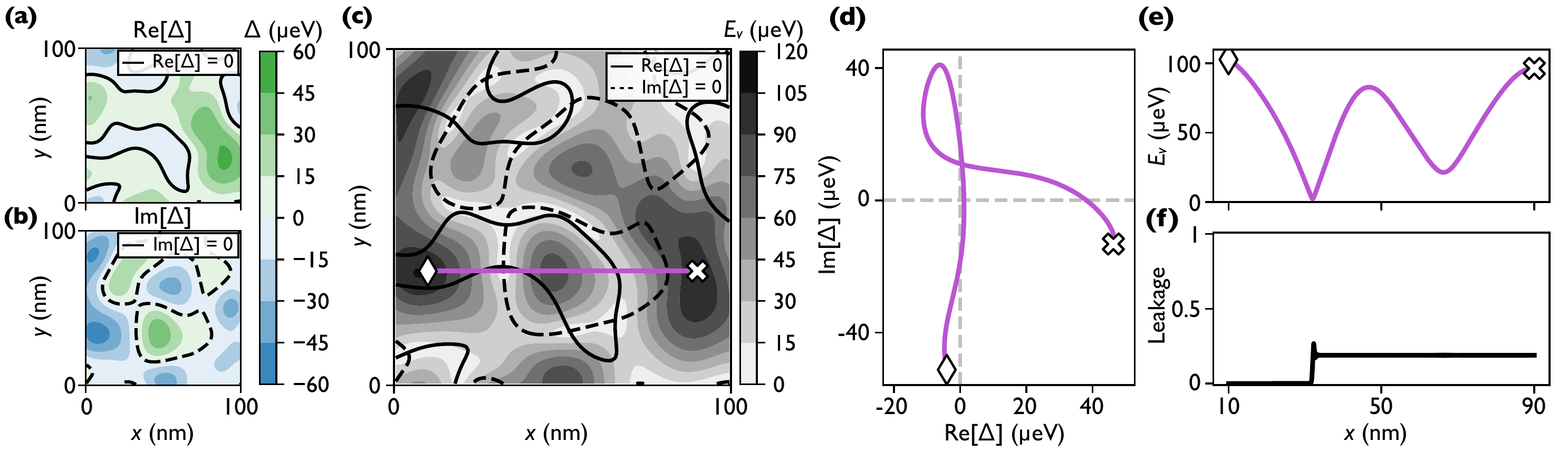}
    \caption{Points of vanishing valley splitting are topologically guaranteed to occur in the disordered regime. 
    (a),(b) Real and imaginary components of the inter-valley coupling $\Delta$, computed across a typical \qtyproduct{100x100}{\nano\meter\squared} region of heterostructure, as described in the main text. 
    Solid and dashed lines highlight contours where $\text{Re}[\Delta] = 0$ and $\text{Im}[\Delta] = 0$, respectively. 
    (c) The valley splitting $E_v = 2|\Delta|$, obtained from (a) and (b), with the same contours superimposed. 
    Correlations are observed between these contours and regions with low $E_v$.
    Intersections between the contours correspond to points of vanishing $E_v$. 
    A sample path across the heterostructure (purple line) passes nearby one such intersection. 
    (d) The inter-valley coupling $\Delta$, and (e) the valley splitting $E_v$, along the same path shown in (c). 
    (f) Leakage to the excited valley state caused by Landau-Zener tunneling, simulated for a dot traveling at a velocity of 1~m/s along the same path. 
    }
    \label{fig:theory_fig}
\end{figure*}

Following Refs.~\cite{losert2023practical,paquelet2022atomic}, we characterize the statistical properties of $\Delta$ in terms of the variance of $\delta \Delta$, given by
\begin{multline} \label{eq:vardelta}
    \sigma_\Delta^2 = \text{Var}[\delta \Delta] \\
    = \frac{1}{\pi a_x a_y} \left[ \frac{a_0^2 \Delta E_c}{8 (X_w - X_s)} \right]^2 \sum_l |\psi_\text{env}(z_l)|^4 \bar X_l (1 - \bar X_l).
\end{multline}
Here, we assume the dot is in the ground state of a lateral harmonic confinement potential with characteristic level spacings $\hbar \omega_{x(y)}$ along the principle $x$($y$) axes, and characteristic length scales $a_{x(y)} = \sqrt{\hbar / m_t \omega_{x(y)}}$, where $m_t = 0.19m_e$ is the transverse effective mass in Si.
The quantity $\Delta E_c$ defines the conduction-band energy offset between the strained quantum well and the strain-relaxed SiGe barriers. 
While the quantum well is conventionally formed of pure Si, we also consider more general situations where the well is formed of SiGe alloy, with a different composition than the barriers.
The variables $X_w$ and $X_s$ then indicate the Si concentrations of the quantum well and the SiGe substrate (i.e., the barriers), respectively. 
Generally, $\bar X_l$ transitions smoothly between $X_s$ and $X_w$, while $X_l$ deviates slightly from $\bar X_l$ due to local fluctuations within a quantum dot.

While full details of the derivation of Eq.~(\ref{eq:vardelta}) are left to Refs. \cite{paquelet2022atomic} and \cite{losert2023practical}, we can provide some physical intuition here. 
%First we note that $\Delta E_c$ depends linearly on the concentration differential $(X_w-X_s)$, to leading order; the ratio of these two quantities therefore represents a characteristic energy scale for any band offset. 
First, we note that the quantum-well confinement energy $U_\text{qw}$ is proportional to both $\Delta E_c$ and the Si concentration $X_l$, normalized by the concentration differential $(X_w - X_s)$.
Moreover, fluctuations related to alloy disorder are expected to vanish in the limit of very large dots, due to large-scale averaging, as captured by the ratio $a_0^2/a_xa_y$. 
When characterizing the effects of alloy disorder, dominant contributions arise from layers with higher wavefunction density. 
(Note that the fourth power of the wavefunction is commonly observed in standard-deviation calculations.) 
Finally, we note that SiGe alloy disorder vanishes in either limit, $\overline X_l \rightarrow 0$ or 1, resulting in the scaling factor $\overline X_l(1-\overline X_l)$.

%Since the dot is finite in extent, the quantum well potential at layer $l$ [$U_\text{qw}(z_l)$ in Eq.~(\ref{eq:delta1d})] is determined from a sum over a finite number of atoms. 
%Since the Ge atoms will be randomly distributed throughout the heterostructure, there is some uncertainty in this sum, leading to the variance described in Eq.~(\ref{eq:vardelta}). 
%Moreover, this uncertainty is larger for layers with Ge concentrations closer to 50\%, so the high-Ge layers in the heterostructure dominate the sum in Eq.~(\ref{eq:vardelta}).}

From Eq.~(\ref{eq:delta1d}), we note that $\Delta$ is a complex quantity, which can be decomposed into its real and imaginary components: $\Delta = \Delta_R + i \Delta_I$.
Under realistic assumptions about the width of the quantum-well interface, $\Delta_R$ and $\Delta_I$ are well described here as Gaussian random variables~\cite{losert2023practical}, each having a variance of $\sigma_\Delta^2 / 2$. 
As we show in the following section, this property leads to the existence of regions of arbitrarily small $E_v$.

Finally we note that Eq.~(\ref{eq:vardelta}) allows us to directly characterize the valley splitting of a given heterostructure as deterministically enhanced vs disordered, based on the crossover between these two regimes, which occurs at $\sqrt{\pi}\sigma_\Delta=2|\Delta_0|$~\cite{losert2023practical}. 
In the deterministic regime, we observe $|\Delta_0| > |\delta \Delta|$ with high probability, and an average valley splitting of $\bar E_v \approx 2|\Delta_0|$. 
(For example, quantum wells with ultra-sharp interfaces exhibit such behavior.)
In contrast, in the disordered regime, we find $|\Delta_0| < |\delta \Delta|$ with high probability, and \cite{losert2023practical}
\begin{equation} \label{eq:disordered_avg_Ev}
    \bar E_v \approx \sqrt{\pi} \sigma_\Delta.
\end{equation}
For conventional heterostructures, like those considered in this work, typical interfaces are not ultra-sharp, and the valley splitting falls into the disordered regime.

\subsection{Valley-coupling landscape and excitations} \label{sec:landscape}

As described above, in the disordered regime, the real and imaginary components of $\Delta$ are independent Gaussian random fields.
We now show that this guarantees the existence of regions with arbitrarily small $E_v$, scattered across a heterostructure.
Figures~\ref{fig:theory_fig}(a) and \ref{fig:theory_fig}(b) illustrate typical instances of $\Re[\Delta]$ and $\Im[\Delta]$ for a 100$\times$\SI{100}{\nano\meter\squared} lateral region of a device.
To compute these landscapes, the heterostructure is modeled atomistically by assigning each atom in the crystal lattice as either Si or Ge.
The probability of choosing Si at a given lattice site in layer $l$ is given by $\bar X_l$.
We then perform the one-dimensional (1D) summation in Eq.~(\ref{eq:delta1d}) via the following procedure.
First we compute the local Si concentration $X_l$ by performing the weighted average described in Eq.~(\ref{eq:deltal}), for a dot with orbital energies $\hbar \omega_x = \hbar \omega_y = $ \SI{2}{\milli\electronvolt}, centered at $(x_0, y_0)$.
Next, we use Eq.~(\ref{eq:qw_pot}) to convert the Si concentration profile $X_l$ to a quantum-well confinement potential $U_\text{qw}(z_l)$.
$\psi_\text{env}(z_l)$ is then computed from $U_\text{qw}(z_l)$ by solving a discretized Schr\"odinger equation, as described in \cref{app:vs_model_supp}.
Equation~(\ref{eq:delta1d}) then gives $\Delta$ as a function of dot position $(x_0, y_0)$, yielding the real and imaginary components shown in Figs.~\ref{fig:theory_fig}(a) and \ref{fig:theory_fig}(b).
The corresponding valley splitting $E_v = 2|\Delta|$ is plotted in Fig.~\ref{fig:theory_fig}(c).

In Figs.~\ref{fig:theory_fig}(a) and \ref{fig:theory_fig}(b), we highlight contours where $\Re[\Delta] = 0$ and $\Im[\Delta] = 0$.
The same contours are also plotted in Fig.~\ref{fig:theory_fig}(c), where they are seen to correlate with regions of low valley splitting.
Points where the contours intersect correspond to zero valley splitting.
Importantly, such points are randomly distributed across the heterostructure and are guaranteed to exist in the disordered regime.
Their spatial distribution is determined by the dot size, and in the disordered regime, we note that this distribution does not depend on the average valley splitting.
Thus, distributions with similar topologies are observed in systems with large average $E_v$.

\begin{figure*}
    \centering
    \includegraphics[width=0.8\textwidth]{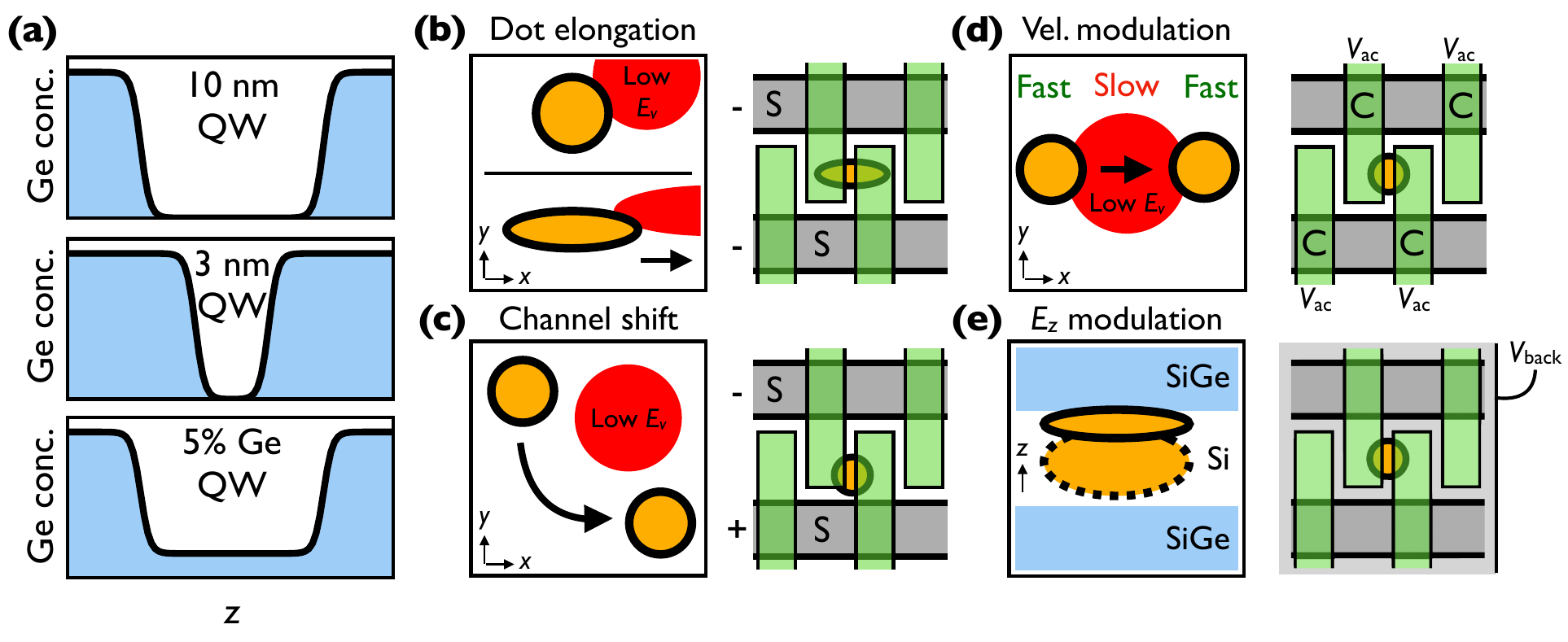}
    \caption{Overview of the heterostructure tuning strategies considered in this work. 
(a) Three different quantum wells, including a conventional \SI{10}{\nano\meter} Si quantum well, a narrow \SI{3}{\nano\meter} Si quantum well, and a \SI{10}{\nano\meter} quantum well with a uniform 5\% Ge concentration in the middle of the well. 
(b)-(e) Four strategies proposed for avoiding valley excitations: 
(b) elongating the dot along the shuttling direction while squeezing the dot in the transverse direction, by applying appropriate voltages to screening gates (S); 
(c) shifting the position of the dot in the channel to avoid low-$E_v$ regions, by modulating the voltages on S gates; 
(d) slowing the dot velocity near $E_v$ minima, by manipulating the ac voltages $V_\text{ac}$ applied to the clavier gates (C); 
(e) modulating the vertical field $E_z$, e.g., by introducing a back gate ($V_\text{back}$).
    }
    \label{fig:model_fig}
\end{figure*}%

We note that a recent spin shuttling experiment has confirmed the presence of such regions of low valley splitting, scattered throughout a 2D heterostructure \cite{volmer2023mapping}, while another shuttling experiment has demonstrated high-fidelity shuttling over a similar length scale, with no indication of low $E_v$ \cite{desmet2024high}.
Given the random nature of the valley splitting landscape, we expect larger-scale shuttling and $E_v$-mapping experiments to provide further clarity on the presence of these valley minima.
We explore the statistics of low-$E_v$ locations further in Appendix~\ref{app:min_Ev_characterization}.

The regions of low valley splitting near $\Delta\approx 0$ are dangerous for spin shuttling, because the electron can potentially tunnel into the excited valley state via a Landau-Zener process, leading to shuttling errors.
This process is illustrated in Figs.~\ref{fig:theory_fig}(c)-\ref{fig:theory_fig}(f).
First in Fig.~\ref{fig:theory_fig}(c), we highlight a shuttling path in purple that passes through a region of low $E_v$. 
In Fig.~\ref{fig:theory_fig}(d), the inter-valley coupling $\Delta$ is plotted along the same path, where it passes very near the origin of the complex plane, $\Delta = 0$.
The valley splitting, $E_v = 2|\Delta|$, is also shown along this path in Fig.~\ref{fig:theory_fig}(e).
In Fig.~\ref{fig:theory_fig}(f), we solve the dynamical evolution of the shuttling electron with regards to the two valley states, assuming the valley splitting shown in Fig.~\ref{fig:theory_fig}(e), using the methods described in Sec.~\ref{sec:sim_dyn}, below.
Specifically, we plot the leakage into the excited valley state, observing a sudden jump near the $\Delta$ minimum, caused by Landau-Zener tunneling.
We expect a similar jump in the shuttling infidelity whenever a shuttling path passes through a region of low valley splitting.
We also note that valley excitations across an avoided crossing require a fluctuating valley phase, in addition to a low valley splitting.
(If the valley phase were constant across a region of low $E_v$, no excitation would occur.)
However, like the valley splitting, the valley phase is also spatially random, as determined by the microscopic alloy disorder at the location of the dot. 
So, we generally expect some degree of valley excitation at these avoided crossings, and our simulations fully account for the effects of valley phase variability.

In the above discussion, we have ignored effects like strain fluctuations and valley-orbit interactions that lift the valley-state degeneracy near points of low $E_v$.
However, these effects are weak, and we expect leakage induced by Landau-Zener processes to remain prevalent in real devices.

Finally, we note that the atomistic method for generating valley-splitting landscapes, described above, is inefficient for determining large-scale statistical properties.
To make this process more efficient, we turn instead to a statistical assignment of valley splittings, using the methods described in \cref{app:random_fields}.
This assignment makes use of the fact that the real and imaginary components of $\Delta$ are Gaussian random variables.
We note that, to a very good approximation, in the disordered regime, the center of these Gaussian distributions is given by $|\Delta_0|\approx 0$~\cite{losert2023practical}.
The full statistical description of $\Delta$ then requires one additional piece of information: the spatial two-point covariance functions for $\text{Re}[\Delta]$ and $\text{Im}[\Delta]$, which were obtained in Ref.~\cite{losert2023practical} as
\begin{multline} \label{eq:2point}
    \text{Cov}(\Re [\Delta ], \Re [\Delta']) = \text{Cov}(\Im [\Delta], \Im [\Delta']) \\ = \frac{1}{2}e^{-\delta_x^2 / 2a_x^2  - \delta_y^2 /2 a_y^2} \sigma_\Delta^2.
\end{multline}
Here, $\Delta$ and $\Delta'$ are separated by the spatial vector $(\delta_x, \delta_y)$ in the $xy$-plane.
Obtaining valley-coupling landscapes with this method yields results like those shown
 in Fig.~\ref{fig:intro_fig}(c), which exhibit the correct statistical properties.

\section{Simulating quantum dynamics}\label{sec:sim_dyn}

\subsection{Physical device}

Figures~\ref{fig:intro_fig}(a) and \ref{fig:intro_fig}(b) schematically illustrate the devices we consider in this work.
During conveyor-mode operation, oscillating voltages are applied to the clavier gates, to produce a moving potential pocket capable of carrying an electron across the device~\cite{seidler2022conveyor}.
Unless otherwise specified, we model this potential pocket as an isotropic harmonic confinement potential with orbital splittings $\hbar \omega_x = \hbar \omega_y = $ \SI{2}{\milli\electronvolt}, and we assume the pocket moves with constant speed along the shuttling trajectory.
The clavier gates generate a vertical electric field that squeezes the electron against the top quantum-well interface.
In some cases, we assume this field can be tuned, for example, by including a back-gate.
However, unless otherwise specified, we consider a fixed vertical field of $E_z = $ \SI{5}{\milli\volt\per\nano\meter}.
In Sec.~\ref{subsec:sim_dyn_tuning}, we consider several additional tuning capabilities that allow us to mitigate the effects of low valley splitting.
These include the ability to vary the shuttling velocity, the position of the electron transverse to the shuttling trajectory, and the dot shape (e.g., with $\omega_x\neq\omega_y$).
In all cases, we consider a total shuttle length of \SI{10}{\micro\meter}, as consistent with a recent architecture proposal involving a medium-range coupler~\cite{Kuenne2023}.

We also consider three types of quantum wells, as illustrated in \cref{fig:model_fig}. 
These include a conventional \SI{10}{\nano\meter} quantum well, and two other wells proposed to give larger average valley splittings: a narrow \SI{3}{\nano\meter} quantum well and a structure containing a uniform 5\% Ge concentration inside the quantum well. 
These are meant to illustrate a range of realistic heterostructures.
We model the quantum-well interfaces using sigmoid functions, as described in \cref{app:vs_model_supp}, and assume interface widths of $\lambda = $ \SI{1}{\nano\meter}, unless otherwise specified.
(Note that our current goal is not to optimize heterostructure parameters, but to characterize schemes for mitigating the effects of small valley splittings.)
As previously noted, the deterministic contribution to the inter-valley coupling can be safely ignored in these heterostructures, with $\Delta_0\approx 0$, since they fall into the disordered regime. 
The key difference between the heterostructures is therefore their $\sigma_\Delta$ values, which are related to the average valley splittings through \cref{eq:disordered_avg_Ev}.
For the heterostructures described above, we obtain the average values $\langle E_v({\SI{10}{\nano\meter}})\rangle\approx\SI{50}{\micro\electronvolt}$, $\langle E_v({\SI{3}{\nano\meter}})\rangle\approx\SI{220}{\micro\electronvolt}$, and $\langle E_v({\mathrm{5\%}})\rangle\approx\SI{360}{\micro\electronvolt}$, respectively.
For our dynamical shuttling simulations, we generate many random valley-splitting landscapes, as described in Sec.~\ref{sec:landscape}, obtaining results like those shown in Fig.~\ref{fig:intro_fig} for the three different heterostructures.

\subsection{Spin-shuttling model}

\subsubsection{Hamiltonian}

We employ a minimal model to investigate decoherence during shuttling.
The model is comprised of two spin and two valley states.
In particular, we ignore the presence of orbitally excited states, which relax quickly and are sufficiently separated in energy (1-2~meV) that they play a much smaller role than the valley excited state~\cite{langrock2023blueprint}.
Here, we first present the model, then provide discussion of some assumptions built into it.
The model is given by
\begin{equation}
H=\frac{E_B}{2}\sigma_z+\bm{\Delta}(x)\cdot\bm{\tau}+\frac{\Delta E_B}{4}[\vu{n}_\Delta(x)\cdot\bm{\tau}]\otimes\sigma_z ,
\label{eq:hamiltonian}
\end{equation}
where $E_B=g\mu_BB$ is the Zeeman energy, $g\approx 2$ is the Land\'{e} $g$ factor, $\mu_B$ is the Bohr magneton, $B$ is the magnetic field along the spin quantization axis, which does not necessarily coincide with the crystallographic $z$ axis, $\Delta E_B= (\delta g)\mu_BB$ is the difference in Zeeman splittings for the two valley states, \mbox{$\bm{\Delta} = (\Delta_R,\Delta_I)^T$} is the inter-valley coupling (which varies from location to location), \mbox{$\vu{n}_\Delta=\bm{\Delta}/\abs{\bm{\Delta}}$} denotes the valley quantization axis, and $\sigma_z$ and $\bm{\tau}=(\tau_x,\tau_y)^T$ are Pauli matrices acting on the spin- and valley-spaces, respectively. 

Several comments are in order for Eq.~(\ref{eq:hamiltonian}).
First, we note that the $\Delta E_B$ term ignores any dependence on the magnitude or angular orientation of the magnetic field, as well as local variations of the applied electric field, all of which can affect $\delta g$.
Moreover, we note that $\delta g$ also varies slightly by location, due to interface steps and random-alloy disorder~\cite{Woods:2023p035418, ruskov2018electron, ferdous2018valley}.
However, we observe almost no dependence of our results on $\Delta E_B$ (see Appendix~\ref{app:vary_k}), and we therefore set it here to a fixed value of $\SI{10}{\mega\hertz}$, reflecting a typical difference in spin-resonance frequencies between the two valley states, as consistent with experimental observations \cite{ferdous2018valley,ruskov2018electron}. 

Second, we note that the model, above, does not include dephasing or relaxation effects, which we now discuss briefly. 
One relaxation process that could affect shuttling in silicon is the spin-valley hot spot, at which the Zeeman and valley energy splittings are nearly equal, giving rise to fast spin relaxation~\cite{hollmann2020large}. 
An electron could potentially encounter many such hot spots when traversing a wildly varying landscape of valley splittings. 
We could suppress the effects of these hot spots by shuttling past them as quickly as possible; however, this has other potential pitfalls.
A simpler approach is to reduce the number of hot spots by operating at low magnetic fields where the Zeeman splitting is much smaller than the average valley splitting \cite{langrock2023blueprint}.
In this work, we consider low external fields of $B=\SI{50}{\milli\tesla}$~\cite{huang2023highfidelity}, which moves the hot spots to below \SI{10}{\micro\electronvolt}.
Although occasional hot spots are still encountered in this regime, Landau-Zener processes are also present, and since they are also detrimental to the shuttling fidelity, it reasonable to ignore the hot spots in favor of Landau-Zener processes.

Spin dephasing of the shuttling electron occurs over a time scale of $T_2^*$, due to the presence of charge or magnetic noise~\cite{langrock2023blueprint}.
In Sec.~\ref{sec:transport_velocity}, we treat these effects phenomenologically, finding that the presence of both dephasing and leakage suggests that there will be an optimal shuttling speed.

Fast valley relaxation processes present interesting opportunities for solving the valley-state leakage problem, which we will investigate in a future publication.
In the present work, we note that experimental measurements of valley relaxation are scarce, but likely of order $\SI{10}{\milli\second}$ for valley splittings of order $\SI{50}{\micro\electronvolt}$ \cite{penthorn2020direct}, which is several orders of magnitude slower than dephasing, and therefore irrelevant.
We note that these measurements were taken in static systems, and there may exist complications for dynamic, shuttled quantum dots.
On the other hand, valley lifetimes scale as the inverse-fifth power of the valley splitting~\cite{langrock2023blueprint, yang2013spin}, so valley and dephasing timescales could become comparable for very large valley splittings of order $\SI{500}{\micro\electronvolt}$ (assuming $\Delta E_B/h=\SI{10}{\mega\hertz}$).
Such large valley splittings may exist in certain heterostructures, but are unlikely to be widespread across a device.
In this work we simply take the conservative approach of assuming no inter-valley relaxation; any nonzero valley relaxation would improve shuttling fidelities beyond what is described here.

\subsubsection{Fidelity metrics} \label{sec:metrics}
 
To quantify the fidelity of our shuttling simulations, we compute the process fidelity, defined as~\cite{wood2018quantification}
\begin{equation}
	F_{\mathrm{process}}(U)=\frac{1}{d_1^2}\abs{\tr{V^\dagger U_{\mathrm{trunc}}}}^2
 \label{eq:fidelity}\,,
\end{equation}
where $V$ is the target unitary in the spin subspace, $U=U_{\mathds{B}\rightarrow\mathds{B}^\prime}U_t$ is the full evolution operator, including the effects of non-adiabatic evolution, $U_{\mathds{B}\rightarrow\mathds{B}^\prime}$ is the transformation matrix from the initial to the instantaneous eigenbasis, $U_t$ is the evolution operator in the initial eigenbasis of our model, and the subscript `trunc’ denotes truncation to the two-dimensional ($d_1=2$) spin subspace of the instantaneous valley ground state. 
Equation~(\ref{eq:fidelity}) compares the evolution of a real shuttling process to an ideal, adiabatic process, while accounting for leakage errors, which can be independently quantified as \cite{wood2018quantification}
\begin{equation}\label{eq:metric_leak}
L=1-\tr{U_{\mathrm{trunc}}^\dagger U_{\mathrm{trunc}}}/d_1\,.
\end{equation}

When applying Eq.~(\ref{eq:fidelity}), we note that $g$-factor fluctuations cause random phases to accumulate during shuttling, even when the system remains in the ground state.
Since such fluctuations are static, they can be compensated in experiments.
It is therefore reasonable to remove these phase shifts from our fidelity estimates.
We do this here by setting $V=\mathds{1}$. 
We then apply a virtual $z$ rotation to $U_\text{trunc}$ and choose the phase of this rotation to maximize the shuttling fidelity.
On the other hand, phase differences between the ground and excited valley states represent true dephasing errors, and cannot be removed.
However, as no valley relaxation is assumed, these errors only directly affect $U_\text{trunc}$ through a (weak) second-order Landau-Zener process, involving valley excitation followed by de-excitation.
Thus, although leakage formally sets a lower bound on the infidelity (defined as $I=1-F$), to a good approximation, we find that $I\approx L$.
Our four-level shuttling fidelity calculations could therefore be replaced by a two-level problem involving just the valley levels.
For better accuracy, we still perform four-level calculations using Eq.~(\ref{eq:fidelity}); however, we expect leakage to be the dominant source of infidelity.

\subsubsection{Computational framework}

The following computational procedure is used in our simulations.
First, we calculate the time evolution of \cref{eq:hamiltonian} using an adapted version of the Python-based software package \texttt{qopt}, described in \cite{teske2022qopt}.
The total propagator of the time evolution is calculated by splitting up the matrix exponential into a product of piecewise-constant Hamiltonians with appropriately small time steps. 
These time steps are chosen in the range of 2-\SI{4}{\pico\second}, depending on the quantum well and the shuttling velocity, to achieve sufficient convergence of the final results.
We check the time step used against test cases
with steps as small as \SI{0.5}{\pico\second}, which indicate that the shape
and the median of the distribution have both converged well forthe coarser time step. The main computational back-end
used to calculate the propagators is given by the JAX
package~\cite{jax2018github} (with 64-bit precision enabled), and its functions \texttt{jax.scipy.linalg.expm} and \texttt{jax.numpy.dot}. 
After obtaining the propagator, the fidelity is computed from \cref{eq:fidelity} and post-optimized as described above, using the Python routine \texttt{jax.scipy.optimize.minimize} to perform phase calibration.

\subsection{Tuning strategies}\label{subsec:sim_dyn_tuning}

In Figs.~\ref{fig:model_fig}(b)-\ref{fig:model_fig}(e), we illustrate the four tuning strategies used in this work to suppress valley excitations when shuttling near points of low valley splitting: (1) elongating the quantum dot along the shuttling trajectory while squeezing it in the transverse direction (this keeps the total dot area fixed, thereby maintaining the same average $E_v$, to allow a fair comparison with other strategies, and ensures that the elongated electron wavefunction sees the same amount of Ge, on average), (2) shifting the lateral position of the dot within the shuttling channel, (3) modifying the shuttling velocity, and (4) varying the vertical electric field of the dot.
The effects of these strategies can be understood intuitively as follows.
Methods (2) and (4) simply avoid the points of low $E_v$.
Method (3) directly suppresses the Landau-Zener tunneling process.
%Method (1) causes the dot to see less change in its average alloy environment for a given velocity, resulting in a lower \emph{effective} velocity.
Method (1) causes three important effects that increase the shuttling fidelity.
First, by elongating the dot along $x$, we reduce the length scale of the shuttling process, relative to the dot size.
This causes the dot to experience fewer local $E_v$ minima, reducing the probability of Landau-Zener valley excitations.
Second, for a fixed shuttling velocity, this length scale modification also reduces the rate of change of $\Delta$, further limiting the probability of valley excitation near $E_v$ minima.
Finally, by squeezing the dot along $y$ by the same factor, we maintain the same average valley splitting, and we enhance our ability to tune $E_v$ by using lateral channel shifts, for a fixed channel width.
In this paper, we consider shifts from an isotropic dot with $\hbar \omega_x = \hbar \omega_y = 2$~\SI{}{\milli\electronvolt} to an elongated dot with $\hbar \omega_x = 1$~\SI{}{\milli\electronvolt} and $\hbar \omega_y = 4$~\SI{}{\milli\electronvolt}.
\Cref{app:squeeze} elaborates further on the physics of the elongation strategy.
The final results obviously depend quantitatively on the imposed parameter constraints. Here we have chosen experimentally reasonable constraints; an exploration of different parameter ranges is given in \cref{app:app_sweeps}.

\section{Results}\label{sec:results}

\subsection{Evolution without applying tuning strategies}

As a baseline, we first characterize spin shuttling across a spatially varying valley-splitting landscape, at a velocity of 5~m/s, without applying any fine-tuning strategies.
Figure~\ref{fig:fig_e_field_var}(a) shows the medians (markers) and 25-75 percentile ranges (bars) of the infidelity, computed according to \cref{eq:fidelity}, for 300 shuttling simulations, each with a different, random valley-splitting landscape.
The results are reported as a function of position along the shuttling trajectory.
(Here, we only show results for the initial \SI{1}{\micro\meter} portion of the trajectory.)
We include results for the \SI{10}{\nano\meter}, \SI{3}{\nano\meter}, and 5\% Ge quantum wells illustrated in \cref{fig:model_fig}(a).
Despite experiencing different average valley splittings, these three heterostructures exhibit similar behaviors, characterized by a rapid increase of the infidelity over short distances, to levels that are incompatible with quantum computing on a sparse-grid architecture~\cite{Kuenne2023}. 
The main contribution to the infidelities observed in these simulations is the Landau-Zener excitation of the upper valley state, caused by momentary dips in $E_v$, as illustrated in Figs.~\ref{fig:theory_fig}(e) and \ref{fig:theory_fig}(f). 
The simplest approach for suppressing such excitations is to uniformly reduce the shuttling velocity, which unfortunately leads to a competition between the shuttling and decoherence timescales.
Other suppression strategies are therefore required, which we now include in our simulations.

\subsection{Electric-field modulation} \label{sec:results_efield}

\begin{figure}
    \centering
    \includegraphics[width=0.49\textwidth]{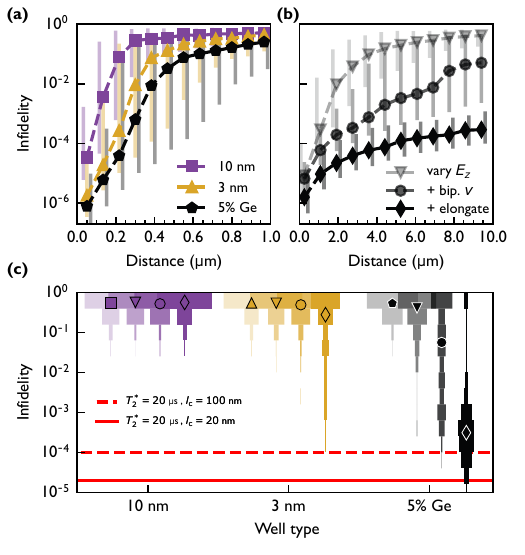}
    \caption{
    A comparison of shuttling infidelities: (a) without, vs (b) with several of the tuning strategies depicted in Figs.~\ref{fig:model_fig}(b)-\ref{fig:model_fig}(e), for an average shuttling velocity of 5~m/s. 
    (a) Infidelity as a function of shuttling distance, for the three quantum wells shown in Fig.~\ref{fig:model_fig}(a).
    (b) Infidelity computed using the following tuning strategies, for the 5\% Ge quantum well: (i) $E_z$ modulation only (gray triangles); (ii) $E_z$ modulation plus bipartite velocity modulation (dark-gray circles); or (iii) $E_z$ modulation, bipartite velocity modulation, and elongated dots (black diamonds).
    For (a) and (b), the markers represent the median values obtained from $300$ different disorder realizations, while the vertical bars show the 25-75 percentiles. 
    (Note the different horizontal scales.)
    (c) Histograms of results like those shown in (a) and (b), for the full shuttling distance of \SI{10}{\micro\meter}, in all three quantum wells (color coded).
    Within each color grouping, the tuning methods are coded with the same marker styles as in (a) and (b).
    Only the 5\% Ge well, with either two or three tuning strategies (last two histogram columns), provides significant improvements of the shuttling fidelity. Red lines are estimates for infidelity due to spin dephasing and correspond to \cref{eq:infidelity_analytical_estimate} evaluated for magnetic noise (solid red line) and charge noise (dashed red line), as explained further in \Cref{sec:transport_velocity}.}
    \label{fig:fig_e_field_var}
\end{figure}%

Local modulation of the vertical electric field $E_z$ causes the electron wavefunction to shift vertically, as illustrated in Fig.~\ref{fig:model_fig}(e), exposing it to a slightly different Ge distribution, and modifying its valley splitting~\cite{losert2023practical}.
It may therefore be beneficial to modify $E_z$ when a low valley-splitting region is encountered.
While any $E_z$ value can be used in the simulations, we adopt some procedural constraints, to make our theoretical methods more compatible with experiments, and to reduce computing times.
First, we restrict the $E_z$ range to lie between 0 and 10~mV/nm.
($E_z<0$ can also be considered, but does not change our results significantly.)
We note that even larger changes in $E_z$ may be possible when using a back gate.
However, the range chosen here includes relatively high fields~\cite{hollmann2020large}, and should therefore be sufficient for evaluating the feasibility of $E_z$ modulations.
Second, we do not allow the field to be adjusted continuously; rather, we assume piece-wise constant $E_z$ values over \SI{1}{\micro\meter} segments.
To optimize the $E_z$ values used in each shuttling segment, for a given valley landscape, we apply a graph traversal algorithm, as described in \cref{app:graph_traversal}.
This path seeks to avoid regions of low valley splittings, while making as few changes to the tuning parameters as possible. 
In real devices with no a-priori knowledge of the valley splitting, such optimization would require either obtaining a high-resolution map of $E_v(\mathbf{r})$~\cite{volmer2023mapping} or applying trial-and-error methods.

Simulation results for infidelity vs shuttling distance are presented in Fig.~\ref{fig:fig_e_field_var}(b) (light-gray triangles) where we show results only for the 5\% Ge quantum well.
We assume a single, fixed velocity of \SI{5}{\meter\per\second}, corresponding to a total shuttling time of $\SI{2}{\micro\second}$, which is slightly shorter than commonly observed $T_2^*$ times of a few microseconds~\cite{mills2022two,noiri2022fast,yoneda2018quantum,struck2020low}.
As indicated here, the $E_z$ modulation procedure provides some improvement over the baseline results shown in Fig.~\ref{fig:fig_e_field_var}(a) (note the different horizontal scales in these two panels); however, the shuttling infidelities remain quite poor over the full shuttling range of \SI{10}{\micro\meter}.

\subsection{Bipartite velocity modulation} \label{sec:results_bipartite}

%As described in \cref{app:app_sweeps}, we settle on a threshold value of \SI{20}{\micro\electronvolt} here, which strikes a balance between the number of trajectory segments that must be tuned vs the final shuttling fidelity.

To further improve the shuttling fidelity, we next consider velocity modulation as a tuning strategy for suppressing Landau-Zener excitations in regions of low valley splitting.
In this case, we adopt the constraint that only two shuttling velocities are allowed (rather than a continuous range): $v_\text{fast}$ and $v_{\mathrm{slow}}=v_{\mathrm{fast}}/5$.
As described in \cref{app:app_sweeps}, the slower velocity is applied whenever $E_v$ falls below a threshold value, defined as $10\%$ of the mean value of $E_v$, or \SI{20}{\micro\electronvolt}, whichever is greater.
These choices strike a balance between sufficiently reducing $v_\text{slow}$ while retaining a reasonable total shuttling velocity.
As a safety margin, we also apply $v_\text{slow}$ within a $\pm\SI{10}{\nano\meter}$ window around these valley-splitting minima, while setting the velocity to $v_\text{fast}$ elsewhere.
On average, since fewer than ten slow-downs occur per trace under these constraints, we still maintain an average velocity of approximately $\SI{5}{\meter\per\second}$, which is nearly equal to $v_\mathrm{fast}$.
Results for such bipartite velocity modulations, combined with $E_z$ modulations, are shown in Fig.~\ref{fig:fig_e_field_var}(b) (dark-gray circles).
By combining tuning strategies in this way, we obtain an approximate order of magnitude improvement in the fidelity for the 5\%~Ge quantum well, as compared to the case where only $E_z$ is modulated. However, the error bars of the infidelity are seen to be quite large.
In Appendix~\ref{app:separated_methods}, we explore the effects of applying different combinations of control strategies. 
In particular, we consider the separate effects of velocity modulation and dot elongation, without including $E_z$ modulation or channel shifting. 
This shows that velocity modulation (and dot elongation) are only effective when used in combination with an evasion strategy (e.g., $E_z$ modulation or channel shifting). 
This can be understood because velocity modulation does not directly address the problem of low valley splitting.

\subsection{Elongated dots} \label{sec:results_squeezed}

Finally, we consider the shuttling of an elongated quantum dot, in which the orbital confinement energy in the shuttling direction $\hbar \omega_x$ is reduced from 2 to \SI{1}{\milli\electronvolt} (elongating the dot along $\hat x$), while increasing the confinement energy in the transverse direction $\hbar \omega_y$ from 2 to \SI{4}{\milli\electronvolt} (squeezing the dot along $\hat y$).
In the current work, we do not explore potential pitfalls of this elongation strategy, although they may occur in some settings~\cite{langrock2023blueprint}.
Results of such elongated-dot simulations, combined with $E_z$ and velocity modulations, are shown in Fig.~\ref{fig:fig_e_field_var}(b) (black diamonds).
In this case, we obtain significant improvement over previous results, by over an order of magnitude for longer shuttling distances.
Figure~\ref{fig:fig_e_field_var}(c) summarizes the results of these simulations, including the base case, for all three types of quantum wells.
It is interesting to note that, while simultaneously applying multiple strategies is found to improve the shuttling fidelity for the 5\% Ge quantum well, much weaker improvements are found for the \SI{10}{\nano\meter} and \SI{3}{\nano\meter} quantum wells.
In \cref{app:Efield_variation}, we show that this tepid response is a consequence of using the $E_z$-modulation strategy, because $E_z$ does not provide effective tuning of $E_v$ for the other two heterostructures.

\subsection{Channel shifting}

\begin{figure*}
    \centering
    \includegraphics[width=1.0\textwidth]{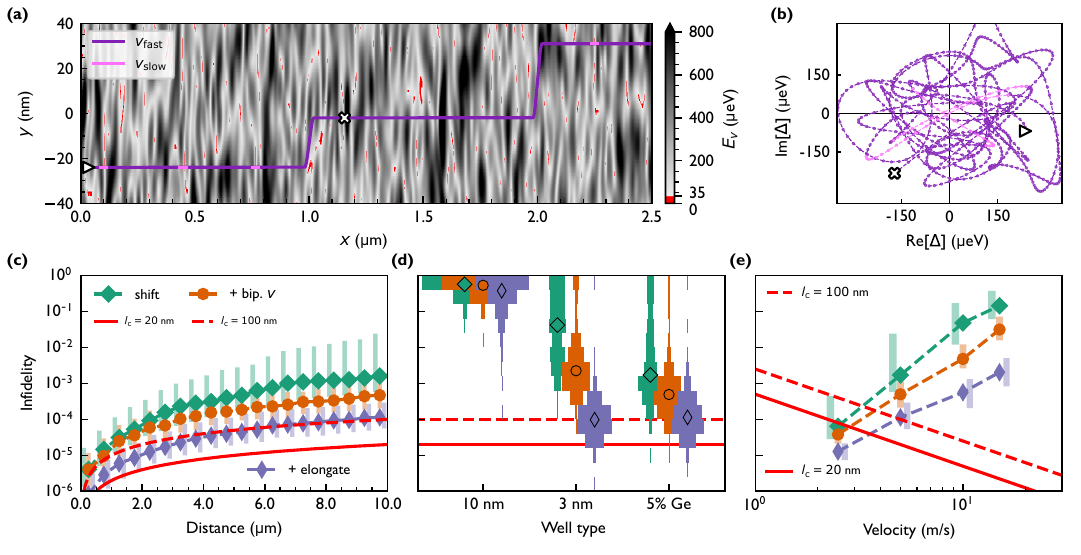}
    \caption{
    Shuttling results involving lateral channel shifting. 
    (a) Typical valley-splitting landscape for a 5\% Ge quantum well, with regions of relatively low valley splitting highlighted in red (10\% or less of the average value of $E_v$, or about \SI{35}{\micro\electronvolt}). 
    An optimized, segmented shuttling trajectory is shown in purple (see main text).
    In the few remaining regions of low valley splitting, the shuttling velocity may be reduced to $v_\text{slow}$ (indicated in pink), to suppress Landau-Zener excitations.
    (b) Inter-valley coupling $\Delta$ along the same path shown in (a).
    (c) Shuttling infidelities computed using the following tuning strategies for the 5\% Ge quantum well and average shuttling velocity 5~m/s: (i) lateral channel shifting only (green diamonds); (ii) channel shifting plus bipartite velocity modulation (rust-colored circles); or (iii) channel shifting, bipartite velocity modulation, and elongated dots (narrow violet diamonds).
    Here, the markers represent the median values obtained from $300$ different disorder realizations, while the vertical bars show the 25-75 percentiles.
    (d) Histograms of results like those shown in (c), for the full shuttling distance of \SI{10}{\micro\meter}, in all three quantum wells.
    Here, the tuning methods are coded by both color and marker styles, as in (c).
    (e) Shuttling infidelities like those shown in (c), for the full shuttling distance of \SI{10}{\micro\meter}, as a function of average shuttling velocity.
    Red lines in panels (c), (d) and (e) show infidelity estimates for magnetic noise (solid red line) or charge noise (dashed red line), based on Eq.~(\ref{eq:infidelity_analytical_estimate}), with the indicated correlation lengths $l_c$.}
    \label{fig:fig_channel_var}
\end{figure*}%

Since the $E_z$ modulation scheme is not found to be effective universally, we also explore the channel-shifting strategy, depicted in Fig.~\ref{fig:model_fig}(c), where the position of the electron is shifted along $\hat y$ to avoid regions of low valley splitting.
We expect this method to be more effective than $E_z$ modulation because $E_v$ typically varies much more as a function of $y$ than as a function of $E_z$ within the parameter constraints we consider.
Moreover, since lateral shifting is not sensitive to the vertical Ge confinement profile, we expect the effectiveness to depend only on the variability of the valley splitting $\sigma_\Delta$, rather than other features of the quantum well.

Similar to $E_z$ modulation, we determine the optimal shuttling trajectory by applying a graph-traversal algorithm. 
In this case, we again consider a full \SI{10}{\micro\meter} shuttling channel formed of piecewise-constant \SI{1}{\micro\meter} segments.
We also constrain the trajectory to lie within the channel width given by $y=\pm 50$~nm.
\Cref{fig:fig_channel_var}(a) illustrates one such optimized shuttling trajectory for the case of a 5\% Ge quantum well, where the valley-splitting landscape is shown as grayscale, with regions of $E_v <$ \SI{35}{\micro\electronvolt} highlighted in red.
Here, the compromises caused by limiting the shifts to \SI{1}{\micro\meter} segments are easy to visualize, because not all red regions can be avoided.
\Cref{fig:fig_channel_var}(b) illustrates the complicated evolution of the inter-valley coupling $\Delta$, following along this same trajectory.
While most of this evolution exhibits a sufficiently large $|\Delta|$, a small portion still approaches $|\Delta|\approx 0$, causing leakage.
Averaging such results over 300 valley-splitting landscapes yields the infidelity results shown as green diamonds in Fig.~\ref{fig:fig_channel_var}(c).
Here we observe immediate improvement over the $E_z$ modulation; we even observe improvement over the best-case results in Fig.~\ref{fig:fig_e_field_var}(b), obtained for the 5\% Ge quantum well.

Taking the same approach as Fig.~\ref{fig:fig_e_field_var}, we now include additional tuning methods in Fig.~\ref{fig:fig_channel_var}(c): the rust-colored circles show the combined results of channel-shifting and bipartite velocity modulation (using the same two velocities as Fig.~\ref{fig:fig_e_field_var}), while the purple diamonds show the combined results of channel-shifting, velocity modulation, and dot elongation. 
In each case, we observe some fidelity improvement.
Results of these different tuning schemes are shown as histograms in Fig.~\ref{fig:fig_channel_var}(d) for all three types of quantum wells.
Here the best results are obtained for the 3~nm and 5\% Ge quantum wells, which have much higher average valley splittings to begin with.
These two quantum wells show similar results when applying multiple tuning strategies, obtaining infidelities consistently below $10^{-3}$, except for a small minority of outlier cases.
Importantly, the 5\% Ge quantum well shows a low average infidelity when applying \emph{only} the channel-shifting strategy, although a significant fraction of results still give poor fidelities.
Finally we note that the 10~nm quantum well -- with a small average valley splitting of \SI{50}{\micro\electronvolt} -- experiences particularly small $E_v$ minima too frequently to be compensated by any tuning method, for any reasonable velocity, for the tuning constraints we have imposed in this work.

\subsection{Transport velocity} \label{sec:transport_velocity}

As previously noted, transport velocity plays an important role in determining the leakage during shuttling, since the probability of Landau-Zener excitations depends exponentially on velocity.
This strong dependence is evident in Fig.~\ref{fig:fig_channel_var}(e), where we plot simulation results like those in Fig.~\ref{fig:fig_channel_var}(d) for the 5\% Ge quantum well, but we now consider a range of velocities.
As in Fig.~\ref{fig:fig_channel_var}(d), we adopt three tuning strategies, using the same color coding as before. 
Here, when bipartite velocity modulations are employed, we note that it is the \emph{average} velocity that is reported on the horizontal axis.
[The results shown in Fig.~\ref{fig:fig_channel_var}(d) correspond to the velocity of 5~m/s in Fig.~\ref{fig:fig_channel_var}(e).]
As expected, we find that lower velocities cause less leakage, and the shuttling fidelity depends sensitively on the choice of velocity.

On the other hand, slower shuttling speeds also increase the risk of decoherence, so the velocity should be carefully chosen.
Although we do not include decoherence directly in our shuttling model, we now provide analytical estimates, to illustrate the emergence of an optimal shuttling velocity. 
In Ref.~\cite{langrock2023blueprint}, it was argued that the main sources of decoherence during shuttling are time-varying Overhauser magnetic fields and low-frequency charge noise, which both cause dephasing of the spin over the time scale $T_2^*$.
The same reference obtains an approximate expression for the noise-induced shuttling infidelity, which we have adapted to the present context (note the slight difference from \cite{langrock2023blueprint}), as discussed in \cref{app:langrock_formula}:
\begin{equation}\label{eq:infidelity_analytical_estimate}
    I\approx \frac{l_cL_s}{(vT_2^*)^2} \, ,
\end{equation}
where $l_c$ denotes the correlation length of the quasistatic noise source, $L_s$=\SI{10}{\micro\meter}  is the shuttling distance, $v$ is the average velocity, and we note that motional narrowing has been taken into account~\cite{struck2024spin}.
\begin{comment}
\red{(Please check this, from Max: to be quantitatively comparable, it should be adapted to the metric used here. When I calculate the infidelity for our metric in contrast to that from the reference, I get an additional factor 1/2. Consider or ignore?) -- MO: The current version (also in the respective figure) is the one re-calculated for our metric, which is more consistent. If needed, a derivation is stored in 'factor0.5.tex'. I have adapted the sentence preceding the formula for now, but that may also be reformulated. }
\end{comment}

In \cref{fig:fig_channel_var}(c)-(e), we include two representative infidelity estimates from Eq.~(\ref{eq:infidelity_analytical_estimate}), obtained using $l_c$=\SI{20}{\nano\meter} (solid red lines), for the case where nuclear spin noise dominates, and $l_c$=\SI{100}{\nano\meter} (dashed red lines), for the case where charge noise dominates~\cite{langrock2023blueprint}.
We also include the same infidelity estimates in Fig.~\ref{fig:fig_e_field_var}(c).
In both cases, we take a cautiously optimistic value of $T_2^*$$\approx$\SI{20}{\micro\second}, as consistent with Refs.~\cite{yoneda2018quantum,yoneda2021coherent,struck2020low}.
This analysis confirms the presence of an optimal velocity value, which depends on the tuning strategies used and the dominant noise source, but generally corresponds to a few m/s.
The analysis also shows that an appropriate choice of tuning strategies and velocities should yield, in principle, shuttling infidelities below $10^{-3}$.

\subsection{Sharp interfaces with steps}
\label{sec:steps}

Up to this point, we have only considered quantum wells with valley splittings that are determined mainly by alloy disorder, since most current devices are expected to fall into this regime~\cite{losert2023practical}.
For completeness, we also briefly consider the opposite regime, where alloy disorder plays a minor role.
The most common quantum-well geometry for this purpose has a super-sharp interface, with an interface width of less than three atomic monolayers.
For such geometries, the valley splitting should be enhanced, and the dominant form of disorder and $E_v$ variability should arise from atomic steps at the quantum-well interface~\cite{losert2023practical}.
To study this situation, we perform simulations of shuttling with super-sharp interfaces and sparse step disorder, as described in \cref{app:perfect_step}.
Our results indicate that high-fidelity shuttling can be achieved in the presence of sparse step disorder, even without applying additional tuning methods.
The fact that step disorder and alloy disorder can obtain such different results highlights the importance of including realistic disorder models that accurately account for random alloys.

\section{Implementing the tuning schemes} \label{sec:scalability}

We close by commenting on the added complexity that comes with the tuning methods proposed here, and their consequences for scalability.
First, we note that dot size and shape in the elongation scheme are closely tied to the predetermined gate-electrode spacing.
Some additional fine-tuning is possible; however, a truly scalable pulsing scheme favors applying the same ac signals to all the clavier gates across a quantum processor~\cite{seidler2022conveyor}, suggesting that the dots (and the gate pitch) should all have a uniform size. 

For the vertical or lateral channel-shifting schemes, we have proposed to divide the channel into smaller segments, which may then be manipulated in two ways, as we now explain.
The most versatile approach involves manipulating each segment independently. 
For vertical shifting, this requires independent control of the clavier gates within each segment, while for lateral shifting it requires independent control of the screening gates in each segment.
While such an approach is highly versatile, it also adds significant overhead to the wiring costs, effectively negating many of the global benefits of our shuttling scheme.
The second approach involves applying tailored shift pulses to \emph{all} the clavier gates or \emph{all} the screening gates. 
This has the advantage of not requiring new control lines but has the disadvantage of affecting all the electrons within the conveyor.
Thus, if multiple electrons are shuttled simultaneously, it would require a more-sophisticated path-traversal algorithm.

Hence, any channel shifting scheme will require some additional wiring costs. 
And, as we demonstrate in this work, we view some degree of channel shifting as necessary for achieving high-fidelity shuttling.
However, this additional wiring is not infeasible. 
For example, if low-resolution control of the dot position within a channel suffices, the screening gates within each segment can be modulated by introducing just a few new DC lines; this represents a moderate increase in wiring complexity, which is not dissimilar from wiring requirements for achieving high-fidelity gates.
Moreover, these screening gate voltages can be locked in after an initial tune-up stage.
Advances in automation techniques (``auto-tuning''), allowing for the simultaneous tune-up of each shuttling channel, will reduce the initial time spent in the tune-up stage.
Altogether, this additional complexity is reasonable.

Similar considerations apply to bipartite velocity modulations, although we note that slow shuttling in a region with high valley splitting is harmless, so multi-electron shuttling is less fraught in this case.
On the other hand, true global control could become a challenge, if a large number of slow-downs are needed.

Finally, we note that all proposed tuning strategies require the valley-splitting landscape to be carefully characterized. 
However, since the valley splitting is a static and materials-dependent property of the device, this characterization only needs to be performed once per shuttling channel.
A recent experiment demonstrates that such mappings can be implemented effectively, within the same shuttling framework~\cite{volmer2023mapping}.

\section{Summary}\label{sec:conclusion}

In this work, we have shown that leakage from the ground valley to the excited valley state is a major source of decoherence for conveyor-mode spin shuttling in Si/SiGe quantum wells.
This leakage is caused by Landau-Zener excitations across a narrow energy gap, as electrons traverse the wildly varying valley-splitting energy landscape caused by alloy disorder.
In turn, leakage causes dephasing of the spin, due to the presence of different $g$-factors in the ground and excited valley states.

Using the most current understanding of random-alloy disorder, we perform simulations of the shuttling evolution within an effective four-level Hamiltonian spanning the spin and valley degrees of freedom.
For quantum wells falling into the ``deterministically enhanced'' valley-splitting regime (e.g., with interfaces narrower than three atomic monolayers), we find that Landau-Zener excitations do not pose a significant challenge for shuttling.
It is hoped that such structures will become available for qubit implementations in the future.

Existing devices are not expected to fall into the deterministically enhanced regime, and our simulations indicate that coherent transport may be unfeasible in common \SI{10}{\nano\meter} quantum wells in this ``disorder-dominated'' regime.
In this case, we have also performed simulations of alternative quantum wells with much higher average valley splittings, including narrow 3~nm quantum wells and quantum wells with a significant concentration of Ge in the middle of the well.
We have also explored a number of tuning strategies, including shifting the location of the electron inside the shuttling channel (either vertically or laterally), to avoid passing through a valley-splitting minimum, slowing down the shuttling velocity when it passes too close to a mininum, and elongating the quantum dot to change its effective velocity.
In our simulations, we have optimized these tuning strategies, and we have also simultaneously applied multiple strategies, obtaining several orders of magnitude improvement in the shuttling fidelity.
Since slower shuttling velocities suppress Landau-Zener excitations but lead to dephasing, we have also optimized the velocity, finding that velocities on the order of several m/s can provide shuttling infidelities below $10^{-3}$.

Finally, we note that the tuning strategies proposed here come with a nontrivial experimental overhead, which must be accounted for in scalable implementations.
The valley-splitting landscape only needs to be mapped out once, however.
In the future, we argue that fidelity-improving strategies like those considered here must be employed in any high-performance shuttling implementation in Si/SiGe quantum wells.

\begin{acknowledgments}
%\red{TODO: mention C. Gorjaew for code contributions?}

This research was sponsored in part by the Army Research Office (ARO) under Awards No.\ W911NF-17-1-0274, W911NF-22-1-0090, and W911NF-23-1-0110. 
The work was performed using the compute resources and assistance of the UW-Madison Center For High Throughput Computing (CHTC) in the Department of Computer Sciences. The CHTC is supported by UW-Madison, the Advanced Computing Initiative, the Wisconsin Alumni Research Foundation, the Wisconsin Institutes for Discovery, and the National Science Foundation, and is an active member of the OSG Consortium, which is supported by the National Science Foundation and the U.S. Department of Energy's Office of Science.
The work was also funded by the German Research Foundation (DFG) within the project 421769186 (SCHR 1404/5-1) and under Germany's ``Excellence Strategy - Cluster of Excellence Matter and Light for Quantum Computing'' (ML4Q) EXC 2004/1 -- 390534769 and by the Federal Ministry of Education and Research under Contract No.\ FKZ: 13N14778.
The views, conclusions, and recommendations contained in this document are those of the authors and are not necessarily endorsed nor should they be interpreted as representing the official policies, either expressed or implied, of the Army Research Office (ARO) or the U.S. Government. The U.S. Government is authorized to reproduce and distribute reprints for Government purposes notwithstanding any copyright notation herein.\par
A method for avoiding problematic spots in the shuttling channel by shifting the shuttling path laterally is covered by a patent family (PCT/EP2023/055058) by the work of the inventors Klos, M.O., K\"unne, H.B., J.D.T., and the patent application, co-owned by RWTH Aachen University and Forschungszentrum J\"ulich GmbH, is currently pending in the designated PCT-states.\par
L.R.S. and H.B. are founders and shareholders of ARQUE Systems GmbH. The other authors declare no competing interest.

\end{acknowledgments}

\appendix

\section{Super-sharp interfaces with atomic step disorder}\label{app:perfect_step}

\begin{figure}
    \centering
    \includegraphics[width=0.45\textwidth]{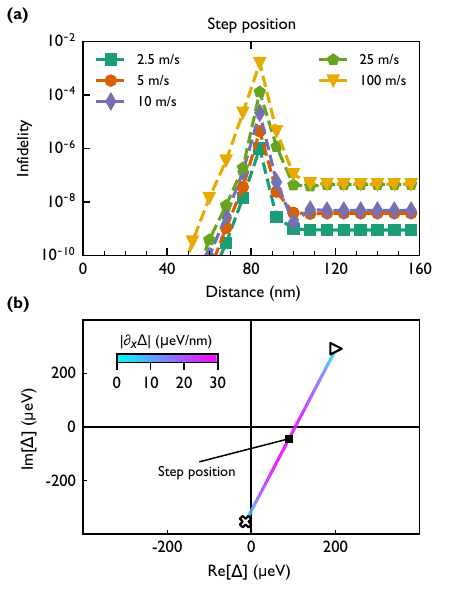}
    \caption{Traversal of a single monoatomic step. (a) Infidelities as a function of shuttling distance, for a quantum well with a perfectly sharp interface containing a single atomic step centered in the middle of the shuttling path, for several different shuttling velocities (color coded).
    Working in the basis of instantaneous eigenstates, we observe an infidelity peak at the center of the Landau-Zener transition, as expected from theory \cite{vitanov1999transition,mullen1989time}.
    The infidelity then remains low after crossing the step, even for velocities up to \SI{100}{\meter\per\second}.
    Compared to the disorder-dominated regime, this represents several orders of magnitude of improvement in the infidelity. 
    (b) The inter-valley coupling $\Delta$ plotted in the complex plane for the same shuttling path. Here, the lowest valley splitting occurs at the step location, with a value of \SI{200}{\micro\electronvolt}. The color code represents $|\partial_x \Delta|$.
    %, in conjunction with the shuttling speed giving the effective Landau-Zener velocity at the step position for the given linear 2D-trajectory of $\Delta$.
    }
    \label{fig:fig_app_step}
\end{figure}%

In the main text, we considered the limit of disorder-dominated quantum wells, for which random-alloy disorder is the main source of variability in the valley splitting. In this Appendix, we consider the opposite, deterministically enhanced valley-splitting limit, which can be achieved, for example, in quantum wells with super-sharp interfaces. In this case, the dominant source of disorder is from single-atom steps at the quantum well interface. We now briefly show that such disorder from sparse steps has a very different effect on shuttling than random-alloy disorder, and that the resulting infidelities are greatly reduced in accordance with much larger minimal $E_v$.

When the QW interfaces are very sharp, monoatomic steps in the QW interfaces are the dominant source of $E_v$ fluctuations.
To study shuttling in this regime, we examine the case of a single atomic step, traversed at different shuttling velocities. 
We assume an orbital splitting of \SI{2}{\milli\electronvolt} and a vertical electric field of \SI{5}{\milli\volt\per\nano\meter} inside a well of width \SI{10}{\nano\meter}, with perfectly sharp interfaces and a single atomic step in the interface.
We use 2D effective mass theory to simulate the inter-valley coupling as the dot moves across the step (see \cref{app:vs_model_supp}).
The evolution of $\Delta$ as the dot traverses the step is shown in \cref{fig:fig_app_step}(b), where at the step position, the dot has a minimum $E_v$ of about \SI{200}{\micro\electronvolt}.
We plot the resulting infidelity as function of shuttling distance in \cref{fig:fig_app_step}(a).
We see spikes in the infidelity near the step position in Fig.~\ref{fig:fig_app_step}(a).
These correspond to the position with maximized $|\partial_x \Delta|$, the rate of change in $\Delta$ as a function of distance, shown in Fig.~\ref{fig:fig_app_step}(b).
However, as the minimal $E_v$ at the step position is much larger than typical minima present in the disordered regime, even velocities an order of magnitude higher than those considered in the main text converge to low infidelity values beyond the step. 
An ideal sharp interface with only occasional single monolayer steps can therefore enable high-fidelity transport even without applying tuning methods.
It should be noted that multiple steps in close vicinity, i.e. on the order of the dot size, may decrease $E_v^\mathrm{min}$ again and induce larger infidelity, as considered in Ref. \cite{langrock2023blueprint}.

%, and $\Delta$ is changing at a rate of about $\SI{30}{\micro\electronvolt\per\nano\meter}$, as indicated by the color scale. 
%Note that the non-constant $\abs{\partial_x \Delta}$, $x$ being the shuttled distance, differentiates this evolution from the standard solution of a Landau-Zener transition.\par
%

%The final infidelity, after peaking around the step position, converges at very low values for all shuttling velocities considered, even up to \SI{100}{\meter\per\second}. This implies a less strict velocity limitation than analyzed in ref. \cite{langrock2023blueprint} with analytical treatment and slightly differing assumptions.

\section{Prefactor of the noise-induced shuttling infidelity}
\label{app:langrock_formula}

Here, we briefly clarify the different prefactor appearing in \cref{eq:infidelity_analytical_estimate}, as compared to Ref.~\cite{langrock2023blueprint}, which we have adapted to match the infidelity metric described in Sec.~\ref{sec:metrics}. For a noise Hamiltonian given by
\begin{equation}
	H_\mathrm{noise}=\frac{\hbar}{2}\Phi(t)\sigma_z\,,\\
 \end{equation}
 our metric can be rewritten as
 \begin{align}
	F&=\frac{1}{d^2}\abs{\Tr\{\underbrace{V^\dagger}_{\mathds{1}}U_\mathrm{noise}(t)\}}^2\\
	&=\frac{1}{d^2}\abs{e^{i\Phi(t)/2}+e^{-i\Phi(t)/2}}^2\,,
 \end{align}
 where $U_\mathrm{noise}(t)$ is the time evolution operator for $H_\mathrm{noise}$. Assuming a Gaussian ensemble average of qubit phases $\Phi$ of zero mean and rms $\delta \Phi$, and with a dimension $d=2$, this evaluates to
 \begin{align}
 \expval{F} &=\frac{1}{d^2}\left(2+2e^{-\delta\Phi^2/2}\right)\\
	\xrightarrow{d=2} \expval{I}=1-\expval{F}&=\frac{1}{2}\left(1-e^{-\delta\Phi^2/2}\right)\approx\frac{\delta\Phi^2}{4}\, ,
\end{align}
which implies an additional factor of $1/2$ compared to Ref.~\cite{langrock2023blueprint}.
 
\section{Parameter choices for two tuning methods}\label{app:app_sweeps}

\begin{figure}
    \centering
    \includegraphics[width=0.45\textwidth]{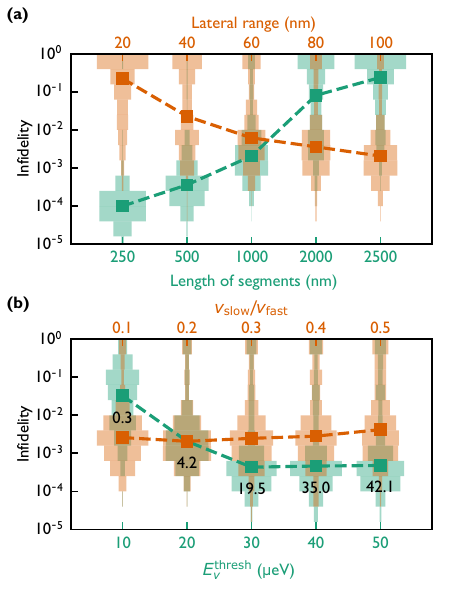}
    \caption{Comparison of different simulation constraint parameters, for two tuning methods. (a) Exploration of shuttling infidelities as we vary parameters used in the channel-shifting simulations. Rust-colored histograms show results for several different lateral shift ranges. 
    Green histograms show results for several different segment lengths. Larger lateral ranges and smaller segment lengths are seen to give higher fidelities. Here, results are obtained for a \SI{10}{\micro\meter} shuttling trajectory, for an average velocity of \SI{5}{\meter\per\second}, in the \SI{3}{\nano\meter} quantum well, and we simultaneously apply bipartite velocity modulation but no dot elongation. 
    (b) Exploration of shuttling infidelities as we vary parameters related to bipartite velocity modulations. Rust-colored histograms show results for several values of the ratio $v_\text{slow} / v_\text{fast}$. Green histograms show results for several values of the threshold valley splitting $E_v^\text{thresh}$, below which the slower velocity is applied. While increasing the ratio $v_{\mathrm{slow}}/v_{\mathrm{fast}}$ has a weak effect on the median infidelity values, it has a stronger effect on the number of very poor infidelity results.
    On the other hand, setting a low valley splitting threshold value causes the infidelity to increase significantly. Black numbers indicate the average number of valley splitting dips below $E_v^\text{min}$ that occur over the full shuttle length. Other parameters are the same as in (a).
    }
    \label{fig:fig_app_tune}
\end{figure}%
 
In this Appendix, we examine how parameter choices for two of the tuning strategies in the main text affect the shuttling infidelities. This discussion is not meant to provide a comprehensive analysis; rather, it is to illustrate some of the compromises that must be considered when making these choices.

We first examine how the segment length and channel width affect shuttling infidelities for the channel-shifting strategy.
In \cref{fig:fig_app_tune}(a), we modify both the length of each segment and the width of the shuttling channel (the ``lateral range'').
We calculate the shuttling infidelities for the \SI{3}{\nano\meter} QW at \SI{10}{\micro\meter} distance for each parameter choice, while simultaneously applying bipartite velocity modulation, with an average shuttling velocity of \SI{5}{\meter\per\second}.
As expected, both parameters have a significant effect on shuttling infidelities of at least several orders of magnitude.
Increasing the lateral range makes it easier to avoid regions with low valley splittings, while smaller segment lengths allow for more frequent adjustments to the optimal path.
For both parameters, the range of parameters considered in \cref{fig:fig_app_tune}(a) does not result in asymptotic behavior of the infidelity; however, the range is imaginable for experimental realizations.
%experimentally feasible and realistic.
On the other hand, the results suggest the existence of thresholds, beyond which the infidelities deteriorate significantly: for channel widths, this occurs below about \SI{60}{\nano\meter}, and for segment lengths, it occurs above \SI{1}{\micro\meter}. 
The threshold for both parameters depends on the dot size, since valley splitting values are essentially uncorrelated when the dot is moved by the distance of a dot diameter.
We therefore expect that smaller dots do not require a wide channel to achieve high shuttling fidelities; however they do require shorter segment lengths.
At the threshold values for these tuning parameters, in addition to increasing median infidelity, we note that the infidelity distribution also becomes problematic, with a much higher portion of infidelities occurring at higher values.
 
Second, we examine parameters related to bipartite velocity tuning.
In \cref{fig:fig_app_tune}(b), we analyze the choice of the ``slow'' velocity used near locations of low valley splitting, $v_\text{slow}$, and we study the choice of threshold valley splitting, $E_v^\text{thresh}$, below which the velocity is reduced.
Again, we consider simulations of the \SI{3}{\nano\meter} QW evaluated at \SI{10}{\micro\meter}, with a fixed average velocity of \SI{5}{\meter\per\second}, and we apply both bipartite velocity modulation and channel shifting techniques.
We see that increasing the ratio $v_\text{slow} / v_\text{fast}$ only slightly increases the median infidelity, but it significantly widens the tails of the distribution at large infidelities. 
Setting a higher threshold $E_v^\text{thresh}$ for velocity switching causes the median infidelity to move to lower infidelities; however, this occurs at the cost of significantly more velocity switches per trace, as indicated by the black numbers accompanying the data points.

\section{Effect of individual tuning methods}\label{app:separated_methods}
\begin{figure}
    \centering
    \includegraphics[width=0.49\textwidth]{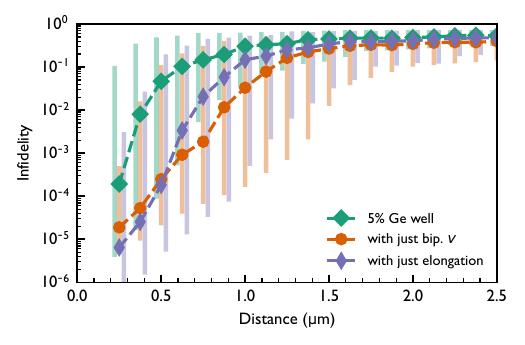}
    \caption{Simulation results showing that velocity modulation and dot elongation do not provide high-fidelity shuttling on their own. Here, we plot the shuttling infidelities as a function of distance for the 5\% Ge quantum well without any modifications (green), then including either velocity modulation (orange) or dot elongation (purple). Note that we use neither channel shifting nor electric field modulation. We find that neither the velocity modulation nor the elongation strategy provides reliable improvements over extended distances, with improvements that are limited to length scales below \SI{2}{\micro\meter}.
    }
    \label{fig:rev_separated}
\end{figure}

In this Appendix, we briefly show that strategies to evade points of low $E_v$ (e.g., modulating the electric field or channel shifting) are a necessary component of high-fidelity shuttling.
Velocity modulation and dot elongation can improve upon these fidelities, but they do not provide high-fidelity shuttling on their own. 
In Fig.~\ref{fig:rev_separated}, we separate out the performance of velocity modulation and dot elongation for the 5\% Ge quantum well. 
While both strategies yield improvements over the unmodified shuttling procedure, after \SI{2}{\micro\meter} no practical improvement is obtained, even for the quantum well with the largest average valley splitting studied here.

\section{Effective-mass theory of valley splitting} \label{app:vs_model_supp}
In this section, we elaborate on the effective-mass envelope-function model for the valley states outlined in Refs.~\cite{paquelet2022atomic, losert2023practical} and referenced in Sec.~\ref{sec:vs_model}.
We start with Eq.~(\ref{eq:delta3d}) in the main text, describing the inter-valley coupling $\Delta$.
For a fully separable system, we can reduce Eq.~(\ref{eq:delta3d}) to a 1D integral over $z$. 
Moreover, following Refs.~\cite{losert2023practical,paquelet2022atomic}, we transform the integral to a sum over atomic monolayers (MLs) to account for the discrete layers of the Si diamond cubic crystal structure, yielding \cref{eq:delta1d} for $\Delta_\text{1D}$ in the main text.
When monoatomic steps are present at the interface, our system is no longer fully separable, and the reduction to 1D is no longer sufficient. 
In this case, we can easily generalize to 2D or 3D. 
For example, to simulate a system with a monoatomic step in the $y$ direction, we can still separate the wavefunction as $\psi_\text{env}(\mathbf{r}) = \psi_\text{env}(x,z) \psi_\text{env}(y)$.
%By integrating over the $y$ direction, we are then left with a 2D problem.
We now perform a weighted average of the Ge distribution in only one ($y$) direction, resulting in a 2D description in the variables $(x,z)$.
Discretizing our system into cells of size $(\Delta x, \Delta z) = (a_0, a_0/4)$ yields the 2D effective mass equation
\begin{equation} \label{eq:delta_2d}
    \Delta_{2D} = \frac{a_0^2}{4} \sum_{j,l} e^{-2 i k_0 z_l} U_\text{qw}(x_j, z_l) |\psi_\text{env}(x_j, z_l)|^2.
\end{equation}
Here, the sum is over indices $j$ and $l$, which label the $x$ and $z$ coordinates of each cell, respectively.

To a very good approximation, the quantum well potential $U_\text{qw}$ is a linear function of the Ge concentration:
\begin{equation} \label{eq:qw_pot}
    U_\text{qw}(\mathbf{r}) = \frac{X_\mathbf{r} - X_s}{X_w - X_s} \Delta E_c
\end{equation}
where $X_\mathbf{r}$ is the Si concentration at position $\mathbf{r}$ in the heterostructure, $X_w$ is the average Si concentration in the quantum well, $X_s$ is the average Si concentration in the SiGe barrier/substrate region, and $\Delta E_c$ is the conduction-band offset of the quantum well.
For 1D systems, we can use $X_\mathbf{r} = X_l$, where $l$ is the layer index.
In this work, unless otherwise stated, we model our interfaces as sigmoid functions, where the average Si concentration at layer $l$ is defined by
\begin{multline} \label{eq:1D_si_concs}
\bar X_l= X_w \\
+\frac{X_s-X_w}{1+\exp[(z_l-z_t)/\tau]}+\frac{X_s-X_w}{1+\exp[(z_b-z_l)/\tau]} , 
\end{multline}
where $z_t$ and $z_b$ label the positions of the top and bottom interfaces, and $\lambda = 4\tau$ is the characteristic interface width.
Unless otherwise specified, we use $\lambda = 4\tau = $ \SI{1}{\nano\meter}.

In the case of monoatomic steps, the Si concentration $X_\mathbf{r}$ adopts some lateral dependence.
In this case, we can define the expected Si concentrations for a system with a step at lateral position $x = 0$:
\begin{equation}
    \bar X_{j,l} = \bar X_l \Theta(x_j \leq 0) + \bar X_{l+1} \Theta(x_j > 0)
\end{equation}
where $\bar X_l$ is given by \cref{eq:1D_si_concs} and $\Theta(\cdot)$ is the Heaviside step function.

Both Eqs.~(\ref{eq:delta1d}) and (\ref{eq:delta_2d}) depend on the envelope function $\psi_\text{env}$. 
To compute $\psi_\text{env}$, we discretize and solve a Schrodinger equation \emph{without} valley physics. 
In the 1D case, we discretize and solve the effective 1D Hamiltonian
\begin{equation}
    H_\text{1D} = -\frac{\hbar^2}{2 m_l} \partial_z^2 + \bar{U}_\text{qw}(z) + U_\phi(z)
    \label{eq:1D}
\end{equation}
where $m_l = 0.916 m_e$ is the longitudinal effective mass of the electron, $m_e$ is the bare electron mass, $U_\phi(z) = e E_z z$ is the potential due to a vertical electric field $E_z$, and $\bar U_\text{qw}$ is the quantum well potential without alloy-disorder-induced fluctuations.
In the 2D case, we discretize and solve the effective 2D Hamiltonian
\begin{multline}
    H_\text{2D} = -\frac{\hbar^2}{2 m_l} \partial_z^2 - \frac{\hbar^2}{2 m_t} \partial_x^2 + \bar{U}_\text{qw}(x, z) + U_\phi(z)  \\
    + \frac{1}{2}m_t \omega_x^2 (x - x_0)^2
    \label{eq:2D}
\end{multline}
where $\omega_x$ is the orbital confinement energy in the $x$-direction, and $x_0$ is the center location of the dot.
In \cref{app:perfect_step}, we considered shuttling across a mono-atomic step, where it was necessary to apply Eq.~(\ref{eq:2D}).
However, in the rest of this work, it is sufficient to use the 1D approximation of Eq.~(\ref{eq:1D}).

To model the conduction-band offset, we follow Ref.~\cite{paquelet2022atomic}:
\begin{multline}
    \Delta E_c = (X_w - X_s) \left[ \frac{X_w}{1-X_s} \Delta E_{\Delta_2}^\text{Si}(X_s) \right. \\
    \left. - \frac{1 - X_w}{X_s} \Delta E_{\Delta_2}^\text{Ge}(X_s)  \right] ,
\end{multline}
where $\Delta E_{\Delta_2}^\text{Si(Ge)}(X)$ are the $\Delta_2$ conduction-band offsets for strained Si (Ge) grown on unstrained $\text{Si}_X \text{Ge}_{1-X}$ substrate, and we approximate these functions as \cite{schaffler1992mobility}
\begin{equation} \begin{split}
    \Delta E_{\Delta_2}^\text{Si}(X) &\approx -0.502(1-X) \; \text{(eV)} \\
    \Delta E_{\Delta_2}^\text{Ge}(X) &\approx 0.743 - 0.625(1-X) \; \text{(eV)}.
\end{split} \end{equation}
Since the crystal lattice itself is composed of discrete atomic sites, the averaged concentration inside a finite-sized dot has an intrinsic uncertainty, given by $X_l = \bar X_l + \delta_l$, where $\bar X_l$ is the mean Si concentration at layer $l$, and $\delta_l$ is the fluctuation for a particular dot.
The Si concentration $X_l$ in layer $l$ can be computed as a weighted average, where the contribution of each atom is weighted by the dot probability density at that atom:
\begin{equation} \label{eq:deltal}
    X_l = \frac{1}{N_l}\sum_{a \in A_l} \mathbbm{1}[a = \text{Si}] |\psi_\text{env}(a)|^2 = \bar X_l + \delta_l
\end{equation}
where the sum is taken over $A_l$, the set of atoms in layer $l$, $\mathbbm{1}[a = \text{Si}]$ is the indicator function that returns 1 if $a$ is Si and 0 otherwise, and $\psi_\text{env}(a)$ is the value of the envelope function at the position of atom $a$.
The normalization constant $N_l = \sum_{a \in A_l} |\psi_\text{env} (a)|^2$.
The inter-valley coupling of Eq.~(\ref{eq:delta1d}) can likewise be broken into fixed and random components $\Delta_0$ and $\delta \Delta$:
\begin{equation} \label{eq:deterministic_vs_disordered}
\begin{split}
    \Delta_0 &= \frac{a_0 \Delta E_c}{4(X_w-X_s)} \sum_l e^{-2 i k_0 z_l} (\bar x_l - x_s) |\psi_\text{env}(z_l)|^2 , \\
    \delta \Delta &= \frac{a_0 \Delta E_c}{4(X_w-X_s)} \sum_l e^{-2 i k_0 z_l} \delta_l |\psi_\text{env}(z_l)|^2.
\end{split}
\end{equation}
In Eq.~(\ref{eq:deterministic_vs_disordered}), $\Delta_0$ is the inter-valley coupling due to larger-scale features of the heterostructure, like the interface width or interface steps.
On the other hand, $\delta \Delta$ is a local fluctuation about $\Delta_0$, caused by alloy disorder.
We can then compute the variance $\sigma_\Delta^2 = \text{Var}[\delta \Delta]$, as given in Eq.~(\ref{eq:vardelta}) in the main text.\par

\section{Generating valley-splitting landscapes} \label{app:random_fields}

To obtain accurate statistics of shuttling fidelities, we need to generate many realistic examples of inter-valley couplings, $\Delta$, which vary spatially across the device. 
That is, we need many examples of $\Delta(x,y)$.
To do so, we use the GSTools python library, which generates spatially random fields \cite{gstools}.
The real and imaginary components of $\Delta$ are generated independently, with variances given by $\sigma_\Delta^2 /2$ and spatial covariances defined in \cref{eq:2point}.

%\red{(The reason why $E_z$ causes a problem, below, is not clearly explained.)}
The above approach works for spatially varying inter-valley couplings, $\Delta(x,y)$.
\begin{comment}
\hl{However, to test the efficacy of modulating the electric field, we also need to generate many sample fields of the form $\Delta(x, E_z)$.
Since $\Delta$ has a non-trivial dependence on the vertical field $E_z$, we cannot generate these fields directly.
Instead, we generate the Si concentrations across the device.
Using GSTools, we create many examples of $X_l(x)$.}
\end{comment}
However, to test the efficacy of modulating the electric field, we also need to generate many sample fields of the form $\Delta(x, E_z)$.
Unlike the relationship between $\Delta$ and $x$, the statistical relationship between $\Delta$ and $E_z$ is not given by a simple covariance equation that can be randomized.
Instead, the effect of $E_z$ depends on the shape of the quantum well interfaces, the quantum well width, and the Ge content in the quantum well.
So, rather than randomly generating the field $\Delta$, we instead generate the local Si concentrations across the device, for each atomic layer of the quantum well. 
Since the dot is finite in extent, as it shuttles across the device, the effective Si concentrations experienced by the dot at each layer in the heterostructure fluctuate slightly. 
Thus, the effective Si concentrations in the quantum well become position-dependent.
We then indicate the Si concentration at layer $l$ and lateral position $x$ by $X_l(x)$.
Using GSTools, we create many examples of $X_l(x)$.
Then, for a given position and vertical field value, we use these local Si concentrations in our effective mass model, outlined above, to compute $\Delta(x, E_z)$. 
Below, we outline how we obtain these spatially fluctuating Si concentrations.

Previous work has shown that $X_l$ can be approximately sampled from a binomial distribution, $X_l \sim \frac{1}{N_\text{eff}} \text{Binom}(N_\text{eff}, \bar X_l)$, where $N_\text{eff} = 4 \pi a_\text{dot}^2 / a_0^2$ \cite{losert2023practical}.
In turn, this is approximately equal to a normal distribution with mean $\bar X_l$ and variance $\sigma_X^2 = N_\text{eff} \bar X_l (1 - \bar X_l)$.
Therefore, $X_l$ can be approximately sampled as
\begin{equation} \label{eq:xl_dist}
    X_l \sim \frac{1}{N_\text{eff}} N(\bar X_l, N_\text{eff} \bar X_l (1 - \bar X_l)) ,
\end{equation}
where $N(\mu, \sigma^2)$ is the normal distribution with mean $\mu$ and variance $\sigma^2$.
The spatial covariance is then given by 
\begin{equation}\label{eq:xl_cov}
    \text{Cov}[X_l, X_l'] = \exp ( -\delta_x^2 / 2 a_x^2 ) \sigma_X^2,
\end{equation}
where $\delta_x$ indicates the distance between two points along the shuttling trajectory.
Equations~(\ref{eq:xl_dist}) and (\ref{eq:xl_cov}) describe the complete spatial statistics of $X_l(x)$, which we can input into GSTools, to generate fluctuating Si concentrations.
%\red{(The $x$ dependence is not clear here:)}
Now, for each position $x$, we have a complete Si concentration profile $X_l(x)$, which we simulate using effective mass theory \cref{eq:delta1d} to compute $\Delta$.
Thus, we can build up sample fields of the form $\Delta(x, E_z)$. 
\section{Dependence on $\Delta E_B$} \label{app:vary_k}

\begin{figure}
    \centering
    \includegraphics[width=0.45\textwidth]{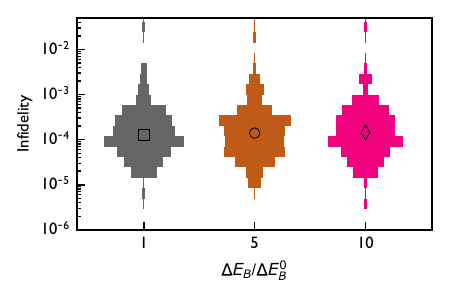}
    \caption{Infidelity histograms for the channel-shifted 5\%~Ge quantum well, including all tuning methods applied, as in \cref{fig:fig_channel_var}(d), as a function of the Zeeman splitting difference $\Delta E_B$ between the ground and excited valley states. Here, $\Delta E_B^0$ is the characteristic value of $\SI{10}{\mega\hertz}$ used in the main text.}
    \label{fig:app_DeltaEB}
\end{figure}

We briefly comment on the dependence of the simulations on the difference in Zeeman splittings $\Delta E_B$ between the ground and excited valley states.
In \cref{fig:app_DeltaEB}, we illustrate the dependence of the shuttling infidelity distribution on $\Delta E_B$ for the highest-fidelity simulations used in this work: shuttling in a 5\% Ge quantum well with all tuning methods applied.
We see that the distribution of infidelities changes only slightly when increasing the value of $\Delta E_B$ by an order of magnitude.
We can understand this result as follows. The metric used to evaluate the fidelity in this work considers, as a conservative assumption, only the population ending up in the ground state \cite{wood2018quantification}. In the absence of fast relaxation dynamics, as discussed in the main text, we do not expect dots in the excited valley state to return to the ground state with significant probability.
Since the energy scale of the $\Delta E_B$ term has little effect on the Landau-Zener transition mechanism, $\Delta E_B$ is not a determining factor for the infidelity results, as confirmed in \cref{fig:app_DeltaEB}.
%This further implies that similar results can be obtained with just a two-state simulation of valley-states, if valley-relaxation is discarded as is done here.

\section{Path selection algorithm}\label{app:graph_traversal}

\begin{figure}
    \centering
    \includegraphics[width=0.3\textwidth]{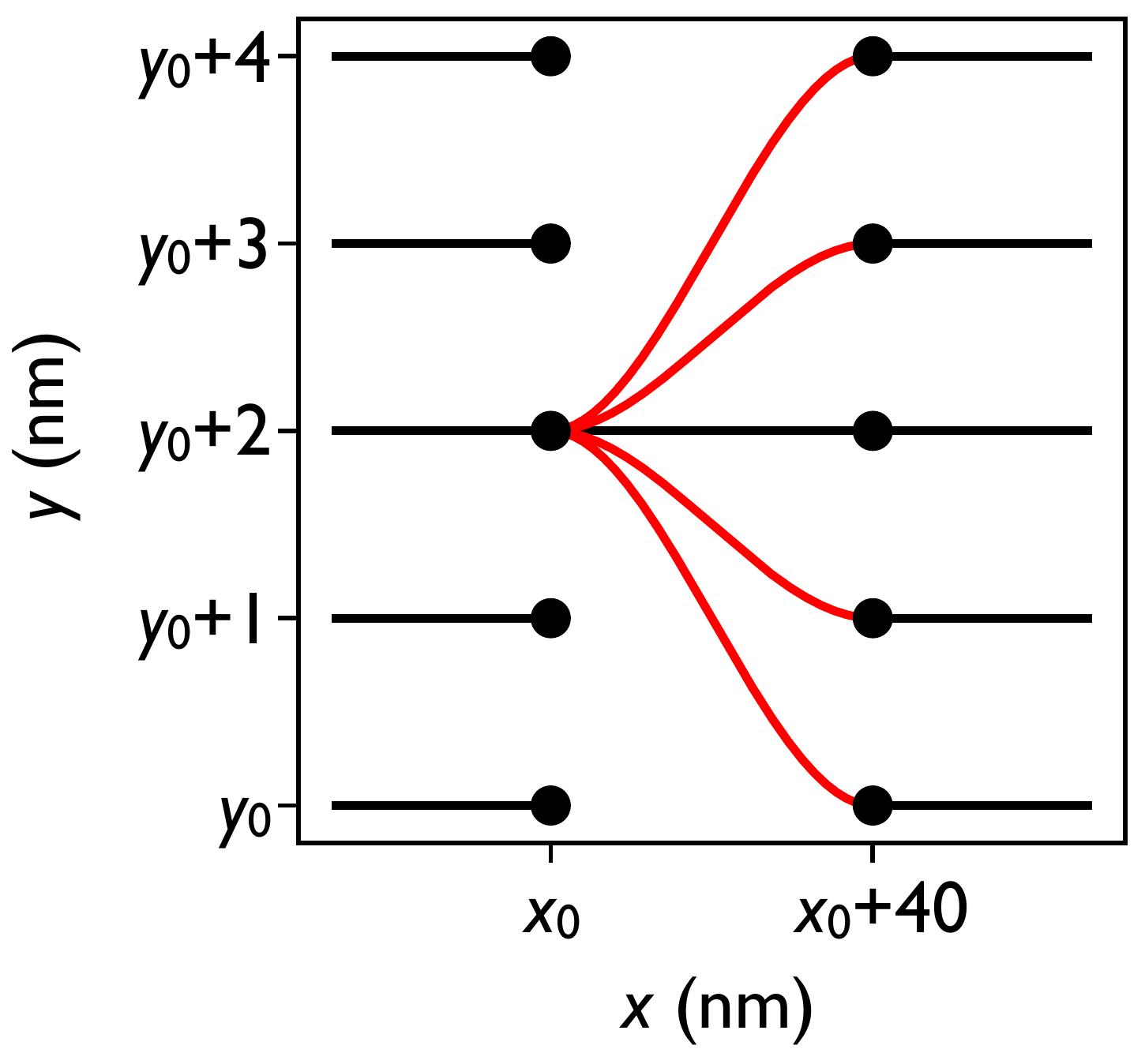}
    \caption{Schematic illustration of the possible paths a dot can take across a device, implemented in our path traversal algorithm. Straight segments of fixed length are connected by transition zones of length \SI{40}{\nano\meter} along the shuttling direction ($\hat x$). The possible transitions from one such straight segment are illustrated, including the option to continue straight ($\Delta p = \Delta y = 0$), shown in black, and the option to modify the parameter ($|\Delta p| = |\Delta y| > 0$), shown in red.}
    \label{fig:channel_shift_schematic}
\end{figure}

To implement segmented channel-shifting strategies, we need to carefully choose our path across the valley splitting landscape, either by modifying the $y$ coordinate of the dot or by modifying the vertical electric field $E_z$.
In this work, we adopt a heuristic graph traversal algorithm to make this path selection.
The valley splitting landscape is discretized into pixels of size \SI{1}{\nano\meter} $\times$ \SI{1}{\nano\meter} for $y$ tuning, or \SI{.1}{\milli\volt/\meter} $\times$ \SI{1}{\nano\meter} for $E_z$-tuning. 
To make use of common graph traversal algorithms, we define a graph representing the possible paths across this discretized landscape. 
%The nodes of the graph represent the values of the tuning parameter in each independent section, and edges the possible changes between the sections.
The edges of the graph represent possible paths the dot could take across the device.
These paths include straight segments of a fixed length with \SI{40}{\nano\meter} transition zones between the segments.
(We set all segment lengths to \SI{1}{\micro\meter} in this work, except briefly in Appendix~\ref{app:app_sweeps}.)
[See Fig.~\ref{fig:fig_channel_var}(a) for an example trajectory, and Fig.~\ref{fig:channel_shift_schematic} for an illustration of the method.]
The parameter being optimized (either $y$ or $E_z$) remains constant along a straight segment (black lines in Fig.~\ref{fig:channel_shift_schematic}), but is modified smoothly and continuously in the transition zones (red lines in Fig.~\ref{fig:channel_shift_schematic}).
Transitions between the optimized segments are heuristically defined as cubic polynomial functions whose derivatives go to zero at the endpoints of the transition.

An ideal path should have the following properties: (1) the minimum valley splitting along each segment should be large enough to avoid Landau-Zener transitions between the valley states, and (2) the transitions between segments should be as short as possible, to avoid increasing the effective shuttling velocity along long, steep transitions.
To achieve paths that globally optimize these two properties, we assign weights $w(e)$ to each edge $e$ according to the following rules, which penalize both low minimum valley splittings ($\min_e E_v$) along a given segment, and transitions with large changes $\Delta p$ in the optimization parameter, where $p=y$ or $E_z$ :
%\red{(The following equation is not dimensionally consistent when $p=y$, so this is confusing. Also, the $N$ notation seems like it should refer to a dimensionless integer, so this is confusing.)}
\begin{equation} \label{eq:path_weights}
    w(e) = 
    \begin{cases}
        100 \cdot \frac{|\Delta p|}{\Delta p^\text{max}} & \min_e E_v \geq T_v, \\
        N +100 \cdot \frac{|\Delta p|}{\Delta p^\text{max}} + \frac{(T_v-\min_e E_v)^2}{1 \SI{}{\micro\electronvolt}^2} & \min_e E_v < T_v ,
    \end{cases}
\end{equation}
where $T_v$ is a threshold value for the valley splitting (measured in \SI{}{\micro\electronvolt}), $\Delta p^\text{max}$ is the maximum variation of $p$ allowed between the segments, and $N > 100$ is a large number chosen such that a transition edge with $\min_e E_v \geq T_v$ has a smaller weight than a straight edge with $\min_e E_v < T_v$.
We note that the exact values of the weights assumed in Eq.~(\ref{eq:path_weights}) are relatively unimportant for our purposes, as long as both low-$E_v$ minima and transitions with large changes $|\Delta p|$ are penalized, relative to paths with large minimum $E_v$ and no transitions.
In this work, we choose $T_v =$ \SI{50}{\micro\electronvolt} and $N=1000$, with $\Delta p^\text{max} =$ \SI{100}{\nano\meter} for the channel-shifting protocol and $\Delta p^\text{max} =$ \SI{10}{\milli\volt\per\nano\meter} for $E$-field modulation. 
After assigning weights to each edge in the graph, the graph traversal algorithm minimizes $w$ to generates an optimized path across a given $E_v$ landscape.

\section{Further characterization of the $E_z$ modulation strategy} \label{app:Efield_variation}
\begin{figure}
    \centering
    \includegraphics[width=0.4\textwidth]{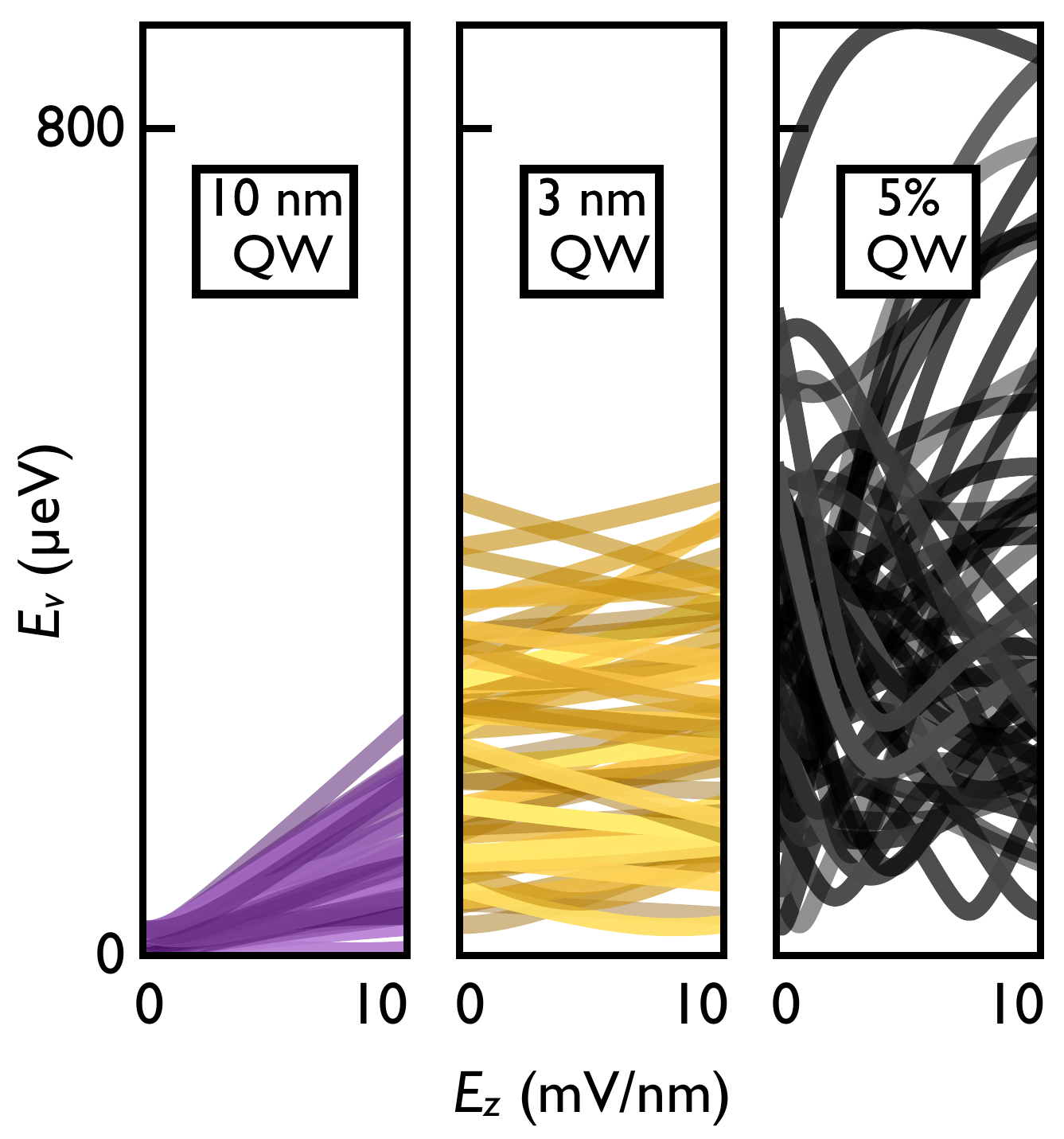}
    \caption{Dependence of valley splitting on the vertical electric field, for the \SI{10}{\nano\meter} quantum well (left), the \SI{3}{\nano\meter} quantum well (middle), and the \SI{5}{\%} Ge quantum well (right).
    Each plot shows the variation of $E_v$ as a function of $E_z$ for 100 instantiations of alloy disorder.}
    \label{fig:efield_adjust_supp}
\end{figure}

\begin{figure}
    \centering
    \includegraphics[width=0.4\textwidth]{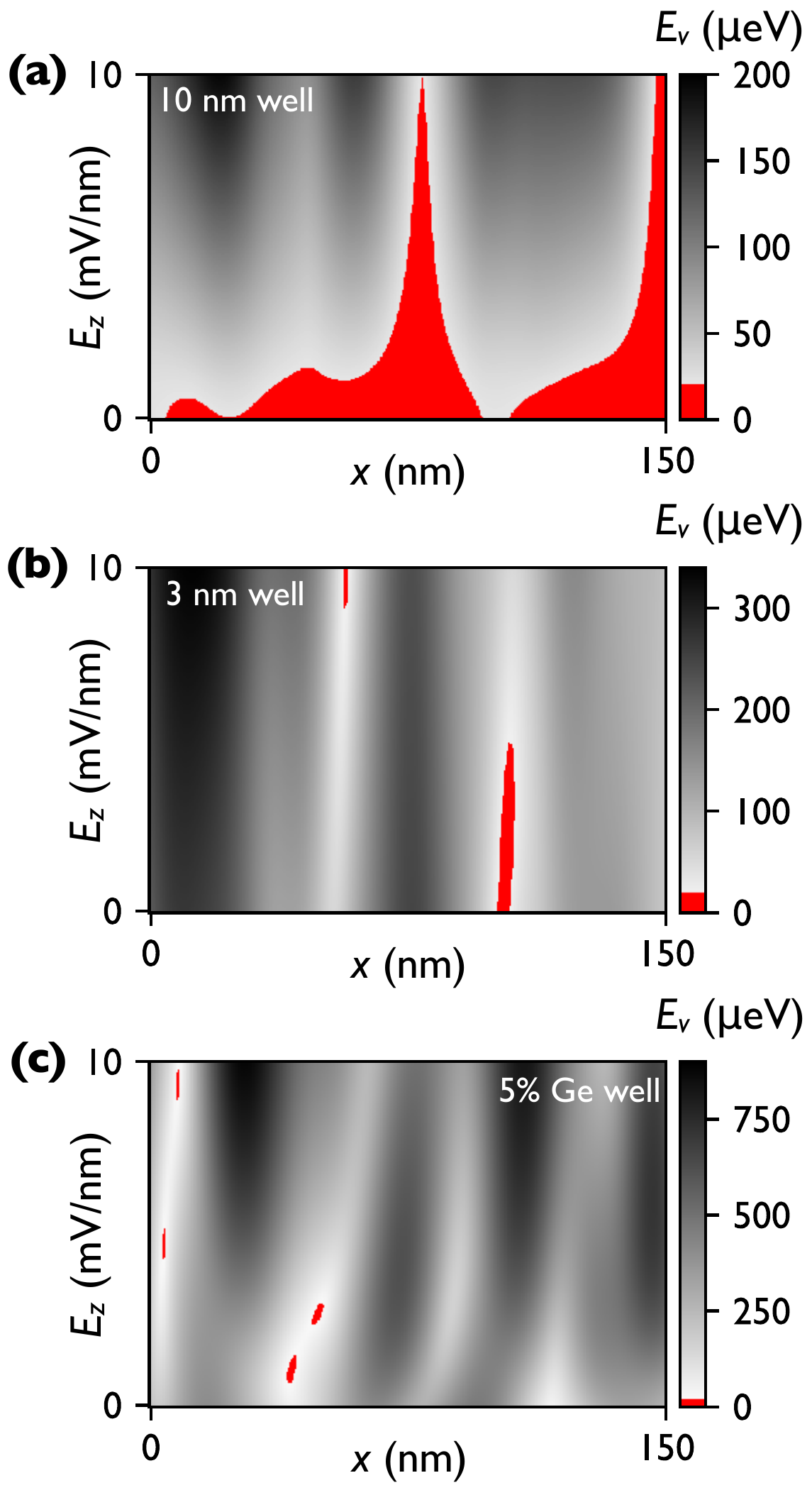}
    \caption{Sample valley splitting landscapes as a function of position, $x$, and vertical field $E_z$, for (a) a \SI{10}{\nano\meter} QW, (b) a \SI{3}{\nano\meter} QW, and (c) a 5\% Ge QW. In all panels, we highlight regions where $E_v < $ \SI{20}{\micro\electronvolt} in red.}
    \label{fig:efield_example_vs}
\end{figure}

In this Appendix, we further examine the performance of the electric-field ($E_z$) modulation strategy.
In \cref{sec:results_efield}, we highlighted that $E_z$ modulation offers improvements in shuttling fidelity for 5\% Ge quantum wells, but not for the other quantum wells analyzed in this work.
We now explain these differences.
First, for devices with sharp interfaces, it is well known that the valley splitting scales linearly with $E_z$, as stronger $E_z$ increases the wavefunction penetration into the top barrier \cite{friesen2007theory}.
For more realistic quantum wells with diffuse interfaces, we expect the average valley splitting to also scale linearly with $E_z$, as increasing $E_z$ forces the wavefunction to overlap with more high-Ge layers \cite{losert2023practical}.
However, for recently demonstrated heterostructures, like narrow quantum wells or quantum wells with a high Ge concentration, $E_v$ has a nontrivial dependence on $E_z$, which we characterize below. 

In \cref{fig:efield_adjust_supp}, we show the variation of $E_v$ as a function of $E_z$ for 100 instantiations of each quantum well.
For the \SI{10}{\nano\meter} quantum well (left), we notice that $E_v$ is largely monotonically increasing with $E_z$, since larger $E_z$ pulls the dot strongly into the top barrier, increasing its overlap with high-Ge layers.

For the \SI{3}{\nano\meter} quantum well (middle), $E_v$ is no longer a monotonic function of $E_z$.
Moreover, since the dot is tightly confined inside the narrow well, the dot position (and thus $E_v$) is not very tunable as a function of $E_z$.
In contrast, for the \SI{5}{\%} Ge quantum well (right), the quantum well is much wider and has strong local fluctuations of the Ge concentration, so small shifts in the wavefunction position can significantly alter $E_v$ as a function of $E_z$.
In this case, if we want to use $E_z$ as a tuning knob to avoid low-$E_v$ regions, this scheme can be very effective.

Figures~\ref{fig:efield_example_vs}(a)-\ref{fig:efield_example_vs}(c) show sample valley splitting landscapes as a function of position, $x$, and vertical field, $E_z$, for the case of (a) a \SI{10}{\nano\meter} quantum well, (b) a \SI{3}{\nano\meter} quantum well, and (c) a 5\% Ge quantum well.
In all plots, regions with $E_v <$ \SI{20}{\micro\electronvolt} are highlighted in red.
As consistent with data in \cref{fig:efield_adjust_supp}, we see in \cref{fig:efield_example_vs}(a) that $E_v$ tends to increase monotonically with $E_z$, for any location in a \SI{10}{\nano\meter} quantum well. 
However, since this quantum well has a relatively low average $E_v$, despite the large vertical field, significant regions of low $E_v$ cannot be avoided.
Indeed, we see in \cref{fig:efield_example_vs}(a) two $x$ locations where $E_v <$ \SI{20}{\micro\electronvolt} regardless of $E_z$.
For the \SI{3}{\nano\meter} well, we have a larger average $E_v$, and therefore fewer zones where $E_v$ is dangerously small. 
However, since $E_v$ is not highly tunable as a function of $E_z$ in these quantum wells, regions of low $E_v$ tend to persist over a wide range of $E_z$.
When $E_z$ is held constant over a distance of \SI{1}{\micro\meter}, these regions are difficult to avoid.

The situation is somewhat improved for the 5\% Ge well in \cref{fig:efield_example_vs}(c). 
First, the large amount of alloy disorder creates much larger average valley splittings.
Additionally, $E_v$ has a non-monotonic dependence on $E_z$, which makes it more likely that we can find an $E_z$ value that avoids all low $E_v$ for a given shuttling trajectory.
We find, however, that taking advantage of such non-monotonicity requires imposing fairly short segments of constant $E_z$.
If segments are too large, one is always likely to encounter low $E_v$ values.
Still, for the \SI{1}{\micro\meter} segments used in the simulations reported in the main text, we find $E_z$ modulation does offer improved fidelities for the 5\% Ge quantum well.

\section{Further characterization of the dot-elongation strategy} \label{app:squeeze}

In this Appendix, we provide further details on the performance of the dot-elongation tuning strategy.
As a reminder, we have considered isotropic dots with orbital splittings $\hbar \omega_x = \hbar \omega_y = $ \SI{2}{\milli\electronvolt}, and ``elongated'' dots with orbital splittings $\hbar \omega_x = $ \SI{1}{\milli\electronvolt} and $\hbar \omega_y = $ \SI{4}{\milli\electronvolt}.
While these choices yield dots with the same area, we find that they yield very different shuttling infidelities. 

\begin{comment}
In \cref{fig:squeezing_performance}(a), we reproduce the shuttling infidelity histograms of \cref{fig:fig_channel_var}(d) for the \SI{3}{\nano\meter} quantum well, which include three different tuning strategies: lateral channel-shifting only (gray), channel-shifting plus bipartite velocity-modulation (brown), or the full combination of channel-shifting, bipartite velocity-modulation, and dot-elongation (blue). 
In addition here, we plot the infidelity histogram for channel shifting plus dot elongation, \textit{without} velocity modulation (pink).
For the velocities and orbital energies considered here, we see that dot elongation can have a stronger impact on the infidelity than velocity modulation.
\red{(Is this because there no velocity-switching happens for this $E_v^\text{min}$ setting? If so, then we need to modify the figure or possibly remove it.)}
In the remainder of this Appendix, we examine the general factors that improve shuttling fidelities for elongated dots, and provide theoretical derivations of the statistical behavior of the valley splitting distributions.
\end{comment}

\begin{figure*}
    \centering
    \includegraphics[width=0.8\textwidth]{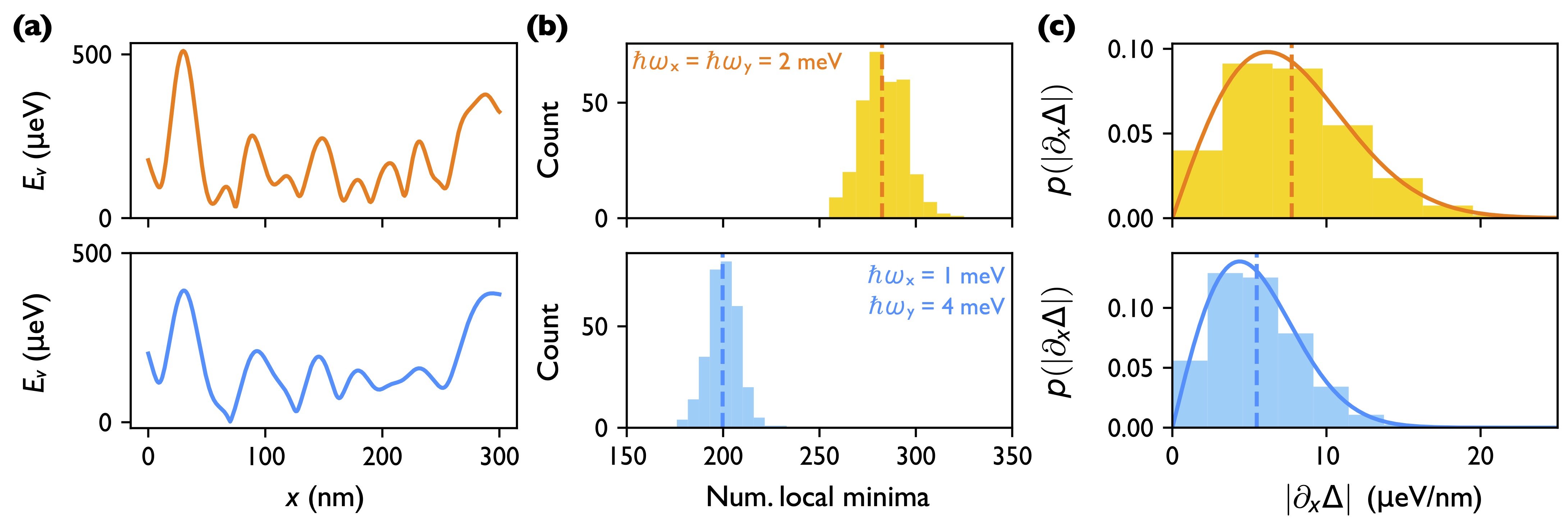}
    \caption{Elongating the dot in the shuttling direction, and squeezing the dot in the transverse direction, reduces the magnitude of $E_v$ fluctuations, significantly boosting shuttling fidelities.
    %(a) Histograms of shuttling infidelities for the \SI{3}{\nano\meter} QW using various modulation strategies, including channel shifting (left, gray), channel shifting plus bipartite velocity modulation (second from left, brown), channel shifting plus dot elongation (second from right, pink), and channel shifting plus bipartite velocity modulation plus dot elongation (right, blue). 
    %The data for channel shifting, shifting plus velocity modulation, and shifting plus velocity modulation plus dot elongation are the same as shown in \cref{fig:fig_channel_var}(d).
    (a) Example $E_v$ traces as a function of position for the \SI{3}{\nano\meter} QW, for isotropic (top, orange) and elongated (bottom, blue) quantum dots.
    (b) Histograms of the number of local $E_v$ minima along a \SI{10}{\micro\meter} shuttling path, for 300 iterations with the \SI{3}{\nano\meter} QW, using isotropic (top, yellow) and elongated (bottom, blue) dots. 
    Dashed lines indicate the expected number of local minima in each case ($E[N] \approx 282$ for isotropic dots and $\approx 200$ for elongated dots), computed with \cref{eq:exp_num_minima_final}. 
    (c) Histograms of $|\partial_x \Delta|$ across the same 300 shuttling trajectories. 
    Solid lines indicate the theoretical probability density functions, computed with \cref{eq:deriv_rayleigh}. 
    Dashed lines indicate the expected mean gradient, $E[|\partial_x \Delta|]$, computed with \cref{eq:exp_gradient}, which are 7.7 for isotropic dots and 5.5 for elongated dots. 
    }
    \label{fig:squeezing_performance}
\end{figure*}

Elongating the dot in the shuttling direction has three main effects on the shuttling procedure.
First, it increases the size of the dot along the shuttling direction, thereby reducing the effective length scale of the shuttling process. 
Since the characteristic length of valley splitting correlations depends entirely on the dot size, this means the moving dot will encounter proportionately fewer regions of low $E_v$ on average.
In \cref{fig:squeezing_performance}(a), we plot the valley splitting for an isotropic (orange) vs an elongated (blue) dot, for the same landscape; here we can clearly see the longer correlation length scale in the blue data, and the larger number of regions with low $E_v$ in the orange data.
To create these plots, we generate an atomistic model of a heterostructure and raster the lateral confinement potential across this heterostructure, computing $\Delta$ for each potential center $x$, using the method outlined in the main text.
In \cref{fig:squeezing_performance}(b), we histogram the number of local $E_v$ minima observed along 300 straight shuttling trajectories, for both isotropic (yellow) and elongated (blue) dots.
To avoid the massive computational overhead of populating \SI{10}{\micro\meter}-wide heterostructures atom-by-atom, we generate these $E_v$ landscapes randomly, using the methods outlined in Appendix~\ref{app:random_fields}.
%, \hl{where again we use the same landscapes for both sets of calculations}. 
Results are shown for a trajectory length of \SI{10}{\micro\meter} in a \SI{3}{\nano\meter} quantum well.
Clearly, there are fewer local minima for elongated dots, leading to fewer opportunities for Landau-Zener excitations.
Here, we also indicate the expected number of local minima in each case [vertical lines in \cref{fig:squeezing_performance}(b)], as derived later in this Appendix. 

The second effect of dot elongation is to increase the tunability of the valley splitting via the channel-shifting technique. 
Just as elongating the dot in the shuttling direction increases the characteristic length scale of valley splitting fluctuations along $\hat x$, narrowing the dot transverse to the shuttling trajectory reduces the characteristic fluctuation length scale along $\hat y$. 
Thus, for a fixed channel width, when we employ the channel-shifting strategy, the path-selection algorithm is effectively able to search over more variations in the $E_v$ landscape, allowing it to identify better shuttling trajectories. 

Finally, the third impact of elongating the dot is to reduce the effective shuttling velocity. 
This is important when passing through a narrow energy gap, because it reduces the probability of Landau-Zener excitations.
To clarify this point, we examine the rate of change of the inter-valley coupling of the moving dot, $\partial_t \Delta$.
Using the chain rule, we have
\begin{equation} \label{eq:partial_delta}
    \partial_t \Delta = v \partial_x \Delta = v \left(\partial_x \Delta_R + i \partial_x \Delta_I \right)
\end{equation}
where $v$ is the shuttling velocity (assumed to be in the $x$-direction) and $\Delta_{R/I}$ refer to the real and imaginary components of $\Delta$.
The rate of change of $\Delta$ is therefore directly related to the spatial derivative $\partial_x \Delta$.
In \cref{fig:squeezing_performance}(c), we plot histograms of $|\partial_x \Delta|$ along 300 shuttling trajectories for the \SI{3}{\nano\meter} quantum well, for both isotropic (yellow) and elongated (blue) dots, using the same set of landscapes as in (b).
%\hl{Again, we use the same set of landscapes for both calculations.}
While both of these distributions exhibit some spread, the average gradient is clearly smaller for the elongated dots. 
Here, we also indicate the theoretically calculated mean gradients (vertical dashed lines) and probability density functions (solid lines) for $|\partial_x \Delta|$, as derived below.

\emph{Calculation of the $\partial_x\Delta$ distributions.}
Since the inter-valley coupling $\Delta$ fluctuates throughout the device, the derivatives on the right-hand-side of \cref{eq:partial_delta} are random variables, and we may evaluate their statistics.
To do so, we compute the variance of $\partial_x \Delta $:
\begin{equation} \label{eq:var_partial_delta}
\begin{split}
    \text{Var}\left[\partial_x \Delta \right] &= 2 \text{Var}\left[ \partial_x \Delta_R \right] \\
    & = 2 \text{Var}\left[ \lim_{\delta_x \rightarrow 0} \frac{1}{\delta_x} \left( \Delta_R(x + \delta_x) - \Delta_R(x) \right) \right] \\
    & = 2 \lim_{\delta_x \rightarrow 0}\frac{1}{\delta_x^2} \left( \text{Var}[\Delta_R(x+\delta_x)] + \text{Var}[\Delta_R(x)] \right. \\
    & \left. \;\;\;\; \;\;\;\; - 2 \text{Cov}[\Delta_R(x + \delta_x), \Delta_R(x) ] \right) \\
    & = 2 \lim_{\delta_x \rightarrow 0}\frac{1}{\delta_x^2} \left( \sigma_\Delta^2 - e^{-\delta_x^2 / 2 a_x^2}\sigma_\Delta^2 \right) \\
    & = \frac{\sigma_\Delta^2 }{a_x^2}
\end{split}
\end{equation}
In the first line of \cref{eq:var_partial_delta}, we use the identity $\text{Var}[A + i B] = \text{Var}[A] + \text{Var}[B]$ and the symmetry between $\Delta_R$ and $\Delta_I$. 
In the second line, we use the definition of a derivative.
In the third line, we interchange the order of the variance and the limit, and we use the identity $\text{Var}[A - B] = \text{Var}[A] + \text{Var}[B] - 2 \text{Cov}[A,B]$.
In the fourth line, we use $\text{Var}[\Delta_R(x + \delta_x)] = \text{Var}[\Delta_R(x)] = \sigma_\Delta^2 / 2$, and we use \cref{eq:2point} to evaluate $\text{Cov}[\Delta_R(x+\delta_x), \Delta_R(x)]$.
Finally, in the fifth line, we evaluate the limit.
As consistent with the central-limit theorem, the quantity $\partial_x\Delta$ is thus seen to be a circular complex Gaussian random variable, centered at the origin, with variance given by \cref{eq:var_partial_delta}.

Since $\partial_x \Delta$ is complex, it is also interesting to evaluate the distribution of $|\partial_x \Delta|$. 
This quantity will have a Rayleigh distribution, whose probability density function is given by
\begin{equation} \label{eq:deriv_rayleigh}
    p_\text{Rayleigh}(z) = \frac{z^2}{\sigma^2} \exp \left( -z^2 / 2 \sigma^2 \right),
\end{equation}
where we define the spread parameter as 
\begin{equation} \label{eq:sigma_gradient}
    \sigma = \frac{1}{2} \text{Var}\left[\partial_x \Delta\right] = \frac{\sigma_\Delta^2}{2 a_x^2}.
\end{equation}
The probability density functions for $|\partial_x \Delta|$ from \cref{eq:deriv_rayleigh} are shown as solid lines in \cref{fig:squeezing_performance}(c).
The expected value of $|\partial_x \Delta|$ is likewise given by
\begin{equation} \label{eq:exp_gradient}
    E[|\partial_x \Delta|] = \sqrt{\frac{\pi}{2}} \sigma .
\end{equation}
Evaluating $E[|\partial_x \Delta|]$ for the elongated and isotropic dots gives the results plotted as vertical dashed lines in \cref{fig:squeezing_performance}(c).

\emph{Estimating the number of valley-splitting minima.} 
We now compute the expected number of valley splitting minima along a straight shuttling trajectory.
Typical results are shown with vertical dashed lines in \cref{fig:squeezing_performance}(b).
We follow the approach of Ref.~\cite{bloomfield2016number}, which we reproduce below for completeness.
Note that we restrict the analysis to just one spatial dimension.

Using the Kac-Rice formula, the number of local minima is given by \cite{bardeen1968statistics}
\begin{equation}
    N = \frac{1}{2} \int_x \; dx \; \delta(\partial_x E_v^2) | \partial_x^2 E_v^2 |,
\end{equation}
where the factor of 1/2 accounts for the fact that half of the extrema points (where $\partial_x E_v^2 = 0$) are minima, and $\delta (\partial_x E_v^2)$ is a delta-function that activates when $E_v^2$ is at an extremum.
Using the identity
\begin{equation} \label{eq:delta_func_id}
    \delta(f(x)) = \sum_i \frac{\delta(x-x_i)}{|f'(x_i)|} \text{, where } f(x_i) = 0 ,
\end{equation}
we see that the remaining factor $|\partial_x^2 E_v^2|$ allows the integral to count the number of extrema in $E_v^2$.
Here, we use $E_v^2$ instead of $E_v$ to simplify the calculation, without changing the results.
Mathematically, the quantity $E_v^2$ is a $\chi^2$ random field with two contributing Gaussian random fields, $E_v^2 = 4\Delta_R^2 + 4\Delta_I^2$.
The derivatives of $E_v^2$ are given by
\begin{equation} \label{eq:derivs_Ev2}
    \begin{split}
        \partial_x E_v^2 &= 8 \Delta_R \partial_x \Delta_R + 8 \Delta_I \partial_x \Delta_I , \\
        \partial_x^2 E_v^2 &= 8 (\partial_x \Delta_R)^2 + 8 \Delta_R \partial_x^2 \Delta_R + 8 (\partial_x \Delta_I)^2 + 8 \Delta_I \partial_x^2 \Delta_I.
    \end{split}
\end{equation}
We compute the expectation value, $E[N]$, by averaging over all possible configurations of the inter-valley coupling:
\begin{multline} \label{eq:exp_num_local_min}
    E[N] = \frac{1}{2} \int \; d\Phi \; P(\Phi) \int_x \; dx \; \\ \times \delta(8 \Delta_R \partial_x \Delta_R + 8 \Delta_I \partial_x \Delta_I)  |\partial_x^2 E_v^2| ,
\end{multline}
where we use $\Phi$ as shorthand notation for the random field configurations of $\Delta_R(x)$, $\Delta_I(x)$, and their derivatives.
More explicitly, the integral element is given by
\begin{equation}
    d\Phi = d\Delta_R d\Delta_I d(\partial_x \Delta_R) d(\partial_x \Delta_I) d(\partial^2_x \Delta_R) d(\partial^2_x \Delta_I) ,
\end{equation}
and the total probability density function is given by
\begin{multline}
    P(\Phi) = P_{\Delta, \partial_x^2 \Delta}(\Delta_R, \partial_x^2 \Delta_R) P_{\Delta, \partial_x^2 \Delta}(\Delta_I, \partial_x^2 \Delta_I) \\ \times
    P_{\partial_x \Delta}(\partial_x \Delta_R) P_{\partial_x \Delta}(\partial_x \Delta_I).
\end{multline}
Note that we do not include higher order derivatives here, since they do not appear in the integrand.
Also note that the random fields $\Delta_R$ and $\Delta_I$ are independent by definition, and any covariance of the form $\langle \Delta_i(x)  \partial_x \Delta_i(x) \rangle$ or $\langle \partial_x^2 \Delta_i(x)  \partial_x \Delta_i(x) \rangle$ must vanish due to the $x \rightarrow -x$ symmetry of the integral, where we use angle brackets $\langle \cdot \rangle$ to denote the expectation value of a quantity over its field configurations.
Thus, we are left with two probability density functions to compute: one for the first derivatives of the field, $P_{\partial_x \Delta}(\partial_x \Delta_j)$, as well as the joint probability density function for the fields and their second derivatives, $P_{\Delta, \partial_x^2 \Delta}(\Delta_j, \partial_x^2 \Delta_j)$.
We showed above that $\partial_x \Delta_R$ and $\partial_x \Delta_I$ are Gaussian random variables with zero mean and variance $\sigma_\Delta^2 / 2 a_x^2$.
We therefore have 
\begin{equation} \label{eq:prob_dist_field_deriv}
    P_{\partial_x\Delta}(\partial_x \Delta_j) = \frac{a_x }{\sqrt{\pi}\sigma_\Delta} \exp \left( -(\partial_x \Delta_j)^2 a_x^2 / \sigma_\Delta^2 \right).
\end{equation}

Finally, we compute $P_{\Delta, \partial_x^2 \Delta}(\Delta_j, \partial_x^2 \Delta_j)$. 
To do this, we need covariances of the form $\langle \Delta_j (x) \partial_x^2 \Delta_j(x) \rangle$. 
By expressing the random fields in the reciprocal basis, 
\begin{equation}
    \Delta_j = \int \frac{dk}{2\pi} e^{ikx} \tilde \Delta_j (k) ,
\end{equation}
we can evaluate
\begin{equation} \label{eq:deriv_correlators}
    \langle \Delta_j (x) \partial_x^2 \Delta_j(x) \rangle = -\frac{1}{2\pi} \int dk \; k^2 P(k) ,
\end{equation}
where the power spectrum $P(k)$ is the Fourier transform of the covariance function $\langle \Delta_j (x) \Delta_j(x') \rangle$, which is provided in \cref{eq:2point}.
Hence, we find 
\begin{equation}
    P(k) = a_x \sigma_\Delta^2 \sqrt{\pi/2} \exp( -a_x^2 k^2 / 2) .
\end{equation}
We then evaluate \cref{eq:deriv_correlators}, obtaining 
\begin{equation} \label{eq:cov_field_second_deriv}
    \langle \Delta_j (x) \partial_x^2 \Delta_j(x) \rangle = - \frac{\sigma_\Delta^2}{2 a_x^2}.
\end{equation}
Using the same technique, we can evaluate the variance as
\begin{equation} \label{eq:var_second_deriv}
    \langle (\partial_x^2 \Delta_j(x))^2 \rangle = \frac{1}{2\pi} \int dk k^4 P(k) = \frac{3 \sigma_\Delta^2}{2 a_x^4}.
\end{equation}
Since the fields $\Delta_j$ and their derivatives are Gaussian random variables with zero mean, we can define the joint probability density function as
\begin{equation} \label{eq:field_second_deriv_joint_pdf}
\begin{split}
    &P_{\Delta, \partial_x^2 \Delta}(\Delta_j, \partial_x^2 \Delta_j) \\
    & = \frac{1}{\sqrt{(2\pi)^2|\mathbf{\Sigma}|}} \exp \left( -\frac{1}{2} \mathbf{v}^T \mathbf{\Sigma}^{-1} \mathbf{v} \right) \\
    & = \frac{a_x^2}{\sqrt{2}\pi \sigma_\Delta^2} \exp \left( - \frac{3 \Delta_j^2 + 2 a_x^2 \Delta_j \partial_x^2 \Delta_j + a_x^4 (\partial_x^2 \Delta_j)^2}{2 \sigma_\Delta^2} \right) ,
\end{split}
\end{equation}
where $\mathbf{v}^T = (\Delta_j, \partial_x^2 \Delta_j)$ and the covariance matrix is given by
\begin{equation}
    \mathbf{\Sigma} = \frac{\sigma_\Delta^2}{2}
    \begin{pmatrix}
        1 & -a_x^{-2}  \\
        -a_x^{-2} & 3 a_x^{-4}  \\
    \end{pmatrix}.
\end{equation}
Here, we used \cref{eq:cov_field_second_deriv} to populate the off-diagonal elements of $\mathbf{\Sigma}$, and \cref{eq:var_second_deriv} to populate the remaining diagonal element.

We are then in a position to evaluate \cref{eq:exp_num_local_min}.
First, we eliminate the $\delta$-function and the integral over $\partial_x(\Delta_I)$ by setting $\partial_x \Delta_I = -\Delta_R \partial_x \Delta_R / \Delta_I $, which yields
\begin{widetext} 
\begin{multline} \label{eq:final_integral}
     E[N] = \frac{1}{2} \int_x \; dx \int \; d\Delta_R \; d\Delta_I \; d(\partial_x \Delta_R) \; d(\partial^2_x \Delta_R) \; d(\partial^2_x \Delta_I) \; \\ \times
     P_{\Delta, \partial_x^2 \Delta}(\Delta_R, \partial_x^2 \Delta_R) P_{\Delta, \partial_x^2 \Delta}(\Delta_I, \partial_x^2 \Delta_I) P_{\partial_x \Delta}(\partial_x \Delta_R) P_{\partial_x \Delta}\left(\frac{-\Delta_R \partial_x \Delta_R}{\Delta_I} \right) |\partial_x^2 E_v^2 | \left( | 8 \Delta_I | \right)^{-1}.
\end{multline}
\end{widetext}
Here, the probability density functions are given in \cref{eq:prob_dist_field_deriv} and \cref{eq:field_second_deriv_joint_pdf}, and the term $|\partial_x^2 E_v^2|$ is given in \cref{eq:derivs_Ev2}.
The final term $(|8\Delta_I|)^{-1}$ comes from the evaluation of the $\delta$-function, where we have used Eq.~(\ref{eq:delta_func_id}).
Finally, we evaluate \cref{eq:final_integral} numerically, obtaining
\begin{equation} \label{eq:exp_num_minima_final}
    E[N] \approx 0.4 \times \frac{x_\text{tot}}{a_x},
\end{equation}
where $x_\text{tot}$ is the total shuttling distance.
The dashed lines indicating $E[N]$ in \cref{fig:squeezing_performance}(b) were computed using \cref{eq:exp_num_minima_final}.
Since $E[N]$ scales as $1/a_x$, we see that reducing the orbital energy from 2 to \SI{1}{\milli\electronvolt}, should cause $E[N]$ to drop by a factor of $1/\sqrt{2}$.

\section{Minimum $E_v$ along a shuttling channel} \label{app:min_Ev_characterization}

\begin{figure*}
    \centering
    \includegraphics[width=0.8\textwidth]{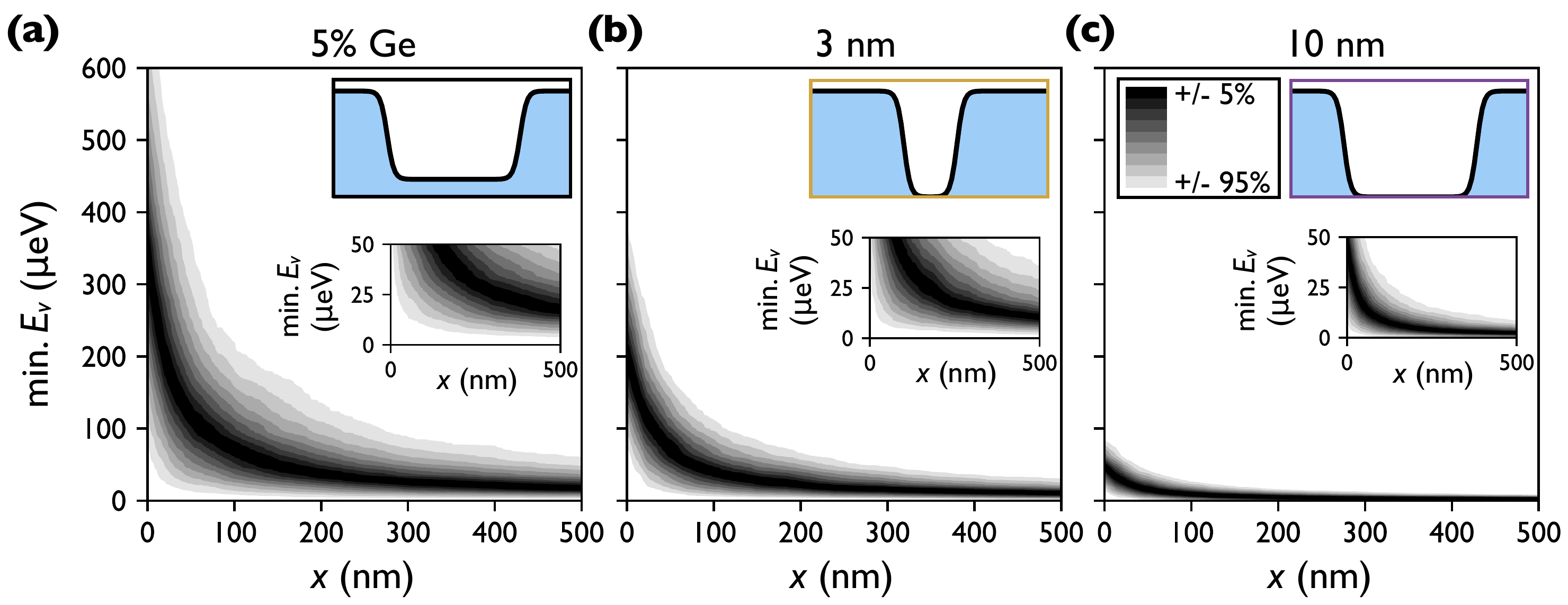}
    \caption{The expected minimum $E_v$ along a 1D shuttling channel quickly falls toward zero as the shuttling channel length increases. (a) For the 5\% Ge quantum well, we plot the distribution of the minimum $E_v$ experienced over 1000 simulated shuttling channels, as the length of the shuttling channel grows. We assume a vertical electric field of $E_z = 5$~\SI{}{\milli\volt\per\nano\meter} and orbital energy splittings of $\hbar \omega_\text{orb} = 2$~\SI{}{\milli\electronvolt}. In (b) and (c) we plot results obtained in the same way for the \SI{3}{\nano\meter} and \SI{10}{\nano\meter} quantum wells, respectively. The insets display the same data over a reduced $y$-axis range. }
    \label{fig:min_Ev_fig}
\end{figure*}

In this Appendix, we compute the expected minimum valley splitting along a 1D shuttling channel to justify our claim that heterostructure modification, alone, is insufficient for achieving high-fidelity shuttling results.
In the disordered regime, even in heterostructures with large average $E_v$, a quantum dot is extremely likely to encounter regions of low $E_v$ during shuttling, for realistically long shuttling channels, as demonstrated in Fig.~\ref{fig:min_Ev_fig}.
For the 5\% Ge quantum well, the \SI{3}{\nano\meter} quantum well, and the \SI{10}{\nano\meter} quantum well, we plot the distribution of the minimum valley splitting experienced along a 1D shuttling channel, as we increase the channel distance. 
These distributions are determined empirically from 1,000 simulations of random valley splitting landscapes, as described in Appendix~\ref{app:random_fields}.
For the \SI{10}{\nano\meter} quantum well, we observe that the expected minimum $E_v$ is below \SI{10}{\micro\electronvolt} for \SI{500}{\nano\meter} shuttling distances. 
Even in high $E_v$ quantum wells, like the 5\% Ge device, we expect a minimum $E_v$ of $\sim$~\SI{20}{\micro\electronvolt} at distances of ~\SI{500}{\nano\meter}.

%\red{That said, for relatively short shuttling trajectories, there is nonetheless significant variation of the minimum valley splitting, especially in QWs with high average $E_v$.
%For shuttling over \SI{200}{\nano\meter}, the minimum $E_v$ in the 5\% Ge QW ranges between 0 and $\sim$~\SI{125}{\micro\electronvolt}, at the 5-95 percentile range.
%Thus, large-scale shuttling and $E_v$ mapping experiments will be critical to determine the true experimental behavior of the valley splitting across a device.}

%\nocite{*}

\bibliography{bibliography}% Produces the bibliography via BibTeX.

\begin{thebibliography}{75}%
\makeatletter
\providecommand \@ifxundefined [1]{%
 \@ifx{#1\undefined}
}%
\providecommand \@ifnum [1]{%
 \ifnum #1\expandafter \@firstoftwo
 \else \expandafter \@secondoftwo
 \fi
}%
\providecommand \@ifx [1]{%
 \ifx #1\expandafter \@firstoftwo
 \else \expandafter \@secondoftwo
 \fi
}%
\providecommand \natexlab [1]{#1}%
\providecommand \enquote  [1]{``#1''}%
\providecommand \bibnamefont  [1]{#1}%
\providecommand \bibfnamefont [1]{#1}%
\providecommand \citenamefont [1]{#1}%
\providecommand \href@noop [0]{\@secondoftwo}%
\providecommand \href [0]{\begingroup \@sanitize@url \@href}%
\providecommand \@href[1]{\@@startlink{#1}\@@href}%
\providecommand \@@href[1]{\endgroup#1\@@endlink}%
\providecommand \@sanitize@url [0]{\catcode `\\12\catcode `\$12\catcode `\&12\catcode `\#12\catcode `\^12\catcode `\_12\catcode `\%12\relax}%
\providecommand \@@startlink[1]{}%
\providecommand \@@endlink[0]{}%
\providecommand \url  [0]{\begingroup\@sanitize@url \@url }%
\providecommand \@url [1]{\endgroup\@href {#1}{\urlprefix }}%
\providecommand \urlprefix  [0]{URL }%
\providecommand \Eprint [0]{\href }%
\providecommand \doibase [0]{https://doi.org/}%
\providecommand \selectlanguage [0]{\@gobble}%
\providecommand \bibinfo  [0]{\@secondoftwo}%
\providecommand \bibfield  [0]{\@secondoftwo}%
\providecommand \translation [1]{[#1]}%
\providecommand \BibitemOpen [0]{}%
\providecommand \bibitemStop [0]{}%
\providecommand \bibitemNoStop [0]{.\EOS\space}%
\providecommand \EOS [0]{\spacefactor3000\relax}%
\providecommand \BibitemShut  [1]{\csname bibitem#1\endcsname}%
\let\auto@bib@innerbib\@empty
%</preamble>
\bibitem [{\citenamefont {Noiri}\ \emph {et~al.}(2022{\natexlab{a}})\citenamefont {Noiri}, \citenamefont {Takeda}, \citenamefont {Nakajima}, \citenamefont {Kobayashi}, \citenamefont {Sammak}, \citenamefont {Scappucci},\ and\ \citenamefont {Tarucha}}]{noiri2022fast}%
  \BibitemOpen
  \bibfield  {author} {\bibinfo {author} {\bibfnamefont {A.}~\bibnamefont {Noiri}}, \bibinfo {author} {\bibfnamefont {K.}~\bibnamefont {Takeda}}, \bibinfo {author} {\bibfnamefont {T.}~\bibnamefont {Nakajima}}, \bibinfo {author} {\bibfnamefont {T.}~\bibnamefont {Kobayashi}}, \bibinfo {author} {\bibfnamefont {A.}~\bibnamefont {Sammak}}, \bibinfo {author} {\bibfnamefont {G.}~\bibnamefont {Scappucci}},\ and\ \bibinfo {author} {\bibfnamefont {S.}~\bibnamefont {Tarucha}},\ }\bibfield  {title} {\bibinfo {title} {Fast universal quantum gate above the fault-tolerance threshold in silicon},\ }\href {https://doi.org/10.1038/s41586-021-04182-y} {\bibfield  {journal} {\bibinfo  {journal} {Nature}\ }\textbf {\bibinfo {volume} {601}},\ \bibinfo {pages} {338} (\bibinfo {year} {2022}{\natexlab{a}})}\BibitemShut {NoStop}%
\bibitem [{\citenamefont {Xue}\ \emph {et~al.}(2022)\citenamefont {Xue}, \citenamefont {Russ}, \citenamefont {Samkharadze}, \citenamefont {Undseth}, \citenamefont {Sammak}, \citenamefont {Scappucci},\ and\ \citenamefont {Vandersypen}}]{Xue:2022p343}%
  \BibitemOpen
  \bibfield  {author} {\bibinfo {author} {\bibfnamefont {X.}~\bibnamefont {Xue}}, \bibinfo {author} {\bibfnamefont {M.}~\bibnamefont {Russ}}, \bibinfo {author} {\bibfnamefont {N.}~\bibnamefont {Samkharadze}}, \bibinfo {author} {\bibfnamefont {B.}~\bibnamefont {Undseth}}, \bibinfo {author} {\bibfnamefont {A.}~\bibnamefont {Sammak}}, \bibinfo {author} {\bibfnamefont {G.}~\bibnamefont {Scappucci}},\ and\ \bibinfo {author} {\bibfnamefont {L.~M.~K.}\ \bibnamefont {Vandersypen}},\ }\bibfield  {title} {\bibinfo {title} {Quantum logic with spin qubits crossing the surface code threshold},\ }\href {https://doi.org/10.1038/s41586-021-04273-w} {\bibfield  {journal} {\bibinfo  {journal} {Nature}\ }\textbf {\bibinfo {volume} {601}},\ \bibinfo {pages} {343} (\bibinfo {year} {2022})}\BibitemShut {NoStop}%
\bibitem [{\citenamefont {Mills}\ \emph {et~al.}(2022)\citenamefont {Mills}, \citenamefont {Guinn}, \citenamefont {Gullans}, \citenamefont {Sigillito}, \citenamefont {Feldman}, \citenamefont {Nielsen},\ and\ \citenamefont {Petta}}]{mills2022two}%
  \BibitemOpen
  \bibfield  {author} {\bibinfo {author} {\bibfnamefont {A.~R.}\ \bibnamefont {Mills}}, \bibinfo {author} {\bibfnamefont {C.~R.}\ \bibnamefont {Guinn}}, \bibinfo {author} {\bibfnamefont {M.~J.}\ \bibnamefont {Gullans}}, \bibinfo {author} {\bibfnamefont {A.~J.}\ \bibnamefont {Sigillito}}, \bibinfo {author} {\bibfnamefont {M.~M.}\ \bibnamefont {Feldman}}, \bibinfo {author} {\bibfnamefont {E.}~\bibnamefont {Nielsen}},\ and\ \bibinfo {author} {\bibfnamefont {J.~R.}\ \bibnamefont {Petta}},\ }\bibfield  {title} {\bibinfo {title} {Two-qubit silicon quantum processor with operation fidelity exceeding 99\%},\ }\href {https://doi.org/10.1126/sciadv.abn5130} {\bibfield  {journal} {\bibinfo  {journal} {Science Advances}\ }\textbf {\bibinfo {volume} {8}},\ \bibinfo {pages} {eabn5130} (\bibinfo {year} {2022})}\BibitemShut {NoStop}%
\bibitem [{\citenamefont {Trifunovic}\ \emph {et~al.}(2012)\citenamefont {Trifunovic}, \citenamefont {Dial}, \citenamefont {Trif}, \citenamefont {Wootton}, \citenamefont {Abebe}, \citenamefont {Yacoby},\ and\ \citenamefont {Loss}}]{Trifunovic:2012p011006}%
  \BibitemOpen
  \bibfield  {author} {\bibinfo {author} {\bibfnamefont {L.}~\bibnamefont {Trifunovic}}, \bibinfo {author} {\bibfnamefont {O.}~\bibnamefont {Dial}}, \bibinfo {author} {\bibfnamefont {M.}~\bibnamefont {Trif}}, \bibinfo {author} {\bibfnamefont {J.~R.}\ \bibnamefont {Wootton}}, \bibinfo {author} {\bibfnamefont {R.}~\bibnamefont {Abebe}}, \bibinfo {author} {\bibfnamefont {A.}~\bibnamefont {Yacoby}},\ and\ \bibinfo {author} {\bibfnamefont {D.}~\bibnamefont {Loss}},\ }\bibfield  {title} {\bibinfo {title} {Long-distance spin-spin coupling via floating gates},\ }\href {https://doi.org/10.1103/PhysRevX.2.011006} {\bibfield  {journal} {\bibinfo  {journal} {Phys. Rev. X}\ }\textbf {\bibinfo {volume} {2}},\ \bibinfo {pages} {011006} (\bibinfo {year} {2012})}\BibitemShut {NoStop}%
\bibitem [{\citenamefont {Braakman}\ \emph {et~al.}(2013)\citenamefont {Braakman}, \citenamefont {Barthelemy}, \citenamefont {Reichl}, \citenamefont {Wegscheider},\ and\ \citenamefont {Vandersypen}}]{Braakman:2013p432}%
  \BibitemOpen
  \bibfield  {author} {\bibinfo {author} {\bibfnamefont {F.~R.}\ \bibnamefont {Braakman}}, \bibinfo {author} {\bibfnamefont {P.}~\bibnamefont {Barthelemy}}, \bibinfo {author} {\bibfnamefont {C.}~\bibnamefont {Reichl}}, \bibinfo {author} {\bibfnamefont {W.}~\bibnamefont {Wegscheider}},\ and\ \bibinfo {author} {\bibfnamefont {L.~M.~K.}\ \bibnamefont {Vandersypen}},\ }\bibfield  {title} {\bibinfo {title} {Long-distance coherent coupling in a quantum dot array},\ }\href {https://doi.org/10.1038/nnano.2013.67} {\bibfield  {journal} {\bibinfo  {journal} {Nature Nanotechnology}\ }\textbf {\bibinfo {volume} {8}},\ \bibinfo {pages} {432} (\bibinfo {year} {2013})}\BibitemShut {NoStop}%
\bibitem [{\citenamefont {Serina}\ \emph {et~al.}(2017)\citenamefont {Serina}, \citenamefont {Kloeffel},\ and\ \citenamefont {Loss}}]{Serina:2017p245422}%
  \BibitemOpen
  \bibfield  {author} {\bibinfo {author} {\bibfnamefont {M.}~\bibnamefont {Serina}}, \bibinfo {author} {\bibfnamefont {C.}~\bibnamefont {Kloeffel}},\ and\ \bibinfo {author} {\bibfnamefont {D.}~\bibnamefont {Loss}},\ }\bibfield  {title} {\bibinfo {title} {Long-range interaction between charge and spin qubits in quantum dots},\ }\href {https://doi.org/10.1103/PhysRevB.95.245422} {\bibfield  {journal} {\bibinfo  {journal} {Phys. Rev. B}\ }\textbf {\bibinfo {volume} {95}},\ \bibinfo {pages} {245422} (\bibinfo {year} {2017})}\BibitemShut {NoStop}%
\bibitem [{\citenamefont {Tosi}\ \emph {et~al.}(2017)\citenamefont {Tosi}, \citenamefont {Mohiyaddin}, \citenamefont {Schmitt}, \citenamefont {Tenberg}, \citenamefont {Rahman}, \citenamefont {Klimeck},\ and\ \citenamefont {Morello}}]{Tosi:2017p450}%
  \BibitemOpen
  \bibfield  {author} {\bibinfo {author} {\bibfnamefont {G.}~\bibnamefont {Tosi}}, \bibinfo {author} {\bibfnamefont {F.~A.}\ \bibnamefont {Mohiyaddin}}, \bibinfo {author} {\bibfnamefont {V.}~\bibnamefont {Schmitt}}, \bibinfo {author} {\bibfnamefont {S.}~\bibnamefont {Tenberg}}, \bibinfo {author} {\bibfnamefont {R.}~\bibnamefont {Rahman}}, \bibinfo {author} {\bibfnamefont {G.}~\bibnamefont {Klimeck}},\ and\ \bibinfo {author} {\bibfnamefont {A.}~\bibnamefont {Morello}},\ }\bibfield  {title} {\bibinfo {title} {Silicon quantum processor with robust long-distance qubit couplings},\ }\href {https://doi.org/10.1038/s41467-017-00378-x} {\bibfield  {journal} {\bibinfo  {journal} {Nature Communications}\ }\textbf {\bibinfo {volume} {8}},\ \bibinfo {pages} {450} (\bibinfo {year} {2017})}\BibitemShut {NoStop}%
\bibitem [{\citenamefont {Samkharadze}\ \emph {et~al.}(2018)\citenamefont {Samkharadze}, \citenamefont {Zheng}, \citenamefont {Kalhor}, \citenamefont {Brousse}, \citenamefont {Sammak}, \citenamefont {Mendes}, \citenamefont {Blais}, \citenamefont {Scappucci},\ and\ \citenamefont {Vandersypen}}]{samkharadze2018strong}%
  \BibitemOpen
  \bibfield  {author} {\bibinfo {author} {\bibfnamefont {N.}~\bibnamefont {Samkharadze}}, \bibinfo {author} {\bibfnamefont {G.}~\bibnamefont {Zheng}}, \bibinfo {author} {\bibfnamefont {N.}~\bibnamefont {Kalhor}}, \bibinfo {author} {\bibfnamefont {D.}~\bibnamefont {Brousse}}, \bibinfo {author} {\bibfnamefont {A.}~\bibnamefont {Sammak}}, \bibinfo {author} {\bibfnamefont {U.}~\bibnamefont {Mendes}}, \bibinfo {author} {\bibfnamefont {A.}~\bibnamefont {Blais}}, \bibinfo {author} {\bibfnamefont {G.}~\bibnamefont {Scappucci}},\ and\ \bibinfo {author} {\bibfnamefont {L.}~\bibnamefont {Vandersypen}},\ }\bibfield  {title} {\bibinfo {title} {Strong spin-photon coupling in silicon},\ }\href {https://doi.org/10.1126/science.aar4054} {\bibfield  {journal} {\bibinfo  {journal} {Science}\ }\textbf {\bibinfo {volume} {359}},\ \bibinfo {pages} {1123} (\bibinfo {year} {2018})}\BibitemShut {NoStop}%
\bibitem [{\citenamefont {Warren}\ \emph {et~al.}(2019)\citenamefont {Warren}, \citenamefont {Barnes},\ and\ \citenamefont {Economou}}]{Warren:2019p161303}%
  \BibitemOpen
  \bibfield  {author} {\bibinfo {author} {\bibfnamefont {A.}~\bibnamefont {Warren}}, \bibinfo {author} {\bibfnamefont {E.}~\bibnamefont {Barnes}},\ and\ \bibinfo {author} {\bibfnamefont {S.~E.}\ \bibnamefont {Economou}},\ }\bibfield  {title} {\bibinfo {title} {Long-distance entangling gates between quantum dot spins mediated by a superconducting resonator},\ }\href {https://doi.org/10.1103/PhysRevB.100.161303} {\bibfield  {journal} {\bibinfo  {journal} {Phys. Rev. B}\ }\textbf {\bibinfo {volume} {100}},\ \bibinfo {pages} {161303} (\bibinfo {year} {2019})}\BibitemShut {NoStop}%
\bibitem [{\citenamefont {Qiao}\ \emph {et~al.}(2021)\citenamefont {Qiao}, \citenamefont {Kandel}, \citenamefont {Fallahi}, \citenamefont {Gardner}, \citenamefont {Manfra}, \citenamefont {Hu},\ and\ \citenamefont {Nichol}}]{Qiao:2021p017701}%
  \BibitemOpen
  \bibfield  {author} {\bibinfo {author} {\bibfnamefont {H.}~\bibnamefont {Qiao}}, \bibinfo {author} {\bibfnamefont {Y.~P.}\ \bibnamefont {Kandel}}, \bibinfo {author} {\bibfnamefont {S.}~\bibnamefont {Fallahi}}, \bibinfo {author} {\bibfnamefont {G.~C.}\ \bibnamefont {Gardner}}, \bibinfo {author} {\bibfnamefont {M.~J.}\ \bibnamefont {Manfra}}, \bibinfo {author} {\bibfnamefont {X.}~\bibnamefont {Hu}},\ and\ \bibinfo {author} {\bibfnamefont {J.~M.}\ \bibnamefont {Nichol}},\ }\bibfield  {title} {\bibinfo {title} {Long-distance superexchange between semiconductor quantum-dot electron spins},\ }\href {https://doi.org/10.1103/PhysRevLett.126.017701} {\bibfield  {journal} {\bibinfo  {journal} {Phys. Rev. Lett.}\ }\textbf {\bibinfo {volume} {126}},\ \bibinfo {pages} {017701} (\bibinfo {year} {2021})}\BibitemShut {NoStop}%
\bibitem [{\citenamefont {Holman}\ \emph {et~al.}(2021)\citenamefont {Holman}, \citenamefont {Rosenberg}, \citenamefont {Yost}, \citenamefont {Yoder}, \citenamefont {Das}, \citenamefont {Oliver}, \citenamefont {McDermott},\ and\ \citenamefont {Eriksson}}]{Holman:2021p137}%
  \BibitemOpen
  \bibfield  {author} {\bibinfo {author} {\bibfnamefont {N.}~\bibnamefont {Holman}}, \bibinfo {author} {\bibfnamefont {D.}~\bibnamefont {Rosenberg}}, \bibinfo {author} {\bibfnamefont {D.}~\bibnamefont {Yost}}, \bibinfo {author} {\bibfnamefont {J.~L.}\ \bibnamefont {Yoder}}, \bibinfo {author} {\bibfnamefont {R.}~\bibnamefont {Das}}, \bibinfo {author} {\bibfnamefont {W.~D.}\ \bibnamefont {Oliver}}, \bibinfo {author} {\bibfnamefont {R.}~\bibnamefont {McDermott}},\ and\ \bibinfo {author} {\bibfnamefont {M.~A.}\ \bibnamefont {Eriksson}},\ }\bibfield  {title} {\bibinfo {title} {3d integration and measurement of a semiconductor double quantum dot with a high-impedance {T}i{N} resonator},\ }\href {https://doi.org/10.1038/s41534-021-00469-0} {\bibfield  {journal} {\bibinfo  {journal} {npj Quantum Information}\ }\textbf {\bibinfo {volume} {7}},\ \bibinfo {pages} {137} (\bibinfo {year} {2021})}\BibitemShut {NoStop}%
\bibitem [{\citenamefont {Wang}\ \emph {et~al.}(2023)\citenamefont {Wang}, \citenamefont {Feng}, \citenamefont {Serrano}, \citenamefont {Gilbert}, \citenamefont {Leon}, \citenamefont {Tanttu}, \citenamefont {Mai}, \citenamefont {Liang}, \citenamefont {Huang}, \citenamefont {Su}, \citenamefont {Lim}, \citenamefont {Hudson}, \citenamefont {Escott}, \citenamefont {Morello}, \citenamefont {Yang}, \citenamefont {Dzurak}, \citenamefont {Saraiva},\ and\ \citenamefont {Laucht}}]{Wang:2023p2208557}%
  \BibitemOpen
  \bibfield  {author} {\bibinfo {author} {\bibfnamefont {Z.}~\bibnamefont {Wang}}, \bibinfo {author} {\bibfnamefont {M.}~\bibnamefont {Feng}}, \bibinfo {author} {\bibfnamefont {S.}~\bibnamefont {Serrano}}, \bibinfo {author} {\bibfnamefont {W.}~\bibnamefont {Gilbert}}, \bibinfo {author} {\bibfnamefont {R.~C.~C.}\ \bibnamefont {Leon}}, \bibinfo {author} {\bibfnamefont {T.}~\bibnamefont {Tanttu}}, \bibinfo {author} {\bibfnamefont {P.}~\bibnamefont {Mai}}, \bibinfo {author} {\bibfnamefont {D.}~\bibnamefont {Liang}}, \bibinfo {author} {\bibfnamefont {J.~Y.}\ \bibnamefont {Huang}}, \bibinfo {author} {\bibfnamefont {Y.}~\bibnamefont {Su}}, \bibinfo {author} {\bibfnamefont {W.~H.}\ \bibnamefont {Lim}}, \bibinfo {author} {\bibfnamefont {F.~E.}\ \bibnamefont {Hudson}}, \bibinfo {author} {\bibfnamefont {C.~C.}\ \bibnamefont {Escott}}, \bibinfo {author} {\bibfnamefont {A.}~\bibnamefont {Morello}}, \bibinfo {author} {\bibfnamefont {C.~H.}\ \bibnamefont {Yang}}, \bibinfo {author} {\bibfnamefont {A.~S.}\ \bibnamefont
  {Dzurak}}, \bibinfo {author} {\bibfnamefont {A.}~\bibnamefont {Saraiva}},\ and\ \bibinfo {author} {\bibfnamefont {A.}~\bibnamefont {Laucht}},\ }\bibfield  {title} {\bibinfo {title} {Jellybean quantum dots in silicon for qubit coupling and on-chip quantum chemistry},\ }\href {https://doi.org/https://doi.org/10.1002/adma.202208557} {\bibfield  {journal} {\bibinfo  {journal} {Advanced Materials}\ }\textbf {\bibinfo {volume} {35}},\ \bibinfo {pages} {2208557} (\bibinfo {year} {2023})}\BibitemShut {NoStop}%
\bibitem [{\citenamefont {Landig}\ \emph {et~al.}(2018)\citenamefont {Landig}, \citenamefont {Koski}, \citenamefont {Scarlino}, \citenamefont {Mendes}, \citenamefont {Blais}, \citenamefont {Reichl}, \citenamefont {Wegscheider}, \citenamefont {Wallraff}, \citenamefont {Ensslin},\ and\ \citenamefont {Ihn}}]{landig2018coherent}%
  \BibitemOpen
  \bibfield  {author} {\bibinfo {author} {\bibfnamefont {A.~J.}\ \bibnamefont {Landig}}, \bibinfo {author} {\bibfnamefont {J.~V.}\ \bibnamefont {Koski}}, \bibinfo {author} {\bibfnamefont {P.}~\bibnamefont {Scarlino}}, \bibinfo {author} {\bibfnamefont {U.}~\bibnamefont {Mendes}}, \bibinfo {author} {\bibfnamefont {A.}~\bibnamefont {Blais}}, \bibinfo {author} {\bibfnamefont {C.}~\bibnamefont {Reichl}}, \bibinfo {author} {\bibfnamefont {W.}~\bibnamefont {Wegscheider}}, \bibinfo {author} {\bibfnamefont {A.}~\bibnamefont {Wallraff}}, \bibinfo {author} {\bibfnamefont {K.}~\bibnamefont {Ensslin}},\ and\ \bibinfo {author} {\bibfnamefont {T.}~\bibnamefont {Ihn}},\ }\bibfield  {title} {\bibinfo {title} {Coherent spin--photon coupling using a resonant exchange qubit},\ }\href {https://doi.org/10.1038/s41586-018-0365-y} {\bibfield  {journal} {\bibinfo  {journal} {Nature}\ }\textbf {\bibinfo {volume} {560}},\ \bibinfo {pages} {179} (\bibinfo {year} {2018})}\BibitemShut {NoStop}%
\bibitem [{\citenamefont {Sigillito}\ \emph {et~al.}(2019)\citenamefont {Sigillito}, \citenamefont {Gullans}, \citenamefont {Edge}, \citenamefont {Borselli},\ and\ \citenamefont {Petta}}]{sigillito2019coherent}%
  \BibitemOpen
  \bibfield  {author} {\bibinfo {author} {\bibfnamefont {A.}~\bibnamefont {Sigillito}}, \bibinfo {author} {\bibfnamefont {M.}~\bibnamefont {Gullans}}, \bibinfo {author} {\bibfnamefont {L.}~\bibnamefont {Edge}}, \bibinfo {author} {\bibfnamefont {M.}~\bibnamefont {Borselli}},\ and\ \bibinfo {author} {\bibfnamefont {J.}~\bibnamefont {Petta}},\ }\bibfield  {title} {\bibinfo {title} {Coherent transfer of quantum information in a silicon double quantum dot using resonant swap gates},\ }\href {https://doi.org/10.1038/s41534-019-0225-0} {\bibfield  {journal} {\bibinfo  {journal} {npj Quantum Information}\ }\textbf {\bibinfo {volume} {5}},\ \bibinfo {pages} {110} (\bibinfo {year} {2019})}\BibitemShut {NoStop}%
\bibitem [{\citenamefont {Fujita}\ \emph {et~al.}(2017)\citenamefont {Fujita}, \citenamefont {Baart}, \citenamefont {Reichl}, \citenamefont {Wegscheider},\ and\ \citenamefont {Vandersypen}}]{Fujita:2017p22}%
  \BibitemOpen
  \bibfield  {author} {\bibinfo {author} {\bibfnamefont {T.}~\bibnamefont {Fujita}}, \bibinfo {author} {\bibfnamefont {T.~A.}\ \bibnamefont {Baart}}, \bibinfo {author} {\bibfnamefont {C.}~\bibnamefont {Reichl}}, \bibinfo {author} {\bibfnamefont {W.}~\bibnamefont {Wegscheider}},\ and\ \bibinfo {author} {\bibfnamefont {L.~M.~K.}\ \bibnamefont {Vandersypen}},\ }\bibfield  {title} {\bibinfo {title} {Coherent shuttle of electron-spin states},\ }\href {https://doi.org/10.1038/s41534-017-0024-4} {\bibfield  {journal} {\bibinfo  {journal} {npj Quantum Information}\ }\textbf {\bibinfo {volume} {3}},\ \bibinfo {pages} {22} (\bibinfo {year} {2017})}\BibitemShut {NoStop}%
\bibitem [{\citenamefont {Mills}\ \emph {et~al.}(2019)\citenamefont {Mills}, \citenamefont {Zajac}, \citenamefont {Gullans}, \citenamefont {Schupp}, \citenamefont {Hazard},\ and\ \citenamefont {Petta}}]{mills2019shuttling}%
  \BibitemOpen
  \bibfield  {author} {\bibinfo {author} {\bibfnamefont {A.}~\bibnamefont {Mills}}, \bibinfo {author} {\bibfnamefont {D.}~\bibnamefont {Zajac}}, \bibinfo {author} {\bibfnamefont {M.}~\bibnamefont {Gullans}}, \bibinfo {author} {\bibfnamefont {F.}~\bibnamefont {Schupp}}, \bibinfo {author} {\bibfnamefont {T.}~\bibnamefont {Hazard}},\ and\ \bibinfo {author} {\bibfnamefont {J.}~\bibnamefont {Petta}},\ }\bibfield  {title} {\bibinfo {title} {Shuttling a single charge across a one-dimensional array of silicon quantum dots},\ }\href {https://doi.org/10.1038/s41467-019-08970-z} {\bibfield  {journal} {\bibinfo  {journal} {Nature Communications}\ }\textbf {\bibinfo {volume} {10}},\ \bibinfo {pages} {1063} (\bibinfo {year} {2019})}\BibitemShut {NoStop}%
\bibitem [{\citenamefont {Yoneda}\ \emph {et~al.}(2021)\citenamefont {Yoneda}, \citenamefont {Huang}, \citenamefont {Feng}, \citenamefont {Yang}, \citenamefont {Chan}, \citenamefont {Tanttu}, \citenamefont {Gilbert}, \citenamefont {Leon}, \citenamefont {Hudson}, \citenamefont {Itoh} \emph {et~al.}}]{yoneda2021coherent}%
  \BibitemOpen
  \bibfield  {author} {\bibinfo {author} {\bibfnamefont {J.}~\bibnamefont {Yoneda}}, \bibinfo {author} {\bibfnamefont {W.}~\bibnamefont {Huang}}, \bibinfo {author} {\bibfnamefont {M.}~\bibnamefont {Feng}}, \bibinfo {author} {\bibfnamefont {C.~H.}\ \bibnamefont {Yang}}, \bibinfo {author} {\bibfnamefont {K.~W.}\ \bibnamefont {Chan}}, \bibinfo {author} {\bibfnamefont {T.}~\bibnamefont {Tanttu}}, \bibinfo {author} {\bibfnamefont {W.}~\bibnamefont {Gilbert}}, \bibinfo {author} {\bibfnamefont {R.}~\bibnamefont {Leon}}, \bibinfo {author} {\bibfnamefont {F.}~\bibnamefont {Hudson}}, \bibinfo {author} {\bibfnamefont {K.}~\bibnamefont {Itoh}}, \emph {et~al.},\ }\bibfield  {title} {\bibinfo {title} {Coherent spin qubit transport in silicon},\ }\href {https://doi.org/10.1038/s41467-021-24371-7} {\bibfield  {journal} {\bibinfo  {journal} {Nature Communications}\ }\textbf {\bibinfo {volume} {12}},\ \bibinfo {pages} {4114} (\bibinfo {year} {2021})}\BibitemShut {NoStop}%
\bibitem [{\citenamefont {Jadot}\ \emph {et~al.}(2021)\citenamefont {Jadot}, \citenamefont {Mortemousque}, \citenamefont {Chanrion}, \citenamefont {Thiney}, \citenamefont {Ludwig}, \citenamefont {Wieck}, \citenamefont {Urdampilleta}, \citenamefont {B{\"a}uerle},\ and\ \citenamefont {Meunier}}]{Jadot:2021p570}%
  \BibitemOpen
  \bibfield  {author} {\bibinfo {author} {\bibfnamefont {B.}~\bibnamefont {Jadot}}, \bibinfo {author} {\bibfnamefont {P.-A.}\ \bibnamefont {Mortemousque}}, \bibinfo {author} {\bibfnamefont {E.}~\bibnamefont {Chanrion}}, \bibinfo {author} {\bibfnamefont {V.}~\bibnamefont {Thiney}}, \bibinfo {author} {\bibfnamefont {A.}~\bibnamefont {Ludwig}}, \bibinfo {author} {\bibfnamefont {A.~D.}\ \bibnamefont {Wieck}}, \bibinfo {author} {\bibfnamefont {M.}~\bibnamefont {Urdampilleta}}, \bibinfo {author} {\bibfnamefont {C.}~\bibnamefont {B{\"a}uerle}},\ and\ \bibinfo {author} {\bibfnamefont {T.}~\bibnamefont {Meunier}},\ }\bibfield  {title} {\bibinfo {title} {Distant spin entanglement via fast and coherent electron shuttling},\ }\href {https://doi.org/10.1038/s41565-021-00846-y} {\bibfield  {journal} {\bibinfo  {journal} {Nature Nanotechnology}\ }\textbf {\bibinfo {volume} {16}},\ \bibinfo {pages} {570} (\bibinfo {year} {2021})}\BibitemShut {NoStop}%
\bibitem [{\citenamefont {Seidler}\ \emph {et~al.}(2022)\citenamefont {Seidler}, \citenamefont {Struck}, \citenamefont {Xue}, \citenamefont {Focke}, \citenamefont {Trellenkamp}, \citenamefont {Bluhm},\ and\ \citenamefont {Schreiber}}]{seidler2022conveyor}%
  \BibitemOpen
  \bibfield  {author} {\bibinfo {author} {\bibfnamefont {I.}~\bibnamefont {Seidler}}, \bibinfo {author} {\bibfnamefont {T.}~\bibnamefont {Struck}}, \bibinfo {author} {\bibfnamefont {R.}~\bibnamefont {Xue}}, \bibinfo {author} {\bibfnamefont {N.}~\bibnamefont {Focke}}, \bibinfo {author} {\bibfnamefont {S.}~\bibnamefont {Trellenkamp}}, \bibinfo {author} {\bibfnamefont {H.}~\bibnamefont {Bluhm}},\ and\ \bibinfo {author} {\bibfnamefont {L.~R.}\ \bibnamefont {Schreiber}},\ }\bibfield  {title} {\bibinfo {title} {Conveyor-mode single-electron shuttling in {S}i/{S}i{G}e for a scalable quantum computing architecture},\ }\href {https://doi.org/10.1038/s41534-022-00615-2} {\bibfield  {journal} {\bibinfo  {journal} {npj Quantum Information}\ }\textbf {\bibinfo {volume} {8}},\ \bibinfo {pages} {100} (\bibinfo {year} {2022})}\BibitemShut {NoStop}%
\bibitem [{\citenamefont {Noiri}\ \emph {et~al.}(2022{\natexlab{b}})\citenamefont {Noiri}, \citenamefont {Takeda}, \citenamefont {Nakajima}, \citenamefont {Kobayashi}, \citenamefont {Sammak}, \citenamefont {Scappucci},\ and\ \citenamefont {Tarucha}}]{noiri2022shuttling}%
  \BibitemOpen
  \bibfield  {author} {\bibinfo {author} {\bibfnamefont {A.}~\bibnamefont {Noiri}}, \bibinfo {author} {\bibfnamefont {K.}~\bibnamefont {Takeda}}, \bibinfo {author} {\bibfnamefont {T.}~\bibnamefont {Nakajima}}, \bibinfo {author} {\bibfnamefont {T.}~\bibnamefont {Kobayashi}}, \bibinfo {author} {\bibfnamefont {A.}~\bibnamefont {Sammak}}, \bibinfo {author} {\bibfnamefont {G.}~\bibnamefont {Scappucci}},\ and\ \bibinfo {author} {\bibfnamefont {S.}~\bibnamefont {Tarucha}},\ }\bibfield  {title} {\bibinfo {title} {A shuttling-based two-qubit logic gate for linking distant silicon quantum processors},\ }\href {https://doi.org/10.1038/s41467-022-33453-z} {\bibfield  {journal} {\bibinfo  {journal} {Nature Communications}\ }\textbf {\bibinfo {volume} {13}},\ \bibinfo {pages} {5740} (\bibinfo {year} {2022}{\natexlab{b}})}\BibitemShut {NoStop}%
\bibitem [{\citenamefont {Boter}\ \emph {et~al.}(2022)\citenamefont {Boter}, \citenamefont {Dehollain}, \citenamefont {Van~Dijk}, \citenamefont {Xu}, \citenamefont {Hensgens}, \citenamefont {Versluis}, \citenamefont {Naus}, \citenamefont {Clarke}, \citenamefont {Veldhorst}, \citenamefont {Sebastiano} \emph {et~al.}}]{boter2022spiderweb}%
  \BibitemOpen
  \bibfield  {author} {\bibinfo {author} {\bibfnamefont {J.~M.}\ \bibnamefont {Boter}}, \bibinfo {author} {\bibfnamefont {J.~P.}\ \bibnamefont {Dehollain}}, \bibinfo {author} {\bibfnamefont {J.~P.~G.}\ \bibnamefont {Van~Dijk}}, \bibinfo {author} {\bibfnamefont {Y.}~\bibnamefont {Xu}}, \bibinfo {author} {\bibfnamefont {T.}~\bibnamefont {Hensgens}}, \bibinfo {author} {\bibfnamefont {R.}~\bibnamefont {Versluis}}, \bibinfo {author} {\bibfnamefont {H.~W.~L.}\ \bibnamefont {Naus}}, \bibinfo {author} {\bibfnamefont {J.~S.}\ \bibnamefont {Clarke}}, \bibinfo {author} {\bibfnamefont {M.}~\bibnamefont {Veldhorst}}, \bibinfo {author} {\bibfnamefont {F.}~\bibnamefont {Sebastiano}}, \emph {et~al.},\ }\bibfield  {title} {\bibinfo {title} {Spiderweb array: a sparse spin-qubit array},\ }\href {https://doi.org/10.1103/PhysRevApplied.18.024053} {\bibfield  {journal} {\bibinfo  {journal} {Physical Review Applied}\ }\textbf {\bibinfo {volume} {18}},\ \bibinfo {pages} {024053} (\bibinfo {year} {2022})}\BibitemShut {NoStop}%
\bibitem [{\citenamefont {K{\"u}nne}\ \emph {et~al.}(2023)\citenamefont {K{\"u}nne}, \citenamefont {Willmes}, \citenamefont {Oberl{\"a}nder}, \citenamefont {Gorjaew}, \citenamefont {Teske}, \citenamefont {Bhardwaj}, \citenamefont {Beer}, \citenamefont {Kammerloher}, \citenamefont {Otten}, \citenamefont {Seidler} \emph {et~al.}}]{Kuenne2023}%
  \BibitemOpen
  \bibfield  {author} {\bibinfo {author} {\bibfnamefont {M.}~\bibnamefont {K{\"u}nne}}, \bibinfo {author} {\bibfnamefont {A.}~\bibnamefont {Willmes}}, \bibinfo {author} {\bibfnamefont {M.}~\bibnamefont {Oberl{\"a}nder}}, \bibinfo {author} {\bibfnamefont {C.}~\bibnamefont {Gorjaew}}, \bibinfo {author} {\bibfnamefont {J.~D.}\ \bibnamefont {Teske}}, \bibinfo {author} {\bibfnamefont {H.}~\bibnamefont {Bhardwaj}}, \bibinfo {author} {\bibfnamefont {M.}~\bibnamefont {Beer}}, \bibinfo {author} {\bibfnamefont {E.}~\bibnamefont {Kammerloher}}, \bibinfo {author} {\bibfnamefont {R.}~\bibnamefont {Otten}}, \bibinfo {author} {\bibfnamefont {I.}~\bibnamefont {Seidler}}, \emph {et~al.},\ }\bibfield  {title} {\bibinfo {title} {The spinbus architecture: Scaling spin qubits with electron shuttling},\ }\href@noop {} {\bibfield  {journal} {\bibinfo  {journal} {arXiv preprint arXiv:2306.16348}\ } (\bibinfo {year} {2023})}\BibitemShut {NoStop}%
\bibitem [{\citenamefont {Zwerver}\ \emph {et~al.}(2023)\citenamefont {Zwerver}, \citenamefont {Amitonov}, \citenamefont {de~Snoo}, \citenamefont {M{\k{a}}dzik}, \citenamefont {Rimbach-Russ}, \citenamefont {Sammak}, \citenamefont {Scappucci},\ and\ \citenamefont {Vandersypen}}]{zwerver2023shuttling}%
  \BibitemOpen
  \bibfield  {author} {\bibinfo {author} {\bibfnamefont {A.~M.~J.}\ \bibnamefont {Zwerver}}, \bibinfo {author} {\bibfnamefont {S.~V.}\ \bibnamefont {Amitonov}}, \bibinfo {author} {\bibfnamefont {S.~L.}\ \bibnamefont {de~Snoo}}, \bibinfo {author} {\bibfnamefont {M.~T.}\ \bibnamefont {M{\k{a}}dzik}}, \bibinfo {author} {\bibfnamefont {M.}~\bibnamefont {Rimbach-Russ}}, \bibinfo {author} {\bibfnamefont {A.}~\bibnamefont {Sammak}}, \bibinfo {author} {\bibfnamefont {G.}~\bibnamefont {Scappucci}},\ and\ \bibinfo {author} {\bibfnamefont {L.~M.~K.}\ \bibnamefont {Vandersypen}},\ }\bibfield  {title} {\bibinfo {title} {Shuttling an electron spin through a silicon quantum dot array},\ }\href {https://doi.org/10.1103/PRXQuantum.4.030303} {\bibfield  {journal} {\bibinfo  {journal} {PRX Quantum}\ }\textbf {\bibinfo {volume} {4}},\ \bibinfo {pages} {030303} (\bibinfo {year} {2023})}\BibitemShut {NoStop}%
\bibitem [{\citenamefont {Langrock}\ \emph {et~al.}(2023)\citenamefont {Langrock}, \citenamefont {Krzywda}, \citenamefont {Focke}, \citenamefont {Seidler}, \citenamefont {Schreiber},\ and\ \citenamefont {Cywi{\'n}ski}}]{langrock2023blueprint}%
  \BibitemOpen
  \bibfield  {author} {\bibinfo {author} {\bibfnamefont {V.}~\bibnamefont {Langrock}}, \bibinfo {author} {\bibfnamefont {J.~A.}\ \bibnamefont {Krzywda}}, \bibinfo {author} {\bibfnamefont {N.}~\bibnamefont {Focke}}, \bibinfo {author} {\bibfnamefont {I.}~\bibnamefont {Seidler}}, \bibinfo {author} {\bibfnamefont {L.~R.}\ \bibnamefont {Schreiber}},\ and\ \bibinfo {author} {\bibfnamefont {{\L}.}~\bibnamefont {Cywi{\'n}ski}},\ }\bibfield  {title} {\bibinfo {title} {Blueprint of a scalable spin qubit shuttle device for coherent mid-range qubit transfer in disordered {S}i/{S}i{G}e/{S}i{O}$_2$},\ }\href {https://doi.org/10.1103/PRXQuantum.4.020305} {\bibfield  {journal} {\bibinfo  {journal} {PRX Quantum}\ }\textbf {\bibinfo {volume} {4}},\ \bibinfo {pages} {020305} (\bibinfo {year} {2023})}\BibitemShut {NoStop}%
\bibitem [{\citenamefont {Struck}\ \emph {et~al.}(2024)\citenamefont {Struck}, \citenamefont {Volmer}, \citenamefont {Visser}, \citenamefont {Offermann}, \citenamefont {Xue}, \citenamefont {Tu}, \citenamefont {Trellenkamp}, \citenamefont {Cywi{\'n}ski}, \citenamefont {Bluhm},\ and\ \citenamefont {Schreiber}}]{struck2024spin}%
  \BibitemOpen
  \bibfield  {author} {\bibinfo {author} {\bibfnamefont {T.}~\bibnamefont {Struck}}, \bibinfo {author} {\bibfnamefont {M.}~\bibnamefont {Volmer}}, \bibinfo {author} {\bibfnamefont {L.}~\bibnamefont {Visser}}, \bibinfo {author} {\bibfnamefont {T.}~\bibnamefont {Offermann}}, \bibinfo {author} {\bibfnamefont {R.}~\bibnamefont {Xue}}, \bibinfo {author} {\bibfnamefont {J.-S.}\ \bibnamefont {Tu}}, \bibinfo {author} {\bibfnamefont {S.}~\bibnamefont {Trellenkamp}}, \bibinfo {author} {\bibfnamefont {{\L}.}~\bibnamefont {Cywi{\'n}ski}}, \bibinfo {author} {\bibfnamefont {H.}~\bibnamefont {Bluhm}},\ and\ \bibinfo {author} {\bibfnamefont {L.~R.}\ \bibnamefont {Schreiber}},\ }\bibfield  {title} {\bibinfo {title} {Spin-{EPR}-pair separation by conveyor-mode single electron shuttling in {S}i/{S}i{G}e},\ }\href {https://doi.org/10.1038/s41467-024-45583-7} {\bibfield  {journal} {\bibinfo  {journal} {Nature Communications}\ }\textbf {\bibinfo {volume} {15}},\ \bibinfo {pages} {1325} (\bibinfo {year} {2024})}\BibitemShut
  {NoStop}%
\bibitem [{\citenamefont {Xue}\ \emph {et~al.}(2024)\citenamefont {Xue}, \citenamefont {Beer}, \citenamefont {Seidler}, \citenamefont {Humpohl}, \citenamefont {Tu}, \citenamefont {Trellenkamp}, \citenamefont {Struck}, \citenamefont {Bluhm},\ and\ \citenamefont {Schreiber}}]{xue2024si}%
  \BibitemOpen
  \bibfield  {author} {\bibinfo {author} {\bibfnamefont {R.}~\bibnamefont {Xue}}, \bibinfo {author} {\bibfnamefont {M.}~\bibnamefont {Beer}}, \bibinfo {author} {\bibfnamefont {I.}~\bibnamefont {Seidler}}, \bibinfo {author} {\bibfnamefont {S.}~\bibnamefont {Humpohl}}, \bibinfo {author} {\bibfnamefont {J.-S.}\ \bibnamefont {Tu}}, \bibinfo {author} {\bibfnamefont {S.}~\bibnamefont {Trellenkamp}}, \bibinfo {author} {\bibfnamefont {T.}~\bibnamefont {Struck}}, \bibinfo {author} {\bibfnamefont {H.}~\bibnamefont {Bluhm}},\ and\ \bibinfo {author} {\bibfnamefont {L.~R.}\ \bibnamefont {Schreiber}},\ }\bibfield  {title} {\bibinfo {title} {Si/{S}i{G}e {Q}u{B}us for single electron information-processing devices with memory and micron-scale connectivity function},\ }\href {https://doi.org/10.1038/s41467-024-46519-x} {\bibfield  {journal} {\bibinfo  {journal} {Nature Communications}\ }\textbf {\bibinfo {volume} {15}},\ \bibinfo {pages} {2296} (\bibinfo {year} {2024})}\BibitemShut {NoStop}%
\bibitem [{\citenamefont {De~Smet}\ \emph {et~al.}(2024)\citenamefont {De~Smet}, \citenamefont {Matsumoto}, \citenamefont {Zwerver}, \citenamefont {Tryputen}, \citenamefont {de~Snoo}, \citenamefont {Amitonov}, \citenamefont {Sammak}, \citenamefont {Samkharadze}, \citenamefont {G{\"u}l}, \citenamefont {Wasserman} \emph {et~al.}}]{desmet2024high}%
  \BibitemOpen
  \bibfield  {author} {\bibinfo {author} {\bibfnamefont {M.}~\bibnamefont {De~Smet}}, \bibinfo {author} {\bibfnamefont {Y.}~\bibnamefont {Matsumoto}}, \bibinfo {author} {\bibfnamefont {A.-M.~J.}\ \bibnamefont {Zwerver}}, \bibinfo {author} {\bibfnamefont {L.}~\bibnamefont {Tryputen}}, \bibinfo {author} {\bibfnamefont {S.~L.}\ \bibnamefont {de~Snoo}}, \bibinfo {author} {\bibfnamefont {S.~V.}\ \bibnamefont {Amitonov}}, \bibinfo {author} {\bibfnamefont {A.}~\bibnamefont {Sammak}}, \bibinfo {author} {\bibfnamefont {N.}~\bibnamefont {Samkharadze}}, \bibinfo {author} {\bibfnamefont {{\"O}.}~\bibnamefont {G{\"u}l}}, \bibinfo {author} {\bibfnamefont {R.~N.}\ \bibnamefont {Wasserman}}, \emph {et~al.},\ }\href@noop {} {\bibinfo {title} {High-fidelity single-spin shuttling in silicon}} (\bibinfo {year} {2024}),\ \Eprint {https://arxiv.org/abs/2406.07267} {arXiv:2406.07267 [cond-mat.mes-hall]} \BibitemShut {NoStop}%
\bibitem [{\citenamefont {Zwanenburg}\ \emph {et~al.}(2013)\citenamefont {Zwanenburg}, \citenamefont {Dzurak}, \citenamefont {Morello}, \citenamefont {Simmons}, \citenamefont {Hollenberg}, \citenamefont {Klimeck}, \citenamefont {Rogge}, \citenamefont {Coppersmith},\ and\ \citenamefont {Eriksson}}]{zwanenburg2013review}%
  \BibitemOpen
  \bibfield  {author} {\bibinfo {author} {\bibfnamefont {F.~A.}\ \bibnamefont {Zwanenburg}}, \bibinfo {author} {\bibfnamefont {A.~S.}\ \bibnamefont {Dzurak}}, \bibinfo {author} {\bibfnamefont {A.}~\bibnamefont {Morello}}, \bibinfo {author} {\bibfnamefont {M.~Y.}\ \bibnamefont {Simmons}}, \bibinfo {author} {\bibfnamefont {L.~C.~L.}\ \bibnamefont {Hollenberg}}, \bibinfo {author} {\bibfnamefont {G.}~\bibnamefont {Klimeck}}, \bibinfo {author} {\bibfnamefont {S.}~\bibnamefont {Rogge}}, \bibinfo {author} {\bibfnamefont {S.~N.}\ \bibnamefont {Coppersmith}},\ and\ \bibinfo {author} {\bibfnamefont {M.~A.}\ \bibnamefont {Eriksson}},\ }\bibfield  {title} {\bibinfo {title} {Silicon quantum electronics},\ }\href {https://doi.org/10.1103/RevModPhys.85.961} {\bibfield  {journal} {\bibinfo  {journal} {Rev. Mod. Phys.}\ }\textbf {\bibinfo {volume} {85}},\ \bibinfo {pages} {961} (\bibinfo {year} {2013})}\BibitemShut {NoStop}%
\bibitem [{\citenamefont {Burkard}\ \emph {et~al.}(2023)\citenamefont {Burkard}, \citenamefont {Ladd}, \citenamefont {Pan}, \citenamefont {Nichol},\ and\ \citenamefont {Petta}}]{burkard2023semiconductor}%
  \BibitemOpen
  \bibfield  {author} {\bibinfo {author} {\bibfnamefont {G.}~\bibnamefont {Burkard}}, \bibinfo {author} {\bibfnamefont {T.~D.}\ \bibnamefont {Ladd}}, \bibinfo {author} {\bibfnamefont {A.}~\bibnamefont {Pan}}, \bibinfo {author} {\bibfnamefont {J.~M.}\ \bibnamefont {Nichol}},\ and\ \bibinfo {author} {\bibfnamefont {J.~R.}\ \bibnamefont {Petta}},\ }\bibfield  {title} {\bibinfo {title} {Semiconductor spin qubits},\ }\href {https://doi.org/10.1103/RevModPhys.95.025003} {\bibfield  {journal} {\bibinfo  {journal} {Reviews of Modern Physics}\ }\textbf {\bibinfo {volume} {95}},\ \bibinfo {pages} {025003} (\bibinfo {year} {2023})}\BibitemShut {NoStop}%
\bibitem [{\citenamefont {Borselli}\ \emph {et~al.}(2011)\citenamefont {Borselli}, \citenamefont {Ross}, \citenamefont {Kiselev}, \citenamefont {Croke}, \citenamefont {Holabird}, \citenamefont {Deelman}, \citenamefont {Warren}, \citenamefont {Alvarado-Rodriguez}, \citenamefont {Milosavljevic}, \citenamefont {Ku}, \citenamefont {Wong}, \citenamefont {Schmitz}, \citenamefont {Sokolich}, \citenamefont {Gyure},\ and\ \citenamefont {Hunter}}]{borselli2011measurement}%
  \BibitemOpen
  \bibfield  {author} {\bibinfo {author} {\bibfnamefont {M.~G.}\ \bibnamefont {Borselli}}, \bibinfo {author} {\bibfnamefont {R.~S.}\ \bibnamefont {Ross}}, \bibinfo {author} {\bibfnamefont {A.~A.}\ \bibnamefont {Kiselev}}, \bibinfo {author} {\bibfnamefont {E.~T.}\ \bibnamefont {Croke}}, \bibinfo {author} {\bibfnamefont {K.~S.}\ \bibnamefont {Holabird}}, \bibinfo {author} {\bibfnamefont {P.~W.}\ \bibnamefont {Deelman}}, \bibinfo {author} {\bibfnamefont {L.~D.}\ \bibnamefont {Warren}}, \bibinfo {author} {\bibfnamefont {I.}~\bibnamefont {Alvarado-Rodriguez}}, \bibinfo {author} {\bibfnamefont {I.}~\bibnamefont {Milosavljevic}}, \bibinfo {author} {\bibfnamefont {F.~C.}\ \bibnamefont {Ku}}, \bibinfo {author} {\bibfnamefont {W.~S.}\ \bibnamefont {Wong}}, \bibinfo {author} {\bibfnamefont {A.~E.}\ \bibnamefont {Schmitz}}, \bibinfo {author} {\bibfnamefont {M.}~\bibnamefont {Sokolich}}, \bibinfo {author} {\bibfnamefont {M.~F.}\ \bibnamefont {Gyure}},\ and\ \bibinfo {author} {\bibfnamefont {A.~T.}\ \bibnamefont
  {Hunter}},\ }\bibfield  {title} {\bibinfo {title} {Measurement of valley splitting in high-symmetry {S}i/{S}i{G}e quantum dots},\ }\href {https://doi.org/10.1063/1.3569717} {\bibfield  {journal} {\bibinfo  {journal} {Appl. Phys. Lett.}\ }\textbf {\bibinfo {volume} {98}},\ \bibinfo {pages} {123118} (\bibinfo {year} {2011})}\BibitemShut {NoStop}%
\bibitem [{\citenamefont {Shi}\ \emph {et~al.}(2011)\citenamefont {Shi}, \citenamefont {Simmons}, \citenamefont {Prance}, \citenamefont {Gamble}, \citenamefont {Friesen}, \citenamefont {Savage}, \citenamefont {Lagally}, \citenamefont {Coppersmith},\ and\ \citenamefont {Eriksson}}]{shi2011tunable}%
  \BibitemOpen
  \bibfield  {author} {\bibinfo {author} {\bibfnamefont {Z.}~\bibnamefont {Shi}}, \bibinfo {author} {\bibfnamefont {C.~B.}\ \bibnamefont {Simmons}}, \bibinfo {author} {\bibfnamefont {J.}~\bibnamefont {Prance}}, \bibinfo {author} {\bibfnamefont {J.~K.}\ \bibnamefont {Gamble}}, \bibinfo {author} {\bibfnamefont {M.}~\bibnamefont {Friesen}}, \bibinfo {author} {\bibfnamefont {D.~E.}\ \bibnamefont {Savage}}, \bibinfo {author} {\bibfnamefont {M.~G.}\ \bibnamefont {Lagally}}, \bibinfo {author} {\bibfnamefont {S.~N.}\ \bibnamefont {Coppersmith}},\ and\ \bibinfo {author} {\bibfnamefont {M.~A.}\ \bibnamefont {Eriksson}},\ }\bibfield  {title} {\bibinfo {title} {Tunable singlet-triplet splitting in a few-electron {Si/SiGe} quantum dot},\ }\href {https://doi.org/10.1063/1.3666232} {\bibfield  {journal} {\bibinfo  {journal} {Appl. Phys. Lett.}\ }\textbf {\bibinfo {volume} {99}},\ \bibinfo {pages} {233108} (\bibinfo {year} {2011})}\BibitemShut {NoStop}%
\bibitem [{\citenamefont {Zajac}\ \emph {et~al.}(2015)\citenamefont {Zajac}, \citenamefont {Hazard}, \citenamefont {Mi}, \citenamefont {Wang},\ and\ \citenamefont {Petta}}]{zajac2015reconfigurable}%
  \BibitemOpen
  \bibfield  {author} {\bibinfo {author} {\bibfnamefont {D.}~\bibnamefont {Zajac}}, \bibinfo {author} {\bibfnamefont {T.}~\bibnamefont {Hazard}}, \bibinfo {author} {\bibfnamefont {X.}~\bibnamefont {Mi}}, \bibinfo {author} {\bibfnamefont {K.}~\bibnamefont {Wang}},\ and\ \bibinfo {author} {\bibfnamefont {J.~R.}\ \bibnamefont {Petta}},\ }\bibfield  {title} {\bibinfo {title} {A reconfigurable gate architecture for {S}i/{S}i{G}e quantum dots},\ }\href {https://doi.org/10.1063/1.4922249} {\bibfield  {journal} {\bibinfo  {journal} {Applied Physics Letters}\ }\textbf {\bibinfo {volume} {106}},\ \bibinfo {pages} {223507} (\bibinfo {year} {2015})}\BibitemShut {NoStop}%
\bibitem [{\citenamefont {Scarlino}\ \emph {et~al.}(2017)\citenamefont {Scarlino}, \citenamefont {Kawakami}, \citenamefont {Jullien}, \citenamefont {Ward}, \citenamefont {Savage}, \citenamefont {Lagally}, \citenamefont {Friesen}, \citenamefont {Coppersmith}, \citenamefont {Eriksson},\ and\ \citenamefont {Vandersypen}}]{scarlino2017dressed}%
  \BibitemOpen
  \bibfield  {author} {\bibinfo {author} {\bibfnamefont {P.}~\bibnamefont {Scarlino}}, \bibinfo {author} {\bibfnamefont {E.}~\bibnamefont {Kawakami}}, \bibinfo {author} {\bibfnamefont {T.}~\bibnamefont {Jullien}}, \bibinfo {author} {\bibfnamefont {D.~R.}\ \bibnamefont {Ward}}, \bibinfo {author} {\bibfnamefont {D.~E.}\ \bibnamefont {Savage}}, \bibinfo {author} {\bibfnamefont {M.~G.}\ \bibnamefont {Lagally}}, \bibinfo {author} {\bibfnamefont {M.}~\bibnamefont {Friesen}}, \bibinfo {author} {\bibfnamefont {S.~N.}\ \bibnamefont {Coppersmith}}, \bibinfo {author} {\bibfnamefont {M.~A.}\ \bibnamefont {Eriksson}},\ and\ \bibinfo {author} {\bibfnamefont {L.~M.~K.}\ \bibnamefont {Vandersypen}},\ }\bibfield  {title} {\bibinfo {title} {Dressed photon-orbital states in a quantum dot: Intervalley spin resonance},\ }\href {https://doi.org/10.1103/PhysRevB.95.165429} {\bibfield  {journal} {\bibinfo  {journal} {Phys. Rev. B}\ }\textbf {\bibinfo {volume} {95}},\ \bibinfo {pages} {165429} (\bibinfo {year} {2017})}\BibitemShut
  {NoStop}%
\bibitem [{\citenamefont {Hollmann}\ \emph {et~al.}(2020)\citenamefont {Hollmann}, \citenamefont {Struck}, \citenamefont {Langrock}, \citenamefont {Schmidbauer}, \citenamefont {Schauer}, \citenamefont {Leonhardt}, \citenamefont {Sawano}, \citenamefont {Riemann}, \citenamefont {Abrosimov}, \citenamefont {Bougeard} \emph {et~al.}}]{hollmann2020large}%
  \BibitemOpen
  \bibfield  {author} {\bibinfo {author} {\bibfnamefont {A.}~\bibnamefont {Hollmann}}, \bibinfo {author} {\bibfnamefont {T.}~\bibnamefont {Struck}}, \bibinfo {author} {\bibfnamefont {V.}~\bibnamefont {Langrock}}, \bibinfo {author} {\bibfnamefont {A.}~\bibnamefont {Schmidbauer}}, \bibinfo {author} {\bibfnamefont {F.}~\bibnamefont {Schauer}}, \bibinfo {author} {\bibfnamefont {T.}~\bibnamefont {Leonhardt}}, \bibinfo {author} {\bibfnamefont {K.}~\bibnamefont {Sawano}}, \bibinfo {author} {\bibfnamefont {H.}~\bibnamefont {Riemann}}, \bibinfo {author} {\bibfnamefont {N.~V.}\ \bibnamefont {Abrosimov}}, \bibinfo {author} {\bibfnamefont {D.}~\bibnamefont {Bougeard}}, \emph {et~al.},\ }\bibfield  {title} {\bibinfo {title} {Large, tunable valley splitting and single-spin relaxation mechanisms in a {S}i/{S}i$_x$ {G}e$_{1-x}$ quantum dot},\ }\href {https://doi.org/10.1103/PhysRevApplied.13.034068} {\bibfield  {journal} {\bibinfo  {journal} {Phys. Rev. Applied}\ }\textbf {\bibinfo {volume} {13}},\ \bibinfo {pages} {034068}
  (\bibinfo {year} {2020})}\BibitemShut {NoStop}%
\bibitem [{\citenamefont {Mi}\ \emph {et~al.}(2017)\citenamefont {Mi}, \citenamefont {P\'eterfalvi}, \citenamefont {Burkard},\ and\ \citenamefont {Petta}}]{mi2017valleyspectroscopy}%
  \BibitemOpen
  \bibfield  {author} {\bibinfo {author} {\bibfnamefont {X.}~\bibnamefont {Mi}}, \bibinfo {author} {\bibfnamefont {C.~G.}\ \bibnamefont {P\'eterfalvi}}, \bibinfo {author} {\bibfnamefont {G.}~\bibnamefont {Burkard}},\ and\ \bibinfo {author} {\bibfnamefont {J.~R.}\ \bibnamefont {Petta}},\ }\bibfield  {title} {\bibinfo {title} {High-resolution valley spectroscopy of {S}i quantum dots},\ }\href {https://doi.org/10.1103/PhysRevLett.119.176803} {\bibfield  {journal} {\bibinfo  {journal} {Phys. Rev. Lett.}\ }\textbf {\bibinfo {volume} {119}},\ \bibinfo {pages} {176803} (\bibinfo {year} {2017})}\BibitemShut {NoStop}%
\bibitem [{\citenamefont {Ferdous}\ \emph {et~al.}(2018)\citenamefont {Ferdous}, \citenamefont {Kawakami}, \citenamefont {Scarlino}, \citenamefont {Nowak}, \citenamefont {Ward}, \citenamefont {Savage}, \citenamefont {Lagally}, \citenamefont {Coppersmith}, \citenamefont {Friesen}, \citenamefont {Eriksson} \emph {et~al.}}]{ferdous2018valley}%
  \BibitemOpen
  \bibfield  {author} {\bibinfo {author} {\bibfnamefont {R.}~\bibnamefont {Ferdous}}, \bibinfo {author} {\bibfnamefont {E.}~\bibnamefont {Kawakami}}, \bibinfo {author} {\bibfnamefont {P.}~\bibnamefont {Scarlino}}, \bibinfo {author} {\bibfnamefont {M.~P.}\ \bibnamefont {Nowak}}, \bibinfo {author} {\bibfnamefont {D.}~\bibnamefont {Ward}}, \bibinfo {author} {\bibfnamefont {D.}~\bibnamefont {Savage}}, \bibinfo {author} {\bibfnamefont {M.}~\bibnamefont {Lagally}}, \bibinfo {author} {\bibfnamefont {S.}~\bibnamefont {Coppersmith}}, \bibinfo {author} {\bibfnamefont {M.}~\bibnamefont {Friesen}}, \bibinfo {author} {\bibfnamefont {M.~A.}\ \bibnamefont {Eriksson}}, \emph {et~al.},\ }\bibfield  {title} {\bibinfo {title} {Valley dependent anisotropic spin splitting in silicon quantum dots},\ }\href {https://doi.org/10.1038/s41534-018-0075-1} {\bibfield  {journal} {\bibinfo  {journal} {npj Quantum Information}\ }\textbf {\bibinfo {volume} {4}},\ \bibinfo {pages} {26} (\bibinfo {year} {2018})}\BibitemShut {NoStop}%
\bibitem [{\citenamefont {Mi}\ \emph {et~al.}(2018)\citenamefont {Mi}, \citenamefont {Kohler},\ and\ \citenamefont {Petta}}]{mi2018landauzener}%
  \BibitemOpen
  \bibfield  {author} {\bibinfo {author} {\bibfnamefont {X.}~\bibnamefont {Mi}}, \bibinfo {author} {\bibfnamefont {S.}~\bibnamefont {Kohler}},\ and\ \bibinfo {author} {\bibfnamefont {J.~R.}\ \bibnamefont {Petta}},\ }\bibfield  {title} {\bibinfo {title} {Landau-{Z}ener interferometry of valley-orbit states in {S}i/{S}i{G}e double quantum dots},\ }\href {https://doi.org/10.1103/PhysRevB.98.161404} {\bibfield  {journal} {\bibinfo  {journal} {Phys. Rev. B}\ }\textbf {\bibinfo {volume} {98}},\ \bibinfo {pages} {161404(R)} (\bibinfo {year} {2018})}\BibitemShut {NoStop}%
\bibitem [{\citenamefont {Neyens}\ \emph {et~al.}(2018)\citenamefont {Neyens}, \citenamefont {Foote}, \citenamefont {Thorgrimsson}, \citenamefont {Knapp}, \citenamefont {McJunkin}, \citenamefont {Vandersypen}, \citenamefont {Amin}, \citenamefont {Thomas}, \citenamefont {Clarke}, \citenamefont {Savage}, \citenamefont {Lagally}, \citenamefont {Friesen}, \citenamefont {Coppersmith},\ and\ \citenamefont {Eriksson}}]{neyens2018substrate}%
  \BibitemOpen
  \bibfield  {author} {\bibinfo {author} {\bibfnamefont {S.~F.}\ \bibnamefont {Neyens}}, \bibinfo {author} {\bibfnamefont {R.~H.}\ \bibnamefont {Foote}}, \bibinfo {author} {\bibfnamefont {B.}~\bibnamefont {Thorgrimsson}}, \bibinfo {author} {\bibfnamefont {T.~J.}\ \bibnamefont {Knapp}}, \bibinfo {author} {\bibfnamefont {T.}~\bibnamefont {McJunkin}}, \bibinfo {author} {\bibfnamefont {L.~M.~K.}\ \bibnamefont {Vandersypen}}, \bibinfo {author} {\bibfnamefont {P.}~\bibnamefont {Amin}}, \bibinfo {author} {\bibfnamefont {N.~K.}\ \bibnamefont {Thomas}}, \bibinfo {author} {\bibfnamefont {J.~S.}\ \bibnamefont {Clarke}}, \bibinfo {author} {\bibfnamefont {D.~E.}\ \bibnamefont {Savage}}, \bibinfo {author} {\bibfnamefont {M.~G.}\ \bibnamefont {Lagally}}, \bibinfo {author} {\bibfnamefont {M.}~\bibnamefont {Friesen}}, \bibinfo {author} {\bibfnamefont {S.~N.}\ \bibnamefont {Coppersmith}},\ and\ \bibinfo {author} {\bibfnamefont {M.~A.}\ \bibnamefont {Eriksson}},\ }\bibfield  {title} {\bibinfo {title} {The critical role of
  substrate disorder in valley splitting in {S}i quantum wells},\ }\href {https://doi.org/10.1063/1.5033447} {\bibfield  {journal} {\bibinfo  {journal} {Appl. Phys. Lett.}\ }\textbf {\bibinfo {volume} {112}},\ \bibinfo {pages} {243107} (\bibinfo {year} {2018})}\BibitemShut {NoStop}%
\bibitem [{\citenamefont {Borjans}\ \emph {et~al.}(2019)\citenamefont {Borjans}, \citenamefont {Zajac}, \citenamefont {Hazard},\ and\ \citenamefont {Petta}}]{borjans2019relaxation}%
  \BibitemOpen
  \bibfield  {author} {\bibinfo {author} {\bibfnamefont {F.}~\bibnamefont {Borjans}}, \bibinfo {author} {\bibfnamefont {D.~M.}\ \bibnamefont {Zajac}}, \bibinfo {author} {\bibfnamefont {T.~M.}\ \bibnamefont {Hazard}},\ and\ \bibinfo {author} {\bibfnamefont {J.~R.}\ \bibnamefont {Petta}},\ }\bibfield  {title} {\bibinfo {title} {Single-spin relaxation in a synthetic spin-orbit field},\ }\href {https://doi.org/10.1103/PhysRevApplied.11.044063} {\bibfield  {journal} {\bibinfo  {journal} {Phys. Rev. Applied}\ }\textbf {\bibinfo {volume} {11}},\ \bibinfo {pages} {044063} (\bibinfo {year} {2019})}\BibitemShut {NoStop}%
\bibitem [{\citenamefont {Oh}\ \emph {et~al.}(2021)\citenamefont {Oh}, \citenamefont {Denisov}, \citenamefont {Chen},\ and\ \citenamefont {Petta}}]{oh2021cryogenfree}%
  \BibitemOpen
  \bibfield  {author} {\bibinfo {author} {\bibfnamefont {S.~W.}\ \bibnamefont {Oh}}, \bibinfo {author} {\bibfnamefont {A.~O.}\ \bibnamefont {Denisov}}, \bibinfo {author} {\bibfnamefont {P.}~\bibnamefont {Chen}},\ and\ \bibinfo {author} {\bibfnamefont {J.~R.}\ \bibnamefont {Petta}},\ }\bibfield  {title} {\bibinfo {title} {Cryogen-free scanning gate microscope for the characterization of {S}i/$\text{Si}_{0.7}\text{Ge}_{0.3}$ quantum devices at milli-kelvin temperatures},\ }\href {https://doi.org/10.1063/5.0056648} {\bibfield  {journal} {\bibinfo  {journal} {AIP Adv.}\ }\textbf {\bibinfo {volume} {11}},\ \bibinfo {pages} {125122} (\bibinfo {year} {2021})}\BibitemShut {NoStop}%
\bibitem [{\citenamefont {Paquelet~Wuetz}\ \emph {et~al.}(2022)\citenamefont {Paquelet~Wuetz}, \citenamefont {Losert}, \citenamefont {Koelling}, \citenamefont {Stehouwer}, \citenamefont {Zwerver}, \citenamefont {Philips}, \citenamefont {Madzik}, \citenamefont {Xue}, \citenamefont {Zheng}, \citenamefont {Lodari} \emph {et~al.}}]{paquelet2022atomic}%
  \BibitemOpen
  \bibfield  {author} {\bibinfo {author} {\bibfnamefont {B.}~\bibnamefont {Paquelet~Wuetz}}, \bibinfo {author} {\bibfnamefont {M.~P.}\ \bibnamefont {Losert}}, \bibinfo {author} {\bibfnamefont {S.}~\bibnamefont {Koelling}}, \bibinfo {author} {\bibfnamefont {L.~E.}\ \bibnamefont {Stehouwer}}, \bibinfo {author} {\bibfnamefont {A.-M.~J.}\ \bibnamefont {Zwerver}}, \bibinfo {author} {\bibfnamefont {S.~G.}\ \bibnamefont {Philips}}, \bibinfo {author} {\bibfnamefont {M.~T.}\ \bibnamefont {Madzik}}, \bibinfo {author} {\bibfnamefont {X.}~\bibnamefont {Xue}}, \bibinfo {author} {\bibfnamefont {G.}~\bibnamefont {Zheng}}, \bibinfo {author} {\bibfnamefont {M.}~\bibnamefont {Lodari}}, \emph {et~al.},\ }\bibfield  {title} {\bibinfo {title} {Atomic fluctuations lifting the energy degeneracy in {S}i/{S}i{G}e quantum dots},\ }\href {https://doi.org/10.1038/s41467-022-35458-0} {\bibfield  {journal} {\bibinfo  {journal} {Nature Communications}\ }\textbf {\bibinfo {volume} {13}},\ \bibinfo {pages} {7730} (\bibinfo {year}
  {2022})}\BibitemShut {NoStop}%
\bibitem [{\citenamefont {Losert}\ \emph {et~al.}(2023)\citenamefont {Losert}, \citenamefont {Eriksson}, \citenamefont {Joynt}, \citenamefont {Rahman}, \citenamefont {Scappucci}, \citenamefont {Coppersmith},\ and\ \citenamefont {Friesen}}]{losert2023practical}%
  \BibitemOpen
  \bibfield  {author} {\bibinfo {author} {\bibfnamefont {M.~P.}\ \bibnamefont {Losert}}, \bibinfo {author} {\bibfnamefont {M.~A.}\ \bibnamefont {Eriksson}}, \bibinfo {author} {\bibfnamefont {R.}~\bibnamefont {Joynt}}, \bibinfo {author} {\bibfnamefont {R.}~\bibnamefont {Rahman}}, \bibinfo {author} {\bibfnamefont {G.}~\bibnamefont {Scappucci}}, \bibinfo {author} {\bibfnamefont {S.~N.}\ \bibnamefont {Coppersmith}},\ and\ \bibinfo {author} {\bibfnamefont {M.}~\bibnamefont {Friesen}},\ }\bibfield  {title} {\bibinfo {title} {Practical strategies for enhancing the valley splitting in {S}i/{S}i{G}e quantum wells},\ }\href {https://doi.org/10.1103/PhysRevB.108.125405} {\bibfield  {journal} {\bibinfo  {journal} {Phys. Rev. B}\ }\textbf {\bibinfo {volume} {108}},\ \bibinfo {pages} {125405} (\bibinfo {year} {2023})}\BibitemShut {NoStop}%
\bibitem [{\citenamefont {Lima}\ and\ \citenamefont {Burkard}(2023)}]{lima2023valley}%
  \BibitemOpen
  \bibfield  {author} {\bibinfo {author} {\bibfnamefont {J.~R.~F.}\ \bibnamefont {Lima}}\ and\ \bibinfo {author} {\bibfnamefont {G.}~\bibnamefont {Burkard}},\ }\bibfield  {title} {\bibinfo {title} {Interface and electromagnetic effects in the valley splitting of {S}i quantum dots},\ }\href {https://doi.org/10.1088/2633-4356/acd743} {\bibfield  {journal} {\bibinfo  {journal} {Mater. Quantum. Technol.}\ }\textbf {\bibinfo {volume} {3}},\ \bibinfo {pages} {025004} (\bibinfo {year} {2023})}\BibitemShut {NoStop}%
\bibitem [{\citenamefont {Peña}\ \emph {et~al.}(2023)\citenamefont {Peña}, \citenamefont {Koepke}, \citenamefont {Dycus}, \citenamefont {Mounce}, \citenamefont {Baczewski}, \citenamefont {Jacobson},\ and\ \citenamefont {Bussmann}}]{pena2023utilizing}%
  \BibitemOpen
  \bibfield  {author} {\bibinfo {author} {\bibfnamefont {L.~F.}\ \bibnamefont {Peña}}, \bibinfo {author} {\bibfnamefont {J.~C.}\ \bibnamefont {Koepke}}, \bibinfo {author} {\bibfnamefont {J.~H.}\ \bibnamefont {Dycus}}, \bibinfo {author} {\bibfnamefont {A.}~\bibnamefont {Mounce}}, \bibinfo {author} {\bibfnamefont {A.~D.}\ \bibnamefont {Baczewski}}, \bibinfo {author} {\bibfnamefont {N.~T.}\ \bibnamefont {Jacobson}},\ and\ \bibinfo {author} {\bibfnamefont {E.}~\bibnamefont {Bussmann}},\ }\href@noop {} {\bibinfo {title} {Utilizing multimodal microscopy to reconstruct {S}i/{S}i{G}e interfacial atomic disorder and infer its impacts on qubit variability}} (\bibinfo {year} {2023}),\ \Eprint {https://arxiv.org/abs/2306.15646} {arXiv:2306.15646 [cond-mat.mtrl-sci]} \BibitemShut {NoStop}%
\bibitem [{\citenamefont {McJunkin}\ \emph {et~al.}(2022)\citenamefont {McJunkin}, \citenamefont {Harpt}, \citenamefont {Feng}, \citenamefont {Losert}, \citenamefont {Rahman}, \citenamefont {Dodson}, \citenamefont {Wolfe}, \citenamefont {Savage}, \citenamefont {Lagally}, \citenamefont {Coppersmith} \emph {et~al.}}]{mcjunkin2022sige}%
  \BibitemOpen
  \bibfield  {author} {\bibinfo {author} {\bibfnamefont {T.}~\bibnamefont {McJunkin}}, \bibinfo {author} {\bibfnamefont {B.}~\bibnamefont {Harpt}}, \bibinfo {author} {\bibfnamefont {Y.}~\bibnamefont {Feng}}, \bibinfo {author} {\bibfnamefont {M.~P.}\ \bibnamefont {Losert}}, \bibinfo {author} {\bibfnamefont {R.}~\bibnamefont {Rahman}}, \bibinfo {author} {\bibfnamefont {J.}~\bibnamefont {Dodson}}, \bibinfo {author} {\bibfnamefont {M.}~\bibnamefont {Wolfe}}, \bibinfo {author} {\bibfnamefont {D.}~\bibnamefont {Savage}}, \bibinfo {author} {\bibfnamefont {M.}~\bibnamefont {Lagally}}, \bibinfo {author} {\bibfnamefont {S.}~\bibnamefont {Coppersmith}}, \emph {et~al.},\ }\bibfield  {title} {\bibinfo {title} {{S}i{G}e quantum wells with oscillating ge concentrations for quantum dot qubits},\ }\href {https://doi.org/10.1038/s41467-022-35510-z} {\bibfield  {journal} {\bibinfo  {journal} {Nature Communications}\ }\textbf {\bibinfo {volume} {13}},\ \bibinfo {pages} {7777} (\bibinfo {year} {2022})}\BibitemShut {NoStop}%
\bibitem [{\citenamefont {Lima}\ and\ \citenamefont {Burkard}(2024)}]{lima2023valley2}%
  \BibitemOpen
  \bibfield  {author} {\bibinfo {author} {\bibfnamefont {J.~R.~F.}\ \bibnamefont {Lima}}\ and\ \bibinfo {author} {\bibfnamefont {G.}~\bibnamefont {Burkard}},\ }\bibfield  {title} {\bibinfo {title} {Valley splitting depending on the size and location of a silicon quantum dot},\ }\href {https://doi.org/10.1103/PhysRevMaterials.8.036202} {\bibfield  {journal} {\bibinfo  {journal} {Phys. Rev. Mater.}\ }\textbf {\bibinfo {volume} {8}},\ \bibinfo {pages} {036202} (\bibinfo {year} {2024})}\BibitemShut {NoStop}%
\bibitem [{\citenamefont {Volmer}\ \emph {et~al.}(2024)\citenamefont {Volmer}, \citenamefont {Struck}, \citenamefont {Sala}, \citenamefont {Chen}, \citenamefont {Oberl{\"a}nder}, \citenamefont {Offermann}, \citenamefont {Xue}, \citenamefont {Visser}, \citenamefont {Tu}, \citenamefont {Trellenkamp}, \citenamefont {Cywi{\'n}ski}, \citenamefont {Bluhm},\ and\ \citenamefont {Schreiber}}]{volmer2023mapping}%
  \BibitemOpen
  \bibfield  {author} {\bibinfo {author} {\bibfnamefont {M.}~\bibnamefont {Volmer}}, \bibinfo {author} {\bibfnamefont {T.}~\bibnamefont {Struck}}, \bibinfo {author} {\bibfnamefont {A.}~\bibnamefont {Sala}}, \bibinfo {author} {\bibfnamefont {B.}~\bibnamefont {Chen}}, \bibinfo {author} {\bibfnamefont {M.}~\bibnamefont {Oberl{\"a}nder}}, \bibinfo {author} {\bibfnamefont {T.}~\bibnamefont {Offermann}}, \bibinfo {author} {\bibfnamefont {R.}~\bibnamefont {Xue}}, \bibinfo {author} {\bibfnamefont {L.}~\bibnamefont {Visser}}, \bibinfo {author} {\bibfnamefont {J.-S.}\ \bibnamefont {Tu}}, \bibinfo {author} {\bibfnamefont {S.}~\bibnamefont {Trellenkamp}}, \bibinfo {author} {\bibfnamefont {{\L}.}~\bibnamefont {Cywi{\'n}ski}}, \bibinfo {author} {\bibfnamefont {H.}~\bibnamefont {Bluhm}},\ and\ \bibinfo {author} {\bibfnamefont {L.~R.}\ \bibnamefont {Schreiber}},\ }\bibfield  {title} {\bibinfo {title} {Mapping of valley splitting by conveyor-mode spin-coherent electron shuttling},\ }\href
  {https://doi.org/10.1038/s41534-024-00852-7} {\bibfield  {journal} {\bibinfo  {journal} {npj Quantum Information}\ }\textbf {\bibinfo {volume} {10}},\ \bibinfo {pages} {61} (\bibinfo {year} {2024})}\BibitemShut {NoStop}%
\bibitem [{\citenamefont {Dodson}\ \emph {et~al.}(2022)\citenamefont {Dodson}, \citenamefont {Ercan}, \citenamefont {Corrigan}, \citenamefont {Losert}, \citenamefont {Holman}, \citenamefont {McJunkin}, \citenamefont {Edge}, \citenamefont {Friesen}, \citenamefont {Coppersmith},\ and\ \citenamefont {Eriksson}}]{Dodson2022}%
  \BibitemOpen
  \bibfield  {author} {\bibinfo {author} {\bibfnamefont {J.~P.}\ \bibnamefont {Dodson}}, \bibinfo {author} {\bibfnamefont {H.~E.}\ \bibnamefont {Ercan}}, \bibinfo {author} {\bibfnamefont {J.}~\bibnamefont {Corrigan}}, \bibinfo {author} {\bibfnamefont {M.~P.}\ \bibnamefont {Losert}}, \bibinfo {author} {\bibfnamefont {N.}~\bibnamefont {Holman}}, \bibinfo {author} {\bibfnamefont {T.}~\bibnamefont {McJunkin}}, \bibinfo {author} {\bibfnamefont {L.~F.}\ \bibnamefont {Edge}}, \bibinfo {author} {\bibfnamefont {M.}~\bibnamefont {Friesen}}, \bibinfo {author} {\bibfnamefont {S.~N.}\ \bibnamefont {Coppersmith}},\ and\ \bibinfo {author} {\bibfnamefont {M.~A.}\ \bibnamefont {Eriksson}},\ }\bibfield  {title} {\bibinfo {title} {How valley-orbit states in silicon quantum dots probe quantum well interfaces},\ }\href {https://doi.org/10.1103/PhysRevLett.128.146802} {\bibfield  {journal} {\bibinfo  {journal} {Phys. Rev. Lett.}\ }\textbf {\bibinfo {volume} {128}},\ \bibinfo {pages} {146802} (\bibinfo {year} {2022})}\BibitemShut
  {NoStop}%
\bibitem [{\citenamefont {Krzywda}\ and\ \citenamefont {Cywi{\'n}ski}(2024)}]{krzywda2024decoherence}%
  \BibitemOpen
  \bibfield  {author} {\bibinfo {author} {\bibfnamefont {J.~A.}\ \bibnamefont {Krzywda}}\ and\ \bibinfo {author} {\bibfnamefont {{\L}.}~\bibnamefont {Cywi{\'n}ski}},\ }\href@noop {} {\bibinfo {title} {Decoherence of electron spin qubit during transfer between two semiconductor quantum dots at low magnetic fields}} (\bibinfo {year} {2024}),\ \Eprint {https://arxiv.org/abs/2405.12185} {arXiv:2405.12185 [cond-mat.mes-hall]} \BibitemShut {NoStop}%
\bibitem [{\citenamefont {Bosco}\ \emph {et~al.}(2024)\citenamefont {Bosco}, \citenamefont {Zou},\ and\ \citenamefont {Loss}}]{bosco2024high}%
  \BibitemOpen
  \bibfield  {author} {\bibinfo {author} {\bibfnamefont {S.}~\bibnamefont {Bosco}}, \bibinfo {author} {\bibfnamefont {J.}~\bibnamefont {Zou}},\ and\ \bibinfo {author} {\bibfnamefont {D.}~\bibnamefont {Loss}},\ }\bibfield  {title} {\bibinfo {title} {High-fidelity spin qubit shuttling via large spin-orbit interactions},\ }\href {https://doi.org/10.1103/PRXQuantum.5.020353} {\bibfield  {journal} {\bibinfo  {journal} {PRX Quantum}\ }\textbf {\bibinfo {volume} {5}},\ \bibinfo {pages} {020353} (\bibinfo {year} {2024})}\BibitemShut {NoStop}%
\bibitem [{\citenamefont {Zhao}\ and\ \citenamefont {Hu}(2018)}]{zhao2018coherent}%
  \BibitemOpen
  \bibfield  {author} {\bibinfo {author} {\bibfnamefont {X.}~\bibnamefont {Zhao}}\ and\ \bibinfo {author} {\bibfnamefont {X.}~\bibnamefont {Hu}},\ }\href@noop {} {\bibinfo {title} {Coherent electron transport in silicon quantum dots}} (\bibinfo {year} {2018}),\ \Eprint {https://arxiv.org/abs/1803.00749} {arXiv:1803.00749 [cond-mat.mes-hall]} \BibitemShut {NoStop}%
\bibitem [{\citenamefont {Ginzel}\ \emph {et~al.}(2020)\citenamefont {Ginzel}, \citenamefont {Mills}, \citenamefont {Petta},\ and\ \citenamefont {Burkard}}]{ginzel2020spin}%
  \BibitemOpen
  \bibfield  {author} {\bibinfo {author} {\bibfnamefont {F.}~\bibnamefont {Ginzel}}, \bibinfo {author} {\bibfnamefont {A.~R.}\ \bibnamefont {Mills}}, \bibinfo {author} {\bibfnamefont {J.~R.}\ \bibnamefont {Petta}},\ and\ \bibinfo {author} {\bibfnamefont {G.}~\bibnamefont {Burkard}},\ }\bibfield  {title} {\bibinfo {title} {Spin shuttling in a silicon double quantum dot},\ }\href {https://doi.org/10.1103/PhysRevB.102.195418} {\bibfield  {journal} {\bibinfo  {journal} {Physical Review B}\ }\textbf {\bibinfo {volume} {102}},\ \bibinfo {pages} {195418} (\bibinfo {year} {2020})}\BibitemShut {NoStop}%
\bibitem [{\citenamefont {Buonacorsi}\ \emph {et~al.}(2020)\citenamefont {Buonacorsi}, \citenamefont {Shaw},\ and\ \citenamefont {Baugh}}]{buonacorsi2020simulated}%
  \BibitemOpen
  \bibfield  {author} {\bibinfo {author} {\bibfnamefont {B.}~\bibnamefont {Buonacorsi}}, \bibinfo {author} {\bibfnamefont {B.}~\bibnamefont {Shaw}},\ and\ \bibinfo {author} {\bibfnamefont {J.}~\bibnamefont {Baugh}},\ }\bibfield  {title} {\bibinfo {title} {Simulated coherent electron shuttling in silicon quantum dots},\ }\href {https://doi.org/10.1103/PhysRevB.102.125406} {\bibfield  {journal} {\bibinfo  {journal} {Physical Review B}\ }\textbf {\bibinfo {volume} {102}},\ \bibinfo {pages} {125406} (\bibinfo {year} {2020})}\BibitemShut {NoStop}%
\bibitem [{\citenamefont {Li}\ \emph {et~al.}(2017)\citenamefont {Li}, \citenamefont {Barnes}, \citenamefont {Kestner},\ and\ \citenamefont {Sarma}}]{li2017intrinsic}%
  \BibitemOpen
  \bibfield  {author} {\bibinfo {author} {\bibfnamefont {X.}~\bibnamefont {Li}}, \bibinfo {author} {\bibfnamefont {E.}~\bibnamefont {Barnes}}, \bibinfo {author} {\bibfnamefont {J.~P.}\ \bibnamefont {Kestner}},\ and\ \bibinfo {author} {\bibfnamefont {S.~D.}\ \bibnamefont {Sarma}},\ }\bibfield  {title} {\bibinfo {title} {Intrinsic errors in transporting a single-spin qubit through a double quantum dot},\ }\href {https://doi.org/10.1103/PhysRevA.96.012309} {\bibfield  {journal} {\bibinfo  {journal} {Physical Review A}\ }\textbf {\bibinfo {volume} {96}},\ \bibinfo {pages} {012309} (\bibinfo {year} {2017})}\BibitemShut {NoStop}%
\bibitem [{\citenamefont {Ruskov}\ \emph {et~al.}(2018)\citenamefont {Ruskov}, \citenamefont {Veldhorst}, \citenamefont {Dzurak},\ and\ \citenamefont {Tahan}}]{ruskov2018electron}%
  \BibitemOpen
  \bibfield  {author} {\bibinfo {author} {\bibfnamefont {R.}~\bibnamefont {Ruskov}}, \bibinfo {author} {\bibfnamefont {M.}~\bibnamefont {Veldhorst}}, \bibinfo {author} {\bibfnamefont {A.~S.}\ \bibnamefont {Dzurak}},\ and\ \bibinfo {author} {\bibfnamefont {C.}~\bibnamefont {Tahan}},\ }\bibfield  {title} {\bibinfo {title} {Electron g-factor of valley states in realistic silicon quantum dots},\ }\href {https://doi.org/10.1103/PhysRevB.98.245424} {\bibfield  {journal} {\bibinfo  {journal} {Physical Review B}\ }\textbf {\bibinfo {volume} {98}},\ \bibinfo {pages} {245424} (\bibinfo {year} {2018})}\BibitemShut {NoStop}%
\bibitem [{\citenamefont {Ungersboeck}\ \emph {et~al.}(2007)\citenamefont {Ungersboeck}, \citenamefont {Dhar}, \citenamefont {Karlowatz}, \citenamefont {Sverdlov}, \citenamefont {Kosina},\ and\ \citenamefont {Selberherr}}]{Ungersboeck2007effect}%
  \BibitemOpen
  \bibfield  {author} {\bibinfo {author} {\bibfnamefont {E.}~\bibnamefont {Ungersboeck}}, \bibinfo {author} {\bibfnamefont {S.}~\bibnamefont {Dhar}}, \bibinfo {author} {\bibfnamefont {G.}~\bibnamefont {Karlowatz}}, \bibinfo {author} {\bibfnamefont {V.}~\bibnamefont {Sverdlov}}, \bibinfo {author} {\bibfnamefont {H.}~\bibnamefont {Kosina}},\ and\ \bibinfo {author} {\bibfnamefont {S.}~\bibnamefont {Selberherr}},\ }\bibfield  {title} {\bibinfo {title} {The effect of general strain on the band structure and electron mobility of silicon},\ }\href {https://doi.org/10.1109/TED.2007.902880} {\bibfield  {journal} {\bibinfo  {journal} {IEEE Transactions on Electron Devices}\ }\textbf {\bibinfo {volume} {54}},\ \bibinfo {pages} {2183} (\bibinfo {year} {2007})}\BibitemShut {NoStop}%
\bibitem [{\citenamefont {Sverdlov}\ and\ \citenamefont {Selberherr}(2008)}]{Sverdlov2008subband}%
  \BibitemOpen
  \bibfield  {author} {\bibinfo {author} {\bibfnamefont {V.}~\bibnamefont {Sverdlov}}\ and\ \bibinfo {author} {\bibfnamefont {S.}~\bibnamefont {Selberherr}},\ }\bibfield  {title} {\bibinfo {title} {Electron subband structure and controlled valley splitting in silicon thin-body {S}{O}{I} {F}{E}{T}s: Two-band k$\cdot$p theory and beyond},\ }\href {https://doi.org/https://doi.org/10.1016/j.sse.2008.06.054} {\bibfield  {journal} {\bibinfo  {journal} {Solid-State Electronics}\ }\textbf {\bibinfo {volume} {52}},\ \bibinfo {pages} {1861} (\bibinfo {year} {2008})},\ \bibinfo {note} {selected Papers from the EUROSOI '08 Conference}\BibitemShut {NoStop}%
\bibitem [{\citenamefont {Adelsberger}\ \emph {et~al.}(2024)\citenamefont {Adelsberger}, \citenamefont {Bosco}, \citenamefont {Klinovaja},\ and\ \citenamefont {Loss}}]{adelsberger2023valleyfree}%
  \BibitemOpen
  \bibfield  {author} {\bibinfo {author} {\bibfnamefont {C.}~\bibnamefont {Adelsberger}}, \bibinfo {author} {\bibfnamefont {S.}~\bibnamefont {Bosco}}, \bibinfo {author} {\bibfnamefont {J.}~\bibnamefont {Klinovaja}},\ and\ \bibinfo {author} {\bibfnamefont {D.}~\bibnamefont {Loss}},\ }\bibfield  {title} {\bibinfo {title} {Valley-free silicon fins caused by shear strain},\ }\href {https://doi.org/10.1103/PhysRevLett.133.037001} {\bibfield  {journal} {\bibinfo  {journal} {Phys. Rev. Lett.}\ }\textbf {\bibinfo {volume} {133}},\ \bibinfo {pages} {037001} (\bibinfo {year} {2024})}\BibitemShut {NoStop}%
\bibitem [{\citenamefont {Woods}\ \emph {et~al.}(2023{\natexlab{a}})\citenamefont {Woods}, \citenamefont {Soomro}, \citenamefont {Joseph}, \citenamefont {Frink}, \citenamefont {Joynt}, \citenamefont {Eriksson},\ and\ \citenamefont {Friesen}}]{woods2023coupling}%
  \BibitemOpen
  \bibfield  {author} {\bibinfo {author} {\bibfnamefont {B.~D.}\ \bibnamefont {Woods}}, \bibinfo {author} {\bibfnamefont {H.}~\bibnamefont {Soomro}}, \bibinfo {author} {\bibfnamefont {E.~S.}\ \bibnamefont {Joseph}}, \bibinfo {author} {\bibfnamefont {C.~C.~D.}\ \bibnamefont {Frink}}, \bibinfo {author} {\bibfnamefont {R.}~\bibnamefont {Joynt}}, \bibinfo {author} {\bibfnamefont {M.~A.}\ \bibnamefont {Eriksson}},\ and\ \bibinfo {author} {\bibfnamefont {M.}~\bibnamefont {Friesen}},\ }\href@noop {} {\bibinfo {title} {Coupling conduction-band valleys in modulated {S}i{G}e heterostructures via shear strain}} (\bibinfo {year} {2023}{\natexlab{a}}),\ \Eprint {https://arxiv.org/abs/2310.18879} {arXiv:2310.18879 [cond-mat.mes-hall]} \BibitemShut {NoStop}%
\bibitem [{\citenamefont {Woods}\ \emph {et~al.}(2023{\natexlab{b}})\citenamefont {Woods}, \citenamefont {Eriksson}, \citenamefont {Joynt},\ and\ \citenamefont {Friesen}}]{Woods:2023p035418}%
  \BibitemOpen
  \bibfield  {author} {\bibinfo {author} {\bibfnamefont {B.~D.}\ \bibnamefont {Woods}}, \bibinfo {author} {\bibfnamefont {M.~A.}\ \bibnamefont {Eriksson}}, \bibinfo {author} {\bibfnamefont {R.}~\bibnamefont {Joynt}},\ and\ \bibinfo {author} {\bibfnamefont {M.}~\bibnamefont {Friesen}},\ }\bibfield  {title} {\bibinfo {title} {Spin-orbit enhancement in {S}i/{S}i{G}e heterostructures with oscillating {G}e concentration},\ }\href {https://doi.org/10.1103/PhysRevB.107.035418} {\bibfield  {journal} {\bibinfo  {journal} {Phys. Rev. B}\ }\textbf {\bibinfo {volume} {107}},\ \bibinfo {pages} {035418} (\bibinfo {year} {2023}{\natexlab{b}})}\BibitemShut {NoStop}%
\bibitem [{\citenamefont {Huang}\ \emph {et~al.}(2024)\citenamefont {Huang}, \citenamefont {Su}, \citenamefont {Lim}, \citenamefont {Feng}, \citenamefont {van Straaten}, \citenamefont {Severin}, \citenamefont {Gilbert}, \citenamefont {Dumoulin~Stuyck}, \citenamefont {Tanttu}, \citenamefont {Serrano}, \citenamefont {Cifuentes}, \citenamefont {Hansen}, \citenamefont {Seedhouse}, \citenamefont {Vahapoglu}, \citenamefont {Leon}, \citenamefont {Abrosimov}, \citenamefont {Pohl}, \citenamefont {Thewalt}, \citenamefont {Hudson}, \citenamefont {Escott}, \citenamefont {Ares}, \citenamefont {Bartlett}, \citenamefont {Morello}, \citenamefont {Saraiva}, \citenamefont {Laucht}, \citenamefont {Dzurak},\ and\ \citenamefont {Yang}}]{huang2023highfidelity}%
  \BibitemOpen
  \bibfield  {author} {\bibinfo {author} {\bibfnamefont {J.~Y.}\ \bibnamefont {Huang}}, \bibinfo {author} {\bibfnamefont {R.~Y.}\ \bibnamefont {Su}}, \bibinfo {author} {\bibfnamefont {W.~H.}\ \bibnamefont {Lim}}, \bibinfo {author} {\bibfnamefont {M.}~\bibnamefont {Feng}}, \bibinfo {author} {\bibfnamefont {B.}~\bibnamefont {van Straaten}}, \bibinfo {author} {\bibfnamefont {B.}~\bibnamefont {Severin}}, \bibinfo {author} {\bibfnamefont {W.}~\bibnamefont {Gilbert}}, \bibinfo {author} {\bibfnamefont {N.}~\bibnamefont {Dumoulin~Stuyck}}, \bibinfo {author} {\bibfnamefont {T.}~\bibnamefont {Tanttu}}, \bibinfo {author} {\bibfnamefont {S.}~\bibnamefont {Serrano}}, \bibinfo {author} {\bibfnamefont {J.~D.}\ \bibnamefont {Cifuentes}}, \bibinfo {author} {\bibfnamefont {I.}~\bibnamefont {Hansen}}, \bibinfo {author} {\bibfnamefont {A.~E.}\ \bibnamefont {Seedhouse}}, \bibinfo {author} {\bibfnamefont {E.}~\bibnamefont {Vahapoglu}}, \bibinfo {author} {\bibfnamefont {R.~C.~C.}\ \bibnamefont {Leon}}, \bibinfo {author}
  {\bibfnamefont {N.~V.}\ \bibnamefont {Abrosimov}}, \bibinfo {author} {\bibfnamefont {H.-J.}\ \bibnamefont {Pohl}}, \bibinfo {author} {\bibfnamefont {M.~L.~W.}\ \bibnamefont {Thewalt}}, \bibinfo {author} {\bibfnamefont {F.~E.}\ \bibnamefont {Hudson}}, \bibinfo {author} {\bibfnamefont {C.~C.}\ \bibnamefont {Escott}}, \bibinfo {author} {\bibfnamefont {N.}~\bibnamefont {Ares}}, \bibinfo {author} {\bibfnamefont {S.~D.}\ \bibnamefont {Bartlett}}, \bibinfo {author} {\bibfnamefont {A.}~\bibnamefont {Morello}}, \bibinfo {author} {\bibfnamefont {A.}~\bibnamefont {Saraiva}}, \bibinfo {author} {\bibfnamefont {A.}~\bibnamefont {Laucht}}, \bibinfo {author} {\bibfnamefont {A.~S.}\ \bibnamefont {Dzurak}},\ and\ \bibinfo {author} {\bibfnamefont {C.~H.}\ \bibnamefont {Yang}},\ }\bibfield  {title} {\bibinfo {title} {High-fidelity spin qubit operation and algorithmic initialization above 1 k},\ }\href {https://doi.org/10.1038/s41586-024-07160-2} {\bibfield  {journal} {\bibinfo  {journal} {Nature}\ }\textbf {\bibinfo {volume}
  {627}},\ \bibinfo {pages} {772} (\bibinfo {year} {2024})}\BibitemShut {NoStop}%
\bibitem [{\citenamefont {Penthorn}\ \emph {et~al.}(2020)\citenamefont {Penthorn}, \citenamefont {Schoenfield}, \citenamefont {Edge},\ and\ \citenamefont {Jiang}}]{penthorn2020direct}%
  \BibitemOpen
  \bibfield  {author} {\bibinfo {author} {\bibfnamefont {N.~E.}\ \bibnamefont {Penthorn}}, \bibinfo {author} {\bibfnamefont {J.~S.}\ \bibnamefont {Schoenfield}}, \bibinfo {author} {\bibfnamefont {L.~F.}\ \bibnamefont {Edge}},\ and\ \bibinfo {author} {\bibfnamefont {H.}~\bibnamefont {Jiang}},\ }\bibfield  {title} {\bibinfo {title} {Direct measurement of electron intervalley relaxation in a {S}i/{S}i-{G}e quantum dot},\ }\href {https://doi.org/10.1103/PhysRevApplied.14.054015} {\bibfield  {journal} {\bibinfo  {journal} {Physical Review Applied}\ }\textbf {\bibinfo {volume} {14}},\ \bibinfo {pages} {054015} (\bibinfo {year} {2020})}\BibitemShut {NoStop}%
\bibitem [{\citenamefont {Yang}\ \emph {et~al.}(2013)\citenamefont {Yang}, \citenamefont {Rossi}, \citenamefont {Ruskov}, \citenamefont {Lai}, \citenamefont {Mohiyaddin}, \citenamefont {Lee}, \citenamefont {Tahan}, \citenamefont {Klimeck}, \citenamefont {Morello},\ and\ \citenamefont {Dzurak}}]{yang2013spin}%
  \BibitemOpen
  \bibfield  {author} {\bibinfo {author} {\bibfnamefont {C.}~\bibnamefont {Yang}}, \bibinfo {author} {\bibfnamefont {A.}~\bibnamefont {Rossi}}, \bibinfo {author} {\bibfnamefont {R.}~\bibnamefont {Ruskov}}, \bibinfo {author} {\bibfnamefont {N.}~\bibnamefont {Lai}}, \bibinfo {author} {\bibfnamefont {F.}~\bibnamefont {Mohiyaddin}}, \bibinfo {author} {\bibfnamefont {S.}~\bibnamefont {Lee}}, \bibinfo {author} {\bibfnamefont {C.}~\bibnamefont {Tahan}}, \bibinfo {author} {\bibfnamefont {G.}~\bibnamefont {Klimeck}}, \bibinfo {author} {\bibfnamefont {A.}~\bibnamefont {Morello}},\ and\ \bibinfo {author} {\bibfnamefont {A.}~\bibnamefont {Dzurak}},\ }\bibfield  {title} {\bibinfo {title} {Spin-valley lifetimes in a silicon quantum dot with tunable valley splitting},\ }\href {https://doi.org/10.1038/ncomms3069} {\bibfield  {journal} {\bibinfo  {journal} {Nature Communications}\ }\textbf {\bibinfo {volume} {4}},\ \bibinfo {pages} {2069} (\bibinfo {year} {2013})}\BibitemShut {NoStop}%
\bibitem [{\citenamefont {Wood}\ and\ \citenamefont {Gambetta}(2018)}]{wood2018quantification}%
  \BibitemOpen
  \bibfield  {author} {\bibinfo {author} {\bibfnamefont {C.~J.}\ \bibnamefont {Wood}}\ and\ \bibinfo {author} {\bibfnamefont {J.~M.}\ \bibnamefont {Gambetta}},\ }\bibfield  {title} {\bibinfo {title} {Quantification and characterization of leakage errors},\ }\href {https://doi.org/10.1103/PhysRevA.97.032306} {\bibfield  {journal} {\bibinfo  {journal} {Physical Review A}\ }\textbf {\bibinfo {volume} {97}},\ \bibinfo {pages} {032306} (\bibinfo {year} {2018})}\BibitemShut {NoStop}%
\bibitem [{\citenamefont {Teske}\ \emph {et~al.}(2022)\citenamefont {Teske}, \citenamefont {Cerfontaine},\ and\ \citenamefont {Bluhm}}]{teske2022qopt}%
  \BibitemOpen
  \bibfield  {author} {\bibinfo {author} {\bibfnamefont {J.~D.}\ \bibnamefont {Teske}}, \bibinfo {author} {\bibfnamefont {P.}~\bibnamefont {Cerfontaine}},\ and\ \bibinfo {author} {\bibfnamefont {H.}~\bibnamefont {Bluhm}},\ }\bibfield  {title} {\bibinfo {title} {qopt: An experiment-oriented software package for qubit simulation and quantum optimal control},\ }\href {https://doi.org/10.1103/PhysRevApplied.17.034036} {\bibfield  {journal} {\bibinfo  {journal} {Physical Review Applied}\ }\textbf {\bibinfo {volume} {17}},\ \bibinfo {pages} {034036} (\bibinfo {year} {2022})}\BibitemShut {NoStop}%
\bibitem [{\citenamefont {Bradbury}\ \emph {et~al.}(2018)\citenamefont {Bradbury}, \citenamefont {Frostig}, \citenamefont {Hawkins}, \citenamefont {Johnson}, \citenamefont {Leary}, \citenamefont {Maclaurin}, \citenamefont {Necula}, \citenamefont {Paszke}, \citenamefont {Vander{P}las}, \citenamefont {Wanderman-{M}ilne},\ and\ \citenamefont {Zhang}}]{jax2018github}%
  \BibitemOpen
  \bibfield  {author} {\bibinfo {author} {\bibfnamefont {J.}~\bibnamefont {Bradbury}}, \bibinfo {author} {\bibfnamefont {R.}~\bibnamefont {Frostig}}, \bibinfo {author} {\bibfnamefont {P.}~\bibnamefont {Hawkins}}, \bibinfo {author} {\bibfnamefont {M.~J.}\ \bibnamefont {Johnson}}, \bibinfo {author} {\bibfnamefont {C.}~\bibnamefont {Leary}}, \bibinfo {author} {\bibfnamefont {D.}~\bibnamefont {Maclaurin}}, \bibinfo {author} {\bibfnamefont {G.}~\bibnamefont {Necula}}, \bibinfo {author} {\bibfnamefont {A.}~\bibnamefont {Paszke}}, \bibinfo {author} {\bibfnamefont {J.}~\bibnamefont {Vander{P}las}}, \bibinfo {author} {\bibfnamefont {S.}~\bibnamefont {Wanderman-{M}ilne}},\ and\ \bibinfo {author} {\bibfnamefont {Q.}~\bibnamefont {Zhang}},\ }\href {http://github.com/google/jax} {\bibinfo {title} {{JAX}: composable transformations of {P}ython+{N}um{P}y programs}} (\bibinfo {year} {2018})\BibitemShut {NoStop}%
\bibitem [{\citenamefont {Yoneda}\ \emph {et~al.}(2018)\citenamefont {Yoneda}, \citenamefont {Takeda}, \citenamefont {Otsuka}, \citenamefont {Nakajima}, \citenamefont {Delbecq}, \citenamefont {Allison}, \citenamefont {Honda}, \citenamefont {Kodera}, \citenamefont {Oda}, \citenamefont {Hoshi} \emph {et~al.}}]{yoneda2018quantum}%
  \BibitemOpen
  \bibfield  {author} {\bibinfo {author} {\bibfnamefont {J.}~\bibnamefont {Yoneda}}, \bibinfo {author} {\bibfnamefont {K.}~\bibnamefont {Takeda}}, \bibinfo {author} {\bibfnamefont {T.}~\bibnamefont {Otsuka}}, \bibinfo {author} {\bibfnamefont {T.}~\bibnamefont {Nakajima}}, \bibinfo {author} {\bibfnamefont {M.~R.}\ \bibnamefont {Delbecq}}, \bibinfo {author} {\bibfnamefont {G.}~\bibnamefont {Allison}}, \bibinfo {author} {\bibfnamefont {T.}~\bibnamefont {Honda}}, \bibinfo {author} {\bibfnamefont {T.}~\bibnamefont {Kodera}}, \bibinfo {author} {\bibfnamefont {S.}~\bibnamefont {Oda}}, \bibinfo {author} {\bibfnamefont {Y.}~\bibnamefont {Hoshi}}, \emph {et~al.},\ }\bibfield  {title} {\bibinfo {title} {A quantum-dot spin qubit with coherence limited by charge noise and fidelity higher than 99.9\%},\ }\href {https://doi.org/10.1038/s41565-017-0014-x} {\bibfield  {journal} {\bibinfo  {journal} {Nature Nanotechnology}\ }\textbf {\bibinfo {volume} {13}},\ \bibinfo {pages} {102} (\bibinfo {year} {2018})}\BibitemShut
  {NoStop}%
\bibitem [{\citenamefont {Struck}\ \emph {et~al.}(2020)\citenamefont {Struck}, \citenamefont {Hollmann}, \citenamefont {Schauer}, \citenamefont {Fedorets}, \citenamefont {Schmidbauer}, \citenamefont {Sawano}, \citenamefont {Riemann}, \citenamefont {Abrosimov}, \citenamefont {Cywi{\'n}ski}, \citenamefont {Bougeard} \emph {et~al.}}]{struck2020low}%
  \BibitemOpen
  \bibfield  {author} {\bibinfo {author} {\bibfnamefont {T.}~\bibnamefont {Struck}}, \bibinfo {author} {\bibfnamefont {A.}~\bibnamefont {Hollmann}}, \bibinfo {author} {\bibfnamefont {F.}~\bibnamefont {Schauer}}, \bibinfo {author} {\bibfnamefont {O.}~\bibnamefont {Fedorets}}, \bibinfo {author} {\bibfnamefont {A.}~\bibnamefont {Schmidbauer}}, \bibinfo {author} {\bibfnamefont {K.}~\bibnamefont {Sawano}}, \bibinfo {author} {\bibfnamefont {H.}~\bibnamefont {Riemann}}, \bibinfo {author} {\bibfnamefont {N.~V.}\ \bibnamefont {Abrosimov}}, \bibinfo {author} {\bibfnamefont {{\L}.}~\bibnamefont {Cywi{\'n}ski}}, \bibinfo {author} {\bibfnamefont {D.}~\bibnamefont {Bougeard}}, \emph {et~al.},\ }\bibfield  {title} {\bibinfo {title} {Low-frequency spin qubit energy splitting noise in highly purified $^{28}${S}i/{S}i{G}e},\ }\href {https://doi.org/10.1038/s41534-020-0276-2} {\bibfield  {journal} {\bibinfo  {journal} {npj Quantum Information}\ }\textbf {\bibinfo {volume} {6}},\ \bibinfo {pages} {40} (\bibinfo {year}
  {2020})}\BibitemShut {NoStop}%
\bibitem [{\citenamefont {Vitanov}(1999)}]{vitanov1999transition}%
  \BibitemOpen
  \bibfield  {author} {\bibinfo {author} {\bibfnamefont {N.~V.}\ \bibnamefont {Vitanov}},\ }\bibfield  {title} {\bibinfo {title} {Transition times in the {L}andau-{Z}ener model},\ }\href {https://doi.org/10.1103/PhysRevA.59.988} {\bibfield  {journal} {\bibinfo  {journal} {Physical Review A}\ }\textbf {\bibinfo {volume} {59}},\ \bibinfo {pages} {988} (\bibinfo {year} {1999})}\BibitemShut {NoStop}%
\bibitem [{\citenamefont {Mullen}\ \emph {et~al.}(1989)\citenamefont {Mullen}, \citenamefont {Ben-Jacob}, \citenamefont {Gefen},\ and\ \citenamefont {Schuss}}]{mullen1989time}%
  \BibitemOpen
  \bibfield  {author} {\bibinfo {author} {\bibfnamefont {K.}~\bibnamefont {Mullen}}, \bibinfo {author} {\bibfnamefont {E.}~\bibnamefont {Ben-Jacob}}, \bibinfo {author} {\bibfnamefont {Y.}~\bibnamefont {Gefen}},\ and\ \bibinfo {author} {\bibfnamefont {Z.}~\bibnamefont {Schuss}},\ }\bibfield  {title} {\bibinfo {title} {Time of zener tunneling},\ }\href {https://doi.org/10.1103/PhysRevLett.62.2543} {\bibfield  {journal} {\bibinfo  {journal} {Phys. Rev. Lett.}\ }\textbf {\bibinfo {volume} {62}},\ \bibinfo {pages} {2543} (\bibinfo {year} {1989})}\BibitemShut {NoStop}%
\bibitem [{\citenamefont {Schaffler}\ \emph {et~al.}(1992)\citenamefont {Schaffler}, \citenamefont {Tobben}, \citenamefont {Herzog}, \citenamefont {Abstreiter},\ and\ \citenamefont {Hollander}}]{schaffler1992mobility}%
  \BibitemOpen
  \bibfield  {author} {\bibinfo {author} {\bibfnamefont {F.}~\bibnamefont {Schaffler}}, \bibinfo {author} {\bibfnamefont {D.}~\bibnamefont {Tobben}}, \bibinfo {author} {\bibfnamefont {H.~J.}\ \bibnamefont {Herzog}}, \bibinfo {author} {\bibfnamefont {G.}~\bibnamefont {Abstreiter}},\ and\ \bibinfo {author} {\bibfnamefont {B.}~\bibnamefont {Hollander}},\ }\bibfield  {title} {\bibinfo {title} {High-electron-mobility {S}i/{S}i{G}e heterostructures: influence of the relaxed {S}i{G}e buffer layer},\ }\href {https://doi.org/10.1088/0268-1242/7/2/014} {\bibfield  {journal} {\bibinfo  {journal} {Semicond Sci Tech}\ }\textbf {\bibinfo {volume} {7}},\ \bibinfo {pages} {260} (\bibinfo {year} {1992})}\BibitemShut {NoStop}%
\bibitem [{\citenamefont {Müller}\ and\ \citenamefont {Schüler}(2023)}]{gstools}%
  \BibitemOpen
  \bibfield  {author} {\bibinfo {author} {\bibfnamefont {S.}~\bibnamefont {Müller}}\ and\ \bibinfo {author} {\bibfnamefont {L.}~\bibnamefont {Schüler}},\ }\href {https://doi.org/10.5281/zenodo.8044720} {\bibinfo {title} {{G}eo{S}tat-{F}ramework/{G}{S}{T}ools: v1.5.0 `{N}ifty {N}eon'}} (\bibinfo {year} {2023})\BibitemShut {NoStop}%
\bibitem [{\citenamefont {Friesen}\ \emph {et~al.}(2007)\citenamefont {Friesen}, \citenamefont {Chutia}, \citenamefont {Tahan},\ and\ \citenamefont {Coppersmith}}]{friesen2007theory}%
  \BibitemOpen
  \bibfield  {author} {\bibinfo {author} {\bibfnamefont {M.}~\bibnamefont {Friesen}}, \bibinfo {author} {\bibfnamefont {S.}~\bibnamefont {Chutia}}, \bibinfo {author} {\bibfnamefont {C.}~\bibnamefont {Tahan}},\ and\ \bibinfo {author} {\bibfnamefont {S.~N.}\ \bibnamefont {Coppersmith}},\ }\bibfield  {title} {\bibinfo {title} {Valley splitting theory of {S}i{G}e/{S}i/{S}i{G}e quantum wells},\ }\href {https://doi.org/10.1103/PhysRevB.75.115318} {\bibfield  {journal} {\bibinfo  {journal} {Phys. Rev. B}\ }\textbf {\bibinfo {volume} {75}},\ \bibinfo {pages} {115318} (\bibinfo {year} {2007})}\BibitemShut {NoStop}%
\bibitem [{\citenamefont {Bloomfield}\ \emph {et~al.}(2016)\citenamefont {Bloomfield}, \citenamefont {Face}, \citenamefont {Guth}, \citenamefont {Kalia}, \citenamefont {Lam},\ and\ \citenamefont {Moss}}]{bloomfield2016number}%
  \BibitemOpen
  \bibfield  {author} {\bibinfo {author} {\bibfnamefont {J.~K.}\ \bibnamefont {Bloomfield}}, \bibinfo {author} {\bibfnamefont {S.~H.~P.}\ \bibnamefont {Face}}, \bibinfo {author} {\bibfnamefont {A.~H.}\ \bibnamefont {Guth}}, \bibinfo {author} {\bibfnamefont {S.}~\bibnamefont {Kalia}}, \bibinfo {author} {\bibfnamefont {C.}~\bibnamefont {Lam}},\ and\ \bibinfo {author} {\bibfnamefont {Z.}~\bibnamefont {Moss}},\ }\href@noop {} {\bibinfo {title} {Number density of peaks in a chi-squared field}} (\bibinfo {year} {2016}),\ \Eprint {https://arxiv.org/abs/1612.03890} {arXiv:1612.03890 [math-ph]} \BibitemShut {NoStop}%
\bibitem [{\citenamefont {{Bardeen}}\ \emph {et~al.}(1986)\citenamefont {{Bardeen}}, \citenamefont {{Bond}}, \citenamefont {{Kaiser}},\ and\ \citenamefont {{Szalay}}}]{bardeen1968statistics}%
  \BibitemOpen
  \bibfield  {author} {\bibinfo {author} {\bibfnamefont {J.~M.}\ \bibnamefont {{Bardeen}}}, \bibinfo {author} {\bibfnamefont {J.~R.}\ \bibnamefont {{Bond}}}, \bibinfo {author} {\bibfnamefont {N.}~\bibnamefont {{Kaiser}}},\ and\ \bibinfo {author} {\bibfnamefont {A.~S.}\ \bibnamefont {{Szalay}}},\ }\bibfield  {title} {\bibinfo {title} {{The Statistics of Peaks of Gaussian Random Fields}},\ }\href {https://doi.org/10.1086/164143} {\bibfield  {journal} {\bibinfo  {journal} {\apj}\ }\textbf {\bibinfo {volume} {304}},\ \bibinfo {pages} {15} (\bibinfo {year} {1986})}\BibitemShut {NoStop}%
\end{thebibliography}%

\end{document}